\documentclass[]{gji}
\usepackage{timet}
\usepackage[colorlinks=True, linkcolor=blue, citecolor=blue, urlcolor=blue]{hyperref}
\usepackage[dvipsnames]{xcolor}
\usepackage{amsmath, amssymb, bbm}
\usepackage{graphicx}
\let\bibhang\relax
\usepackage[]{natbib}
\usepackage{bm}
\usepackage{textcomp}
\usepackage{float}

\usepackage{scalerel}
\usepackage{multirow}
\usepackage{subfig}
\usepackage{mathtools}

\DeclarePairedDelimiter\floor{\lfloor}{\rfloor}

\title[Fast seismic inference using machine learning]
  {Towards fast machine-learning-assisted Bayesian posterior inference of microseismic event location and source mechanism}
\author[D. Piras, A. Spurio Mancini, A. M. G. Ferreira, B. Joachimi, M. Hobson]
  {D. Piras,$^1$\thanks{\href{mailto:d.piras@ucl.ac.uk}{d.piras@ucl.ac.uk}}
  A. Spurio Mancini,$^{2,1,4}$
  A. M. G. Ferreira,$^3$
  B. Joachimi$^1$ and
  M. P. Hobson$^4$\\
  $^1$Department of Physics and Astronomy, University College London, Gower Street, London, WC1E 6BT, UK \\
  $^2$Mullard Space Science Laboratory, University College London, Holmbury St. Mary, Dorking, Surrey, RH5 6NT, UK\\
  $^3$Dept. of Earth Sciences, Faculty of Mathematical \& Physical Sciences, University College London, WC1E 6BT, UK \\
  $^4$Astrophysics Group, Cavendish Laboratory, J. J. Thomson Avenue, Cambridge, CB3 0HE, UK
  }

\begin{document}

\label{firstpage}

\maketitle

\begin{summary}
Bayesian inference applied to microseismic activity monitoring allows the accurate location of microseismic events from recorded seismograms and the estimation of the associated uncertainties. However, the forward modelling of these microseismic events, which is necessary to perform Bayesian source inversion, can be prohibitively expensive in terms of computational resources. A viable solution is to train a surrogate model based on machine learning techniques, to emulate the forward model and thus accelerate Bayesian inference. 
In this paper, we substantially enhance previous work, which considered only sources with isotropic moment tensors. We train a machine learning algorithm on the power spectrum of the recorded pressure wave and show that the trained emulator allows complete and fast event locations for $\textit{any}$ source mechanism. Moreover, we show that our approach is computationally inexpensive, as it can be run in less than 1 hour on a commercial laptop, while yielding accurate results using less than $10^4$ training seismograms. We additionally demonstrate how the trained emulators can be used to identify the source mechanism through the estimation of the Bayesian evidence. Finally, we demonstrate that our approach is robust to real noise as measured in field data. This work lays the foundations for efficient, accurate future joint determinations of event location and moment tensor, and associated uncertainties, which are ultimately key for accurately characterising human-induced and natural earthquakes, and for enhanced quantitative seismic hazard assessments.

\end{summary}


\section{Introduction}
\label{sec:introduction}
Underground human activity, including fluid injection in rocks and mining operations, can cause microseismic events \citep{Majer07, Ellsworth13}. The monitoring of both human-induced and natural microseismicity is critical for understanding seismic hazard \citep[][and references therein]{Brueckl08, Shapiro10, Mukuhira16, Das17}. 
Accurate seismic event locations in space and in time are of paramount importance for reliable seismic monitoring efforts, and are mainly obtained from seismograms recorded on land and/or at the seafloor.

%



Various methods for locating seismic events are available in the literature, dating back to the work of \citet{Geiger10}, and up to today \citep[see e.g.][and references therein, for a recent review]{Vasco19}. One of the most common approaches relies on using the Eikonal equation to determine the theoretical travel time of first seismic arrivals  \citep[see e.g.][]{Noack17, Smith20}, which is compared with real travel times through a direct grid search or more sophisticated inverse modelling techniques \citep{Wuestfeld18}. More accurate source location estimations can be obtained exploiting methods that use the full waveform, even though they generally require heavier computational resources (\citealt{Song11, Li13, Angus14, Li16, Willacy19, Vasco19}; see also \citealt{Li20} for a recent review of waveform-based inversion methods).

A Bayesian approach can also be adopted to solve the location inverse problem \citep{Lomax00, Tarantola05, Staehler14, Staehler16, Pugh16}. In this framework, a posterior distribution of the model parameters (i.e., the event's location and/or the moment tensor) is estimated and used to determine the optimal model parameters and their associated uncertainty.
Markov Chain Monte Carlo \citep[MCMC; see e.g.][for a review]{Craiu14} and nested sampling \citep{Skilling06} techniques are among those employed to sample the posterior distribution. However, these approaches become prohibitive when dealing with a high number of parameters, or when the forward model is computationally expensive to simulate \citep[see e.g.][]{Rajaratnam15, Conrad16, Alsing18}. For these reasons, being able to cheaply and accurately simulate the theoretical waveforms of microseismic events given their location has become paramount in recent years. The forward modelling of microseismic events typically involves solving the elastic wave equation in a wide frequency range given a 3-D heterogeneous density and velocity model for the propagating medium \citep{Das17}, which can be prohibitively expensive.

Machine learning generative models have gained considerable attention in recent years, with applications to many fields ranging from computer vision \citep{Goodfellow14, Gulrajani16} to astrophysics \citep{Auld07, Auld08, SpurioMancini21}, as well as climate science, nuclear physics and drug selection \citep[see e.g.][]{Kasim20, Chenthamarakshan20}. These advances have been enabled by both an increased accessibility to computational resources, as well as by a significant growth in the amount of available data.

In seismology, machine learning techniques have been successfully applied to a wide range of problems \citep[see e.g.][for a recent review]{Bergen19}. For instance, \citet{Zheng18} used a recurrent neural network to pick the arrival times of microseismic (or acoustic) events, while \citet{Zhu18}, \citet{Ross18} and \citet{Zhu19} trained convolutional neural networks (CNNs) to measure P- and S-wave arrival times and determine high-level features with high precision, often outperforming the measurements performed manually or semi-automatically by expert seismologists. \citet{Mousavi20} further proposed a deep-learning model for simultaneous earthquake detection and phase picking based on an attention mechanism, while \citet{Li22} trained multiple CNNs both to detect whether there is an acoustic emission data event in some recorded data, and to pick the arrival time of the P-wave.

Looking in particular at generative models, \citet{Das17} developed an optimised approach to simulate microseismic event propagation that, for each event location and given a physical model for the propagating medium, produces the corresponding seismogram in $\mathcal{O}(1 \ \rm{h})$ using a Tesla graphics processing unit (GPU) and the software \textsc{k-wave} \citep{Treeby14}. Subsequently, \citet{Das18} and \citet{ASM20} (D18 and SM20 hereafter, respectively) showed the limitations of this direct approach, and presented an alternative whereby the mapping is learnt using machine learning techniques. In particular, D18 showed how Gaussian Processes \citep[GPs,][]{Rasmussen05} can be used to learn an accurate surrogate model, while SM20 demonstrated the effectiveness of a variety of machine learning algorithms as emulators, and showed how their surrogate model yields an accurate estimate of the posterior distribution of an event's location in a fraction of the time required by the D18 method. 

D18 and SM20, however, only applied their methodologies to  microseismic events with an isotropic source mechanism. It is well known that any source mechanism can be mathematically decomposed into three components: isotropic (ISO), double couple (DC), and compensated linear vector dipole (CLVD) \citep[see e.g.][]{Knopoff70, Vavrycuk01, Vavrycuk05, Vavricuk15}. The pure ISO source is associated with implosive or explosive forces, while the pure DC source is caused by shear faulting. 

In this paper, we present an approach that aims at learning the direct mapping between event locations and seismograms for any microseismic source type (ISO, DC and CLVD). We show that it is sufficient to consider the power spectrum of the recorded pressure waves in order to distinguish different seismograms (as already investigated e.g. in \citealt{Pratt99a, Pratt99b, Tao13,Jakobsen15}), and we train a simple machine learning algorithm to learn this mapping efficiently. Moreover, we demonstrate how our method allows the accurate inference of the posterior distribution of the coordinates of a single source in $\mathcal{O}(0.1 \ \rm{h})$ on a commercial laptop, thus paving the way for fast and computationally cheap joint determinations of event locations and seismic moment tensors.
Finally, we show how we can use the trained emulators to identify the source mechanism through Bayesian evidence estimation, thus demonstrating the versatility of our Bayesian approach.

Regarding the structure of this paper,  in Sect.~\ref{sec:data} we describe the data we use in this work. In Sect.~\ref{sec:pgi} we explain our inversion approach, describe what preprocessing steps we perform and recall the details of the generative method we employ. We show its performance at both training and inference time in Sect.~\ref{sec:results}, where we also compare our approach to standard arrival time analysis \citep{Lomax09} and include an experiment with real noise from Groningen field data \citep{Smith19thesis}, thus showing the robustness of our method. Finally, in Sect.~\ref{sec:conclusions} we discuss our results and provide an outlook on possible extensions of this work.

\section{Data}
\label{sec:data}

\begin{figure*}
    \centering
    \includegraphics[scale=0.07]{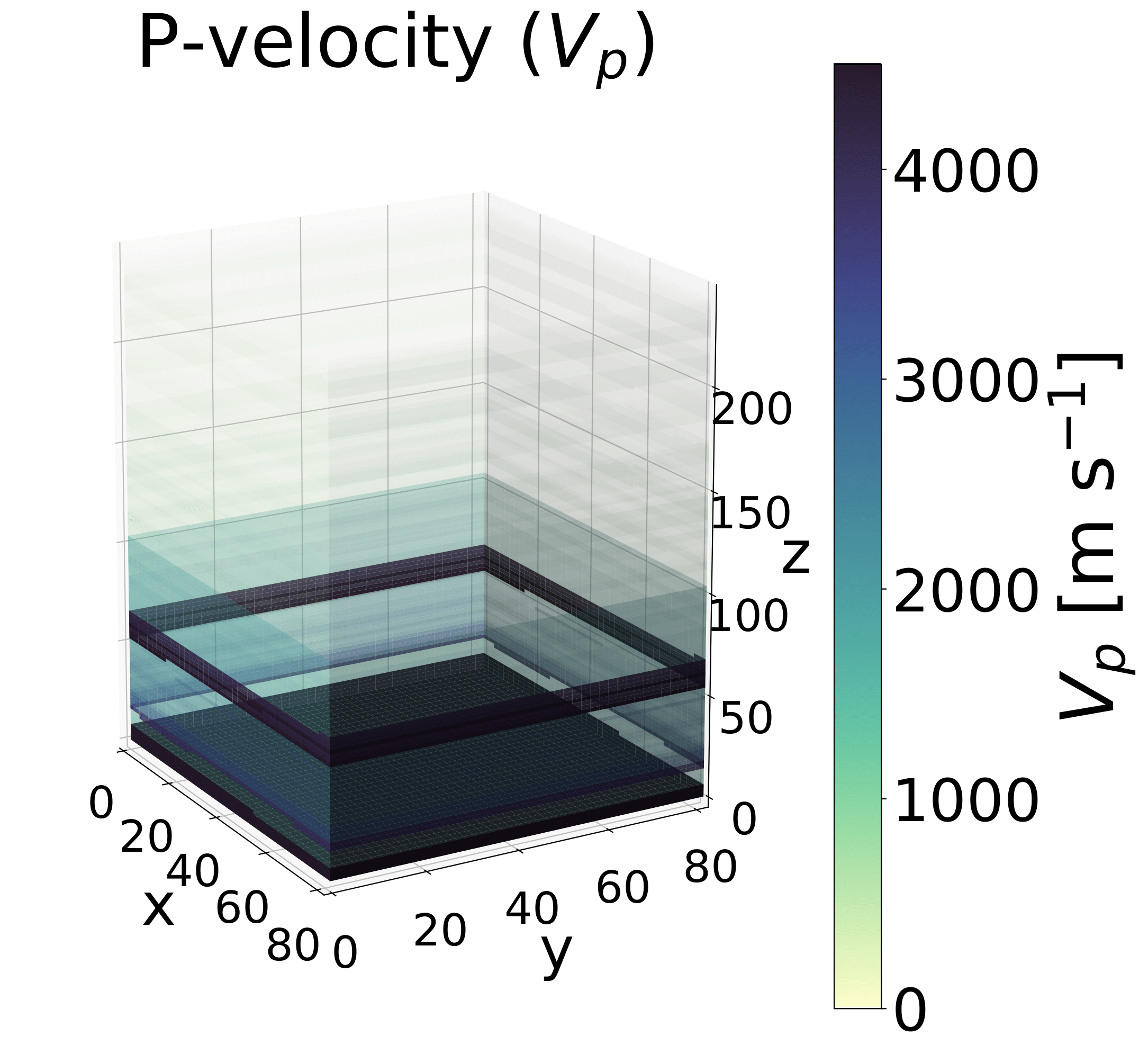}
    \includegraphics[scale=0.195]{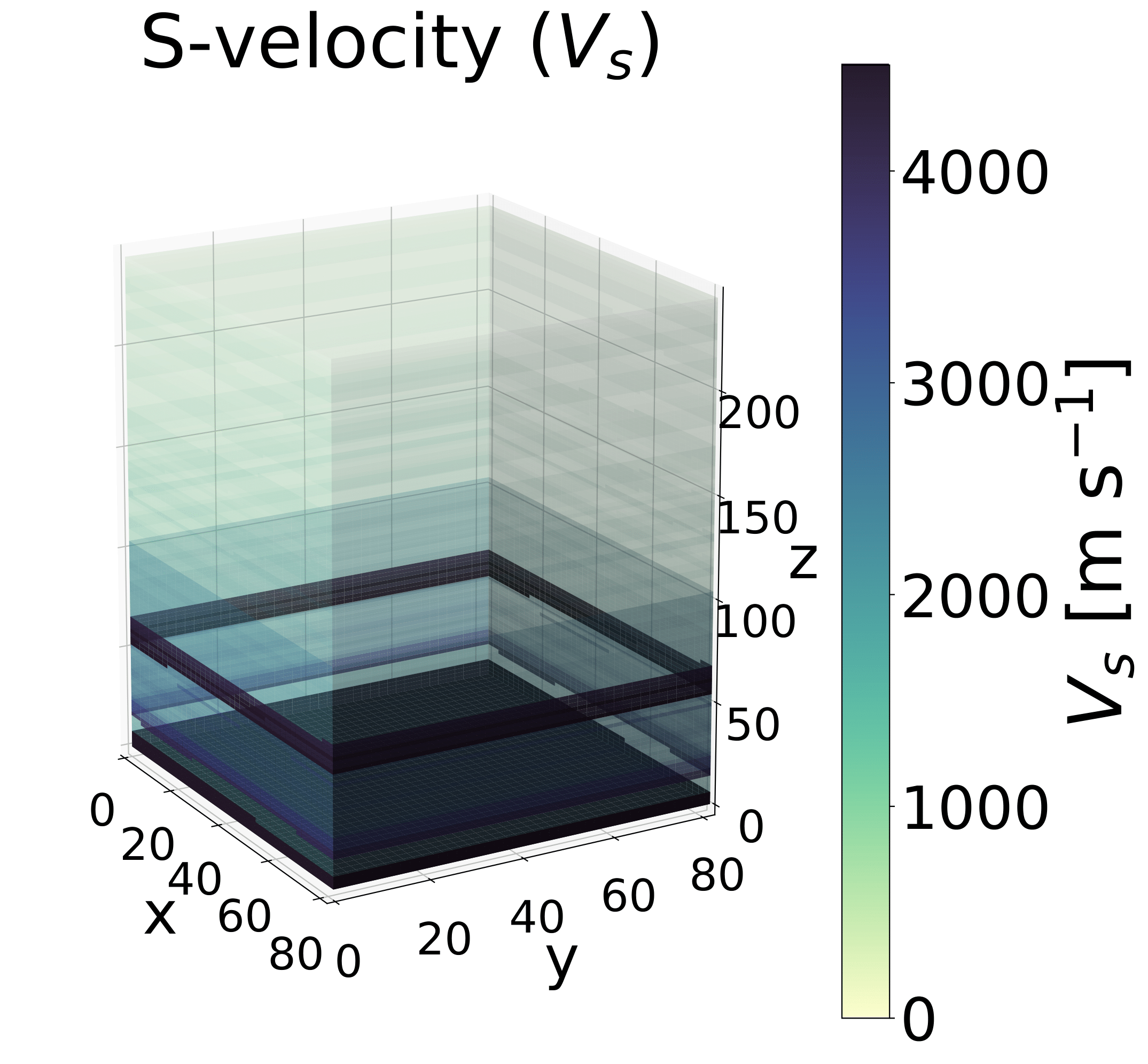}
    \includegraphics[scale=0.07]{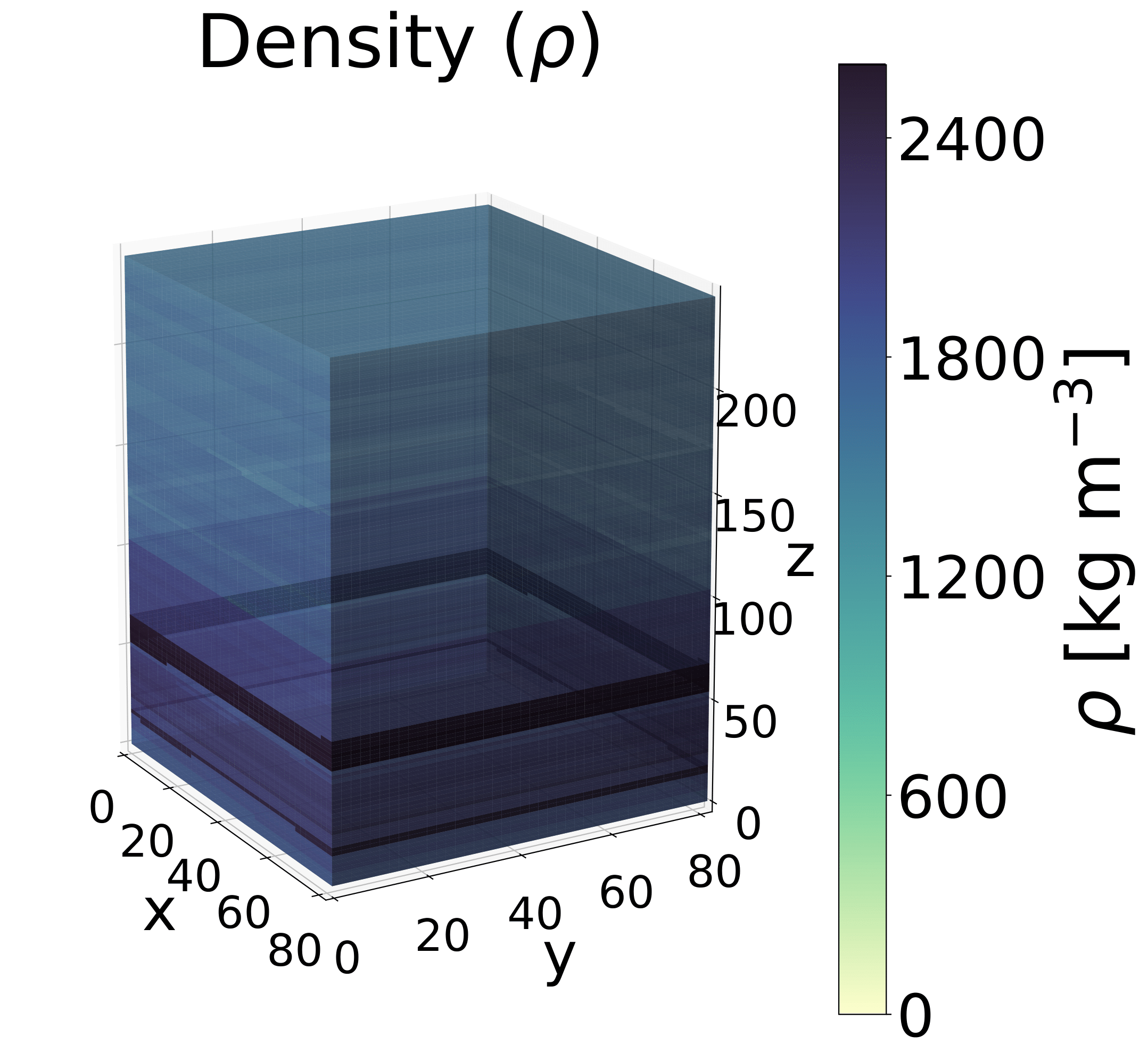}
    \caption{P-wave velocity ($V_p$), S-wave velocity ($V_s$) and density ($\rho$) models of the simulated domain we consider in this work. The models are specified as 3-D grids of voxels, with size 81 $\times$ 81 $\times$ 301 points, corresponding to a real geological model of size 1 km $\times$ 1 km $\times$ 3 km. We observe that our model has a layered structure, with variation along the vertical dimension more marked than along horizontal planes. The plots were adapted from figure 1 in \citet{ASM20}.}
    \label{fig:velocity_model}
\end{figure*}

In this work, we consider the same data framework as D18 and SM20, starting from 3-D heterogeneous density and velocity models for the propagating medium, which we show in Fig.~\ref{fig:velocity_model}. The model, which is discretised on a 3-D grid of voxels, specifies the values of the density $\rho$ of the propagating medium, as well as the propagation velocities for P- and S-waves ($V_p$, $V_s$). We assume that sensors are placed at the seabed, and that they record both pressure and three-component particle velocity of the propagating medium (even though we will use only the former, as we explain later on). As anticipated, our aim is to apply our method to any source mechanism, so we will consider a more general generation procedure than previous work. Unlike D18 and SM20, who only considered isotropic sources, we take the microseismic moment tensor to be one of three types, which we denote as $\textbf{M}_{\rm{ISO}}$, $\textbf{M}_{\rm{DC}}$ and $\textbf{M}_{\rm{CLVD}}$. Following \citet{Vavrycuk05} and \citet{Li15}, we define these quantities as:

\begin{align}
    &\textbf{M}_{\rm{ISO}}=
    \begin{bmatrix}
    M_{11} & 0 & 0  \\
    0 & M_{22} & 0  \\
    0 & 0 & M_{33}  
    \end{bmatrix} \label{eq:M_ISO}, \\ 
    &\textbf{M}_{\rm{DC}}=
    \begin{bmatrix}
    0 & M_{12} & 0  \\
    M_{21} & 0 & 0  \\
    0 & 0 & 0  
    \end{bmatrix}, \label{eq:M_DC} \\
    &\textbf{M}_{\rm{CLVD}}=
    \begin{bmatrix}
    M_{11} & 0 & 0  \\
    0 & M_{22} & 0  \\
    0 & 0 & -2M_{33}  
    \end{bmatrix},
\label{eq:M_CLVD}
\end{align}
where each $M_{ij}$ represents a different couple of forces. We additionally assume $M_{11}=M_{22}=M_{33}=M_{12}=M_{21}=1$ MPa, which is a realistic assumption following \citet{Collettini02}; we anticipate that while in this work we fix the $M_{ij}$ coefficients, we will explore the joint determination of moment tensor components and event coordinates in future work. In practice, isotropic (ISO) events are characterised by a single (explosive or implosive) P-wave, while double couple (DC) events are linked to shear stress and are characterised by both a P- and S-wave, with comparable amplitudes. Similarly to isotropic events, compensated linear vector dipole (CLVD) events display an often dominant principal wave, whose amplitude is however much smaller than the isotropic one, and a non-negligible S-wave.

\begin{figure}
    \centering
    \includegraphics[width=\columnwidth, keepaspectratio]{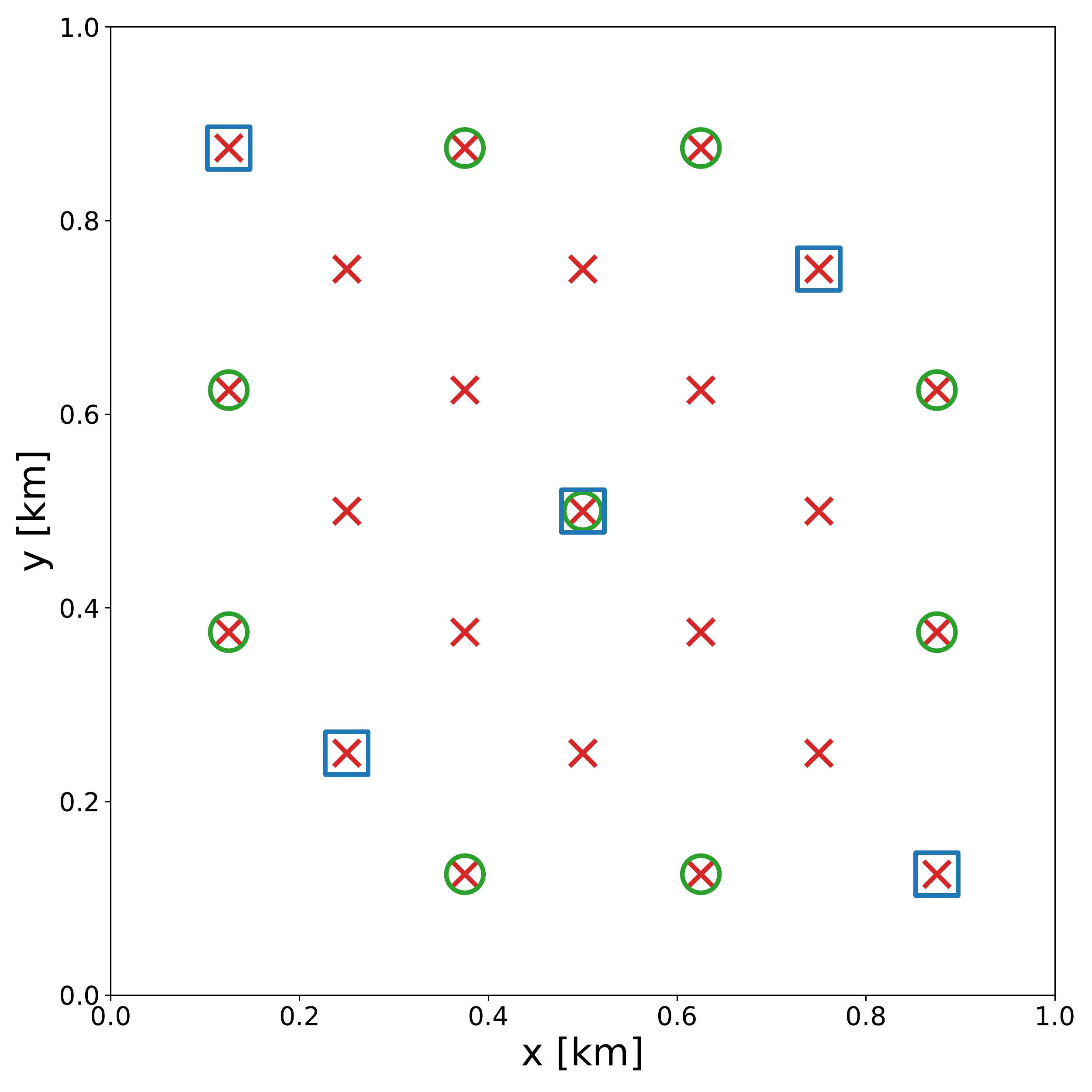}
    \caption{Projection of the positions of the 23 receivers on the $x$-$y$ plane; $z=2.43$ km corresponds to the seabed, where all sensors lie. Each receiver records the acoustic pressure wave and particle velocity generated by a microseismic event below the seabed. The red crosses indicate all 23 receivers, while green circles and blue squares (9 and 5 receivers, respectively) refer to subsets of receivers we used to test the robustness of our method, as shown in Sect.~\ref{sec:inference_results}.}
    \label{fig:rec_pos}
\end{figure}

We generate forward simulations using the GPU implementation of \citet{Das17}, employing different types of GPUs with memory size ranging from 5 GB to 11 GB. The precise GPU cards used to generate the data are Tesla K20, Tesla C2075, GeForce RTX 2080 Ti and GeForce GTX 1080 Ti, which carry a different number of CUDA cores each, ranging from 500 to 5000. We observe that the speed of the generation scales linearly with the number of cores available; however, given that we produce the data on a shared cluster, we cannot always choose which card to employ. Note that the software with which we work is optimised for GPUs, and therefore cannot be expected to scale similarly when running on Central Processing Units (CPUs) or Tensor Processing Units (TPUs). 

For each source type, we produce 10\,000 events corresponding to different source locations, which are randomly sampled using Latin Hypercube Sampling on a 3-D grid of 81 $\times$ 81 $\times$ 301 points, corresponding to a real geological model of size 1 km $\times$ 1 km $\times$ 3 km. The total time to generate these data using the hardware specified above is about 150 h for each source type. 
We consider 23 receivers in total, whose position is depicted in Fig.~\ref{fig:rec_pos}, even though we stress that we will not focus on finding the optimal geometry of the sensors in this work. Each receiver records each of the 10\,000 events independently, thus producing a set of 10\,000 waveforms for each source type at each receiver. In Sect.~\ref{sec:inference_results} we will present the results of a full analysis of the dependence of the posterior distribution on the number of training points, the number of receivers and the noise level, in order to demonstrate the robustness of our approach.

\begin{figure*}
    \centering
    \includegraphics[width=\textwidth]{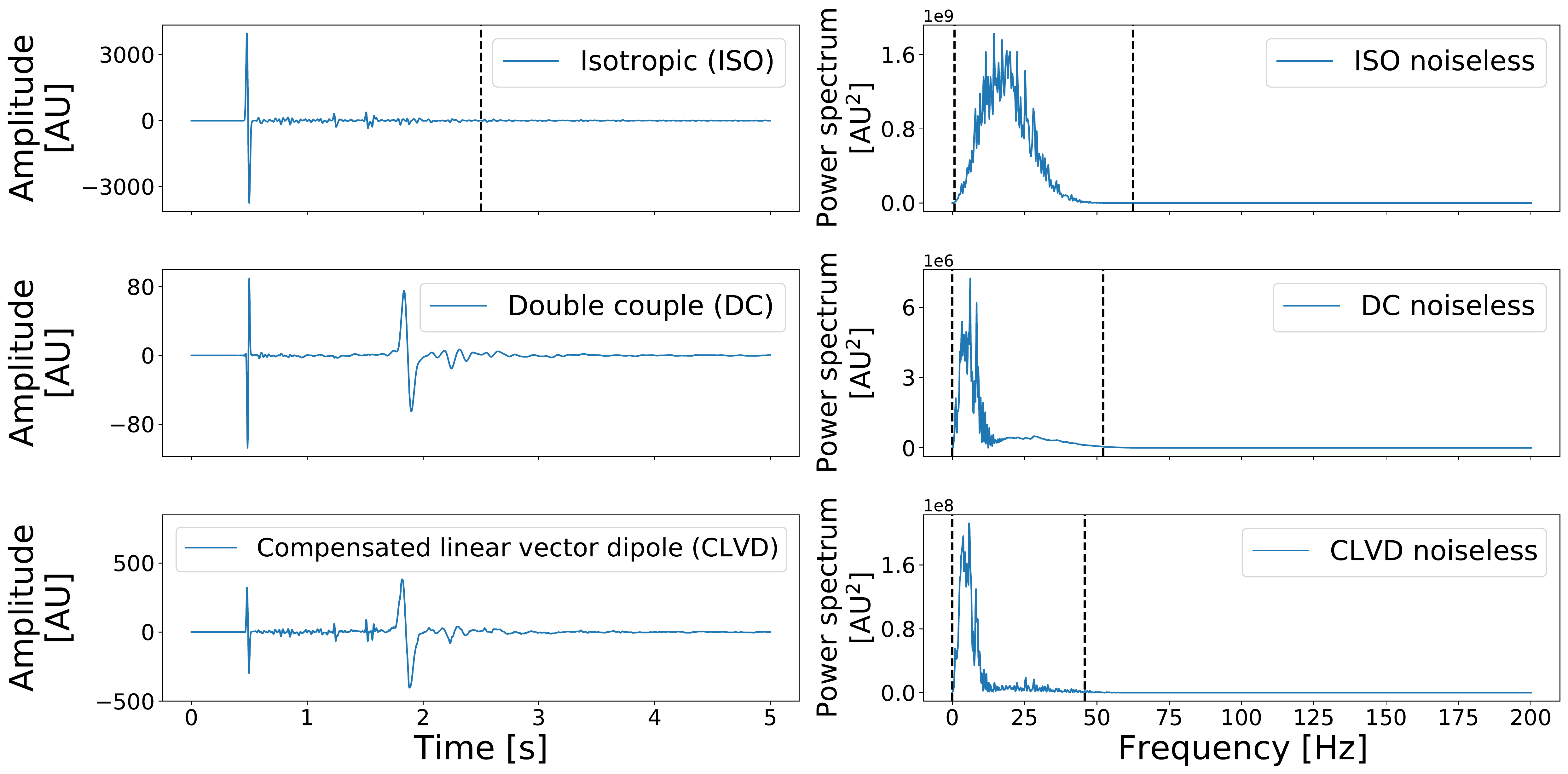}
    \caption{\textit{Left column:} Example acoustic pressure wave for each different type of moment tensor: isotropic (ISO), double couple (DC) and compensated linear vector dipole (CLVD). These seismograms correspond to a source location of $(x,y,z)=(0.55 \ \rm{km},0.73 \ \rm{km},1.8 \ \rm{km})$ as recorded by a receiver in $(x,y,z)=(0.13 \ \rm{km},0.38 \ \rm{km},2.43 \ \rm{km})$. The seismograms' amplitude is measured in arbitrary units of pressure. Note the different scales for each source mechanism. The vertical dashed black line in the top panel indicates a cut we perform for the isotropic sources only, based on the left panel of Fig.~\ref{fig:cuts}. \textit{Right column:} The corresponding power spectra, calculated as described in Sect.~\ref{sec:preprocessing}. No noise is added when training the emulator, while some noise is introduced when doing inference, as described in Sec.~\ref{sec:inference}. The vertical dashed lines indicate a frequency cut we perform to further reduce the number of features and to be robust to noise, based on the right panel of Fig.~\ref{fig:cuts}.}
    \label{fig:data_comparison}
\end{figure*}

The time interval (sampling rate) for the solution of the elastic wave equation is $0.5$ ms (2 kHz), and the total length of each seismic event is $5$ s. After generation, all seismograms are downsampled to a time resolution of $2.5$ ms (400 Hz) to reduce computational storage. In this way, each seismic trace is ultimately a time series composed of $N_t = 2001$ time samples: we show an example for each source mechanism in the left column of Fig.~\ref{fig:data_comparison}. 
Finally, note that we consider the seismograms to be noiseless at training time, while some noise is added to the simulated recorded event when performing inference on the coordinates' posterior distribution, as we are working in a Bayesian framework, detailed in Sect.~\ref{sec:inference}. We will discuss and show the effect of the noise level in Sect.~\ref{sec:preprocessing} and Sect.~\ref{sec:inference_results}, respectively.

\section{Inversion approach}
\label{sec:pgi}
\begin{figure*}
    \centering
    \includegraphics[width=\textwidth]{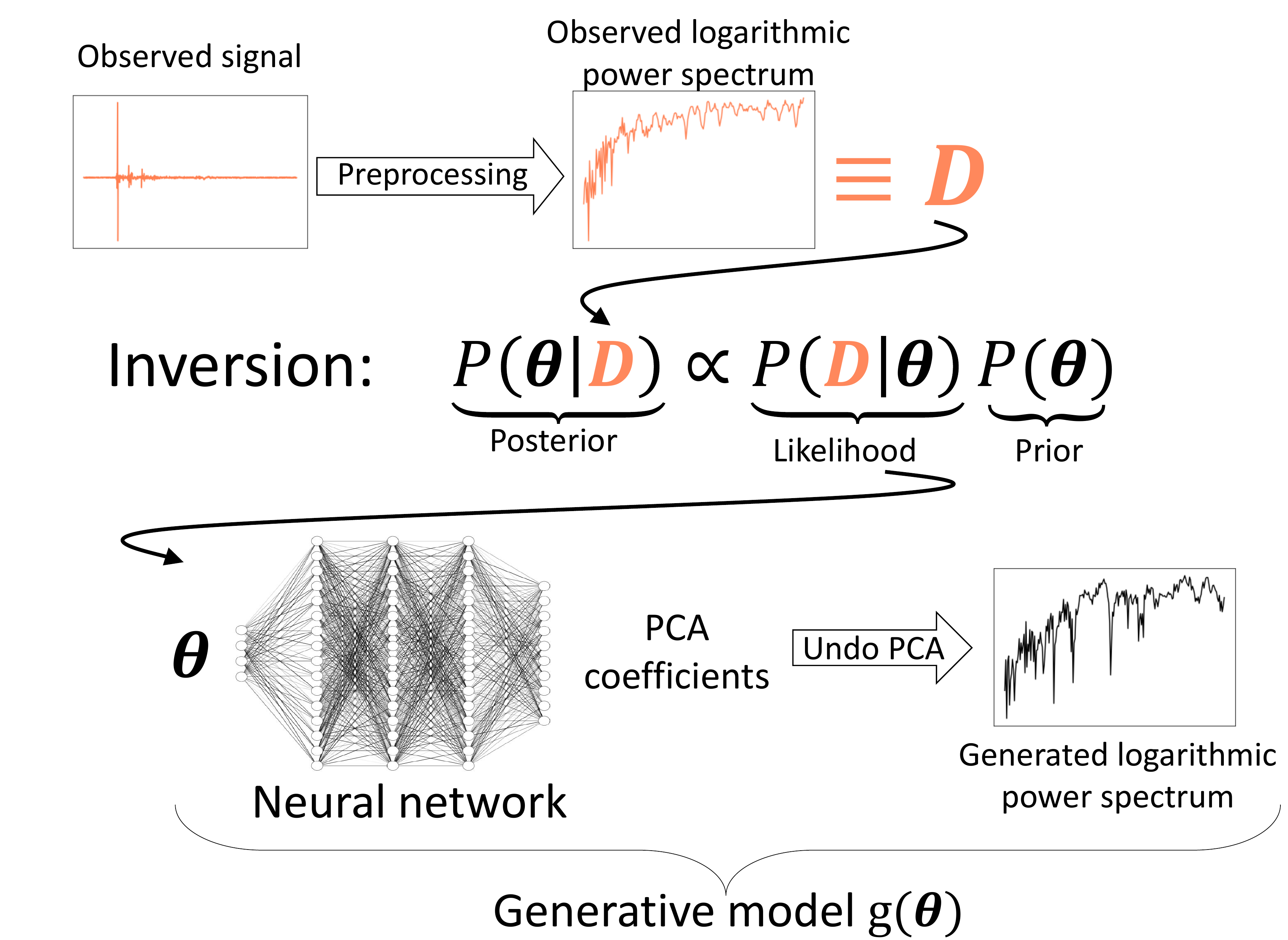}
    \caption{Workflow of our inversion approach. The observed event, indicated in coral in the top left corner, is preprocessed according to the steps described in Sect.~\ref{sec:preprocessing} in order to obtain a logarithmic power spectrum. Our goal is to obtain the posterior distribution of the coordinates $\theta$ of the event's location beneath the surface, as detailed in Sect.~\ref{sec:inference}. In order to sample from the posterior distribution, we specify a data likelihood, and use a generative model $g(\boldsymbol{\theta})$ to accelerate the evaluation of the likelihood. As our generative model, which we describe in Sect.~\ref{sec:gen_model}, we employ a feedforward neural network that maps source coordinates to the principal components of the logarithmic power spectra. The neural network sketch has been drawn using \textsc{NN-SVG} \citep{LeNail19}. Note that the form of Bayes' theorem in this picture is a simplified version of Eq.~\ref{eq:bayes}.}
    \label{fig:workflow}
\end{figure*}

Our goal is to perform Bayesian inference on the source location of a microseismic event. In order to do so, we will employ a neural network (NN) as the forward model to parametrise the mapping between the coordinates and the principal components of the power spectra of the seismograms. Our inversion approach is summarised in Fig.~\ref{fig:workflow}.


\subsection{Inference}
\label{sec:inference}
We recall here the basic assumptions of our Bayesian analysis. Given a set of parameters $\boldsymbol{\theta}$ (the coordinates, in our case), their posterior distribution given some data $\boldsymbol{D}$ and some hypothesis $\mathcal{H}$ can be written using Bayes' theorem \citep[see e.g.][]{Bishop06}:
\begin{align}
\label{eq:bayes}
    \mathrm{Pr} \left( \boldsymbol{\theta} | \boldsymbol{D}, \mathcal{H} \right) = \frac{ \mathrm{Pr} \left( \boldsymbol{D} | \boldsymbol{\theta}, \mathcal{H}\right) \mathrm{Pr} \left( \boldsymbol{\theta} | \mathcal{H} \right)  }{ \mathrm{Pr} \left( \boldsymbol{D} | \mathcal{H}\right) } \ ,
\end{align}
which expresses the posterior distribution $\mathrm{Pr} \left( \boldsymbol{\theta} | \boldsymbol{D}, \mathcal{H} \right)$ as the product of the likelihood $\mathrm{Pr} \left( \boldsymbol{D} | \boldsymbol{\theta}, \mathcal{H} \right)$ and the parameters' prior $\mathrm{Pr} \left( \boldsymbol{\theta} | \mathcal{H}\right)$, divided by the evidence $\mathrm{Pr} \left( \boldsymbol{D} | \mathcal{H}\right)$. For the purposes of inference, we will ignore this last term, as it is just a normalisation factor independent of $\boldsymbol{\theta}$; however, in Sect.~\ref{sec:evidence} we show how the evidence can be used to perform model selection, which is another advantage of working in a Bayesian framework. In order to sample the posterior distribution of the source coordinates, we employ nested sampling \citep{Skilling06}, as implemented in \textsc{PyMultiNest}\footnote{\url{https://github.com/JohannesBuchner/PyMultiNest}}~\citep{Buchner14}, the Python interface to \textsc{MultiNest} \citep{Feroz08}. We choose nested sampling over Metropolis-Hastings sampling or other MCMC techniques as it generally converges faster \citep{Allison14} and provides an estimate of the evidence. For the prior, we assume a uniform distribution in the range of the physical model ($[0, 1] \times [0, 1] \times [0, 2.43]$, with units in kilometres).

\begin{figure}
    \centering
    \includegraphics[width=\columnwidth, keepaspectratio]{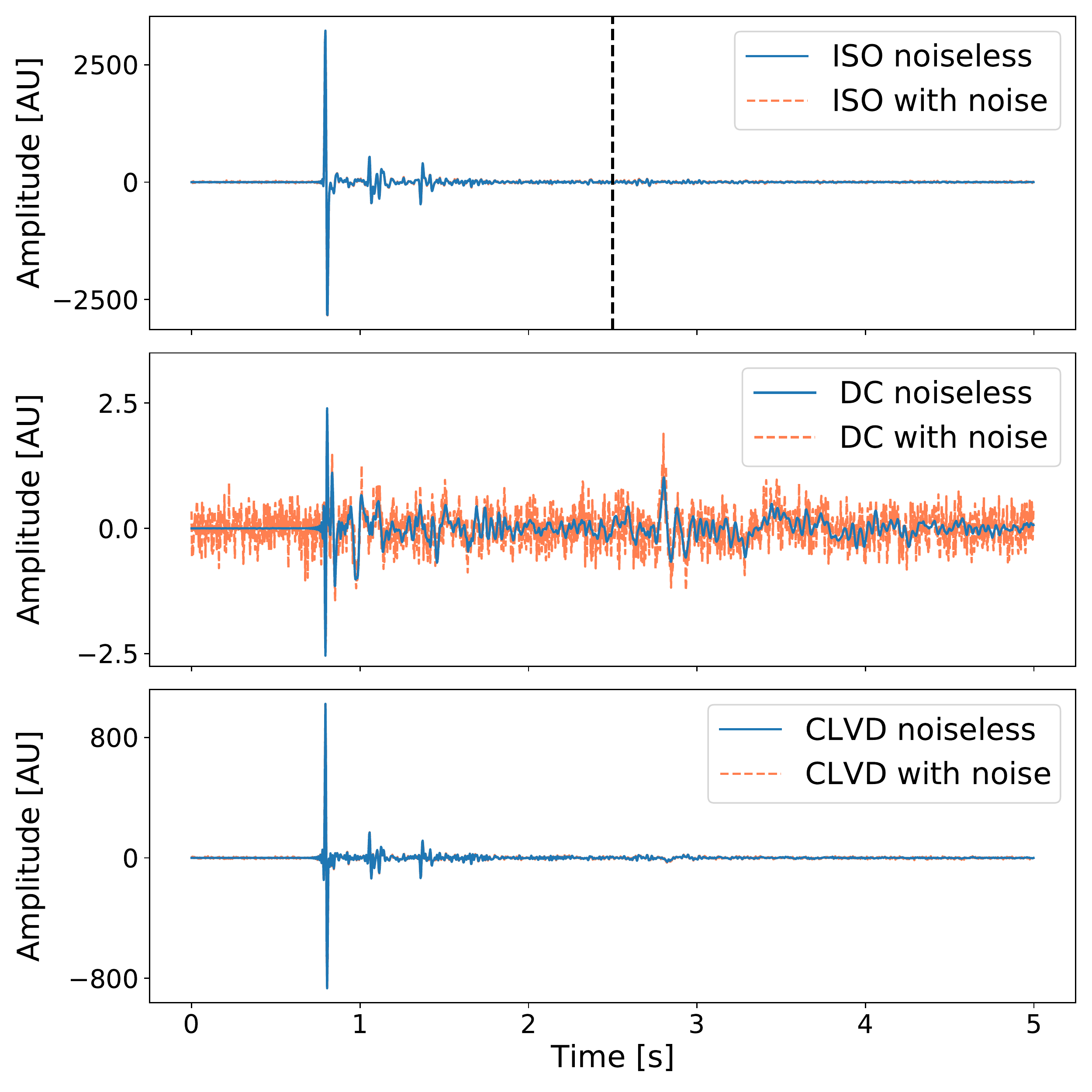}
    \caption{Comparison of the signal without (solid blue) and with (dashed coral) noise for each source mechanism: isotropic (ISO), double couple (DC) and compensated linear vector dipole (CLVD). As explained in Sect.~\ref{sec:inference}, after training the emulator on the noiseless traces, we add Gaussian noise to the observed signal to infer its coordinates. In this case, the seismogram corresponds to event 3 in Table~\ref{tab:inference} as recorded by a receiver in $(x,y,z)=(0.13 \ \rm{km},0.38 \ \rm{km},2.43 \ \rm{km})$, and the signal-to-noise ratio (SNR) is 33 dB; we explore higher and lower SNR values in Sect.~\ref{sec:inference_results} and Fig.~\ref{fig:compare_noise}. The vertical dashed black line in the top panel indicates a cut we perform for the isotropic sources only, based on the left panel of Fig.~\ref{fig:cuts}.}
    \label{fig:compare_all_noises}
\end{figure}

To perform Bayesian inference on a given seismogram, we first randomly choose an event's coordinates from the test set. For this set of coordinates, we simulate the observation of a microseismic event for each source mechanism, and generate the noiseless trace as it would be recorded by each of the 23 receivers. We add random Gaussian noise to each component of the noiseless trace. We define the signal-to-noise ratio (SNR) as \citep{Li18, Zhang20}:

\begin{align}
\label{eq:snr}
    \mathrm{SNR} = 10\log_{10}{\frac{\sum^{\rm{N}}_i \sum^{\rm{N_{keep}}}_j s^2_{ij}}{\sum^{\rm{N}}_i \sum^{\rm{N_{keep}}}_j (s_{ij} - \tilde{s}_{ij})^2}} \ ,
\end{align}
where $s_{ij}$ refers to the $j$-th sample of the $i$-th trace and $\tilde{s}_{ij}$ to the corresponding noisy trace, $\rm{N}=8000$ is the number of training data and $\rm{N_{keep}}=2001$ (1000 in the ISO case) is the number of time samples. As we explain in Sect.~\ref{sec:gen_model}, we consider 8000 events out of the total 10\,000 for training, and reserve the remainder for validation and testing purposes. Following \citet{Li18}, we set SNR$=33$ dB, which corresponds to a standard deviation of the Gaussian noise of $\sigma$ = 10.0, 0.3 and 3.5 for ISO, DC and CLVD, respectively, in the same arbitrary units as the seismograms' amplitude. We show examples of noiseless and noisy signals in Fig.~\ref{fig:compare_all_noises}.

We note that the choice of Gaussian noise has a quantifiable consequence on the power spectra of the signals, as white noise has an expected constant power in Fourier space \citep[see e.g.][]{Haykin01, Papoulis02}. This is reflected on the right-hand side of Fig.~\ref{fig:cuts}, where the mean noisy signal is shifted up with respect to the noiseless signals due to the noise addition. We argue that in general any information about the noise power (even if more complicated than Gaussian noise) can be easily accounted for when preprocessing the data: knowing the behaviour of the noise signal in the frequency domain allows one to fully take into account its effect on the relevant signal. For this reason, we decide to transform our data into the Fourier space, as we describe in Sect.~\ref{sec:preprocessing}. We will show in Sect.~\ref{sec:inference_results} that this choice makes our proposed approach robust to noise; moreover, our approach lends itself to the extension to coloured noise, which is most likely in realistic seismic data, especially for events with a low SNR \citep{Liu17}.

The noisy seismogram is further preprocessed as described in Sect.~\ref{sec:preprocessing}. At each likelihood evaluation of \textsc{PyMultiNest}, the proposed coordinates are mapped to the predicted preprocessed seismograms by means of the generative model $g( \boldsymbol{\theta})$ described in Sect.~\ref{sec:gen_model}: by evaluating the likelihood in multiple points of the prior space, \textsc{PyMultiNest} can sample from the posterior distribution of the coordinates, thus yielding the required credibility regions in parameter space. Note that, similarly to D18 and SM20, we assume a Gaussian likelihood, i.e.\ we write: 
\begin{align}
\label{eq:glike}
    \mathrm{Pr} \left( \boldsymbol{D} | \boldsymbol{\theta}, \mathcal{H}\right) \propto \exp \left( -\frac{1}{2} \left(\boldsymbol{D}-g(\boldsymbol{\theta})\right)^T \boldsymbol{C}^{-1} \left(\boldsymbol{D}-g(\boldsymbol{\theta})\right)  \right) \ ,
\end{align}
where $\boldsymbol{C}$ indicates the covariance matrix of the preprocessed seismograms, estimated from the training data.

We note that our choice of a Gaussian likelihood comes without loss of generality as our method can be easily extended to more complicated likelihood models. It is worth stressing that adding Gaussian noise to the seismograms does not necessarily imply that the distribution of the preprocessed seismograms will also be Gaussian; however, we verified experimentally that the distribution of each preprocessed seismogram component is unimodal and symmetric, thus supporting our assumption.


\subsection{Preprocessing}
\label{sec:preprocessing}
Learning a mapping between coordinates and seismograms directly would be hard for at least two reasons. First, each signal has features with different amplitudes: this means that e.g.\ a neural network (which will be described in detail in Sect.~\ref{sec:gen_model}) would likely just focus on the main peak and ignore the other components, thus losing useful information for the source location purpose. Secondly, given the complexity of the seismograms and the high number of features, the amount of data required to train an accurate emulator without overfitting would be at least an order of magnitude higher than what we consider in this work \citep[see e.g.][and references therein]{Bishop06, Zhu15}. 

In this sense, we $\textit{have to}$ preprocess the seismograms in order to extract only the relevant information that is needed to locate an event, while discarding all the noisy or redundant features of the signal. Both D18 and SM20 showed the importance of preprocessing, employing GPs in order to select only the components of each seismogram that are essential for inference. However, their methods fail on more complicated sources like the ones we consider in this work. While it could be argued that employing more training data could improve the results, it is also well-known that GPs do not scale well with the number of training points \citep{Liu18}, thus it is likely that the D18 method would struggle to generalise to more complicated source mechanisms. Additionally, the method proposed in SM20 is applied directly to the complicated seismic traces like the ones in the left panel of Fig.~\ref{fig:data_comparison}, thus making it more difficult for any algorithm to capture the most useful features for event location. Therefore, we follow a different preprocessing procedure, based on translating the data to the Fourier domain, and we outline the steps in the next paragraphs.

\begin{figure*}
    \centering
    \includegraphics[width=\textwidth]{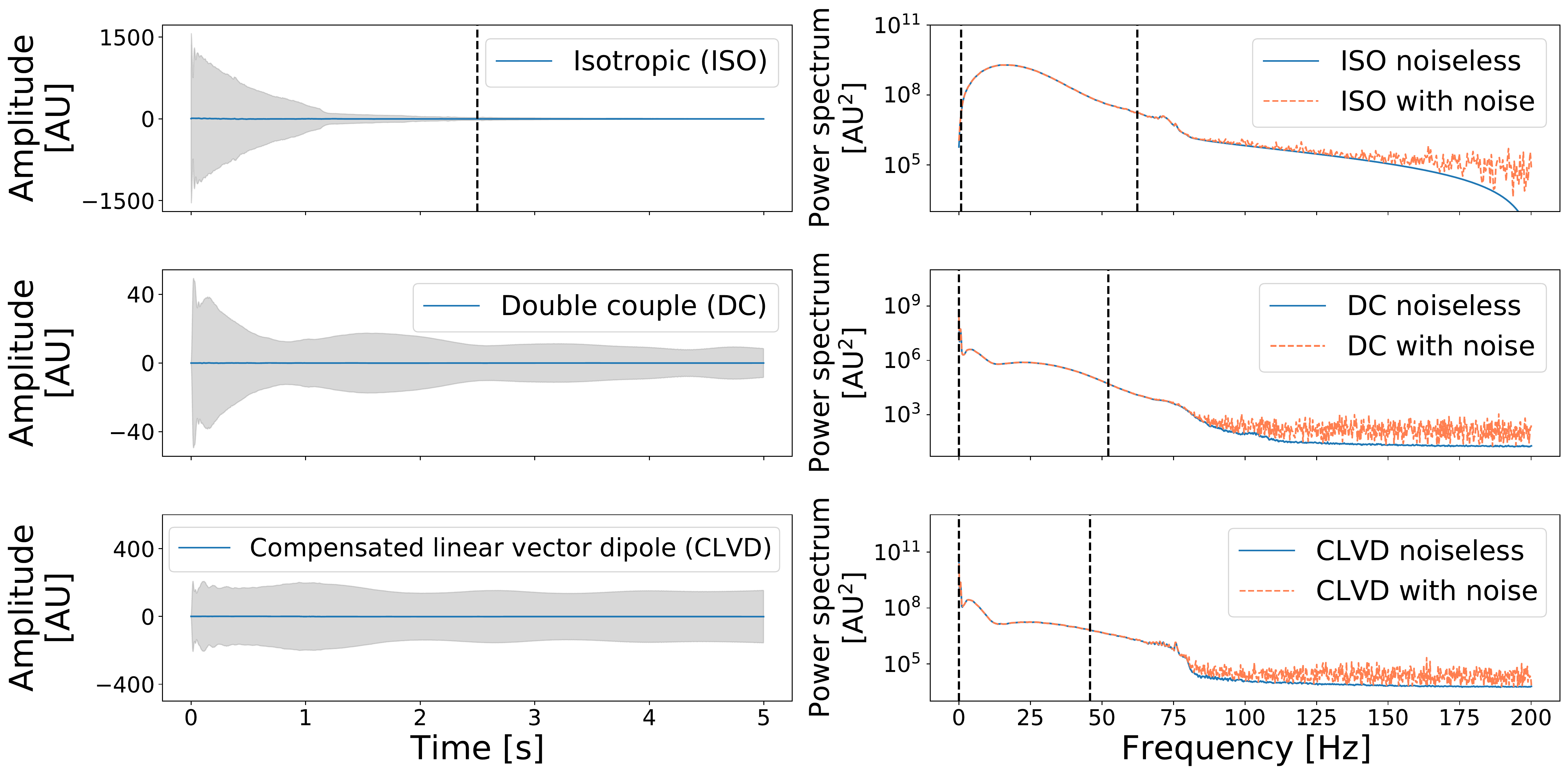}
    \caption{\textit{Left column:} Mean (blue line) and standard deviation (grey area) of all the seismic traces in the training set, for each source mechanism: isotropic (ISO), double couple (DC) and compensated linear vector dipole (CLVD). The seismograms' amplitude is measured in arbitrary units of pressure. The training set is made of 8000 traces for each source mechanism. We cut the isotropic sources at $2.5$ s, as indicated by the vertical dashed black line. \textit{Right column:} Noiseless (solid blue) and noisy (dashed coral) mean of the corresponding power spectra, calculated as described in Sect.~\ref{sec:preprocessing}. We consider a signal-to-noise ratio of 33 dB, as described in Sect.~\ref{sec:inference}. We filter the power spectra between the vertical dashed lines selecting only the frequencies where the ratio between the noiseless and noisy signals is more than 99$\%$.}
    \label{fig:cuts}
\end{figure*}

The left panel of Fig.~\ref{fig:cuts} shows the mean and standard deviation of all seismograms in the training set (8000 seismic traces): based on these distributions, we keep all samples of the DC and CLVD signals, and only keep the first half of the ISO traces since they sharply vanish after about one third of the trace. In other words, we keep only $\rm{N_{keep}}=1000$ time samples for the ISO traces, and $\rm{N_{keep}}=\rm{N_t}=2001$ time samples for the other mechanisms. 

The next preprocessing step we implement is applying the one-dimensional discrete Fourier Transform \citep{Cooley65} to each seismogram, using the version of \textsc{NumPy}\footnote{\url{https://numpy.org/doc/stable/reference/routines.fft.html}}. Since the amplitudes at each time sample are real numbers, the Fourier Transform returns $(\floor{\rm{N_{keep}}/2}+1)$ frequency components: this means that for DC and CLVD sources we are left with 1001 components (501 in the ISO case) in the Fourier domain. We then take the square of the absolute value of these complex numbers: this is usually referred to as a power spectrum. In the right column of Fig.~\ref{fig:data_comparison} we report the power spectra corresponding to the individual seismograms in the left column of the same figure. We further take the decimal logarithm of the power spectra at each frequency value, and refer to it as logarithmic power spectra in the rest of the paper.

As anticipated in Sect.~\ref{sec:inference}, we shall add some noise to the observed seismogram whose source location coordinates will be inferred. In the right panel of Fig.~\ref{fig:cuts} we show the mean power spectra for each source mechanism with and without noise.
We calculate the ratio between the noiseless and the noisy signals, and filter out the frequencies for which this ratio is less than 0.99. We experimented with different thresholds, and chose 0.99 as a good balance between retaining enough features to locate a seismogram and being insensitive to noise. In other words, we additionally cut each power spectrum in the ranges $[1$ Hz, $62.4$ Hz$]$, $[0$ Hz, $52.2$ Hz$]$ and $[0$ Hz, $45.8$ Hz$]$ for ISO, DC and CLVD, respectively, to keep the parts of each signal that are less affected by noise. We observe that translating the seismograms to the Fourier domain has allowed us to obtain smoother signals, as well as to reduce the number of features by a factor of $10$. Moreover, this allows our proposed method to be robust to noise: any effect due to noise can be translated into some information of the noise power, and hence readily accounted for in the analysis, e.g.\ by selecting a band-limited frequency signal as we discussed in Sect.~\ref{sec:inference}.

To further reduce the number of features, we apply principal component analysis (PCA). PCA is a standard linear compression technique where the data is projected along the eigenvectors of the data covariance matrix. Considering only the components that carry more variance (the so-called `principal components', corresponding to the largest eigenvalues), it is possible to reduce the number of features while maintaining the relevant information for inference \citep{Bishop06}. We fit PCA to the training data, and use it to compress the whole dataset. After applying PCA to the logarithmic power spectra, we retain 10 principal components for each signal when training the generative model; note that inference is done at the level of the logarithmic power spectra instead. We verified experimentally that varying the number of retained PCA components does not impact the final results significantly.

\subsection{Generative model}
\label{sec:gen_model}
As our generative model $g(\boldsymbol{\theta})$, we choose to employ a feedforward neural network (NN). A feedforward NN is a set of subsequent layers, each made of a certain number of neurons, that allow for the parametrisation of any measurable function between finite dimensional spaces \citep{Hornik89}. Each neuron is associated to a weight, and each layer is additionally associated to a bias (i.e.,\ an offset): weights and biases constitute the parameters that we wish to learn. Additionally, activation functions can be introduced after each layer to model non-linear mappings. Training the neural network consists of feeding some data through all the layers, and then updating the value of the parameters in order to optimise a chosen loss function.

We employ a neural network made of three layers with 256 neurons each, to provide enough flexibility to the parametrisation without consuming too much memory. We set Leaky ReLU \citep{Maas13} as the activation function for all layers except the last one, where we keep a linear activation function. We recall here that Leaky ReLU acts on the output of each layer $\mathbf{O}$ as follows:
\begin{align}
    \rm{Leaky ReLU}(\mathbf{O}) = \bigg \{\begin{array}{lr}
    \mathbf{O} & \mathrm{if} \, \mathbf{O} > 0 \\
    \alpha \mathbf{O} & \mathrm{otherwise} \\
    \end{array} \ ,
\end{align}
where we set the hyperparameter $\alpha = 0.3$; Leaky ReLU is usually preferred over the standard ReLU (Rectified Linear Unit) because of non-vanishing gradients \citep{Kolen01}. We also experimented with the ELU (Exponential Linear Unit, \citealt{Clevert16}) activation function, and found no significant improvements in the overall results with respect to using Leaky ReLU. We choose the Mean Squared Error (MSE) between the network output and the principal components of the training data as our loss function to minimise.


For each source mechanism and each receiver, we train the emulator using 8000 traces; we reserve 1000 seismograms for validation purposes and 1000 seismograms for testing purposes. We remark that we train a single neural network for each receiver: for a given source mechanism and underground location, the training data for each receiver includes the waveforms as observed by that particular station. In this way, we are able to include the information on the receiver's position in our generative model.

To train the neural network, we use the Adam optimiser \citep{Kingma14} with default parameters; moreover, we choose a learning rate of $0.001$ and a batch size of $256$: the former controls the step size of the parameters' update, while the latter indicates the number of training points that are fed through the network at each iteration. We additionally set a patience of $50$ to early-stop \citep{Yao07} based on the validation loss: this means that if the loss calculated on the validation set has not decreased in the last 50 epochs, we stop training and take the model corresponding to the minimum validation loss as the best model, as we interpret the model to have achieved its minimum error on unseen data\footnote{We also tested a dynamic learning rate decreasing by a factor of ten every time the validation loss did not decrease for 50 epochs, but found no significant improvement over a constant learning rate of 0.001, which we therefore chose for our analysis.}. We report the typical trend for the training and validation loss curves in Fig.~1 in the supplementary material.

The test set is used to randomly sample events on which we perform our Bayesian inversion analysis, as described in Sect.~\ref{sec:inference}. We also explore the behaviour of the posterior distribution as a function of the number of training events, number of receivers and noise scale, which we show in Sect.~\ref{sec:inference_results}. However, we do not perform a full grid search amongst the hyperparameters (e.g.\ number of layers, number of neurons, activation function and learning rate), as we observe the results are not significantly affected by them; we defer a more complete grid search to future work. 


\begin{figure*}%
\subfloat[][ISO]{\includegraphics[scale=0.2]{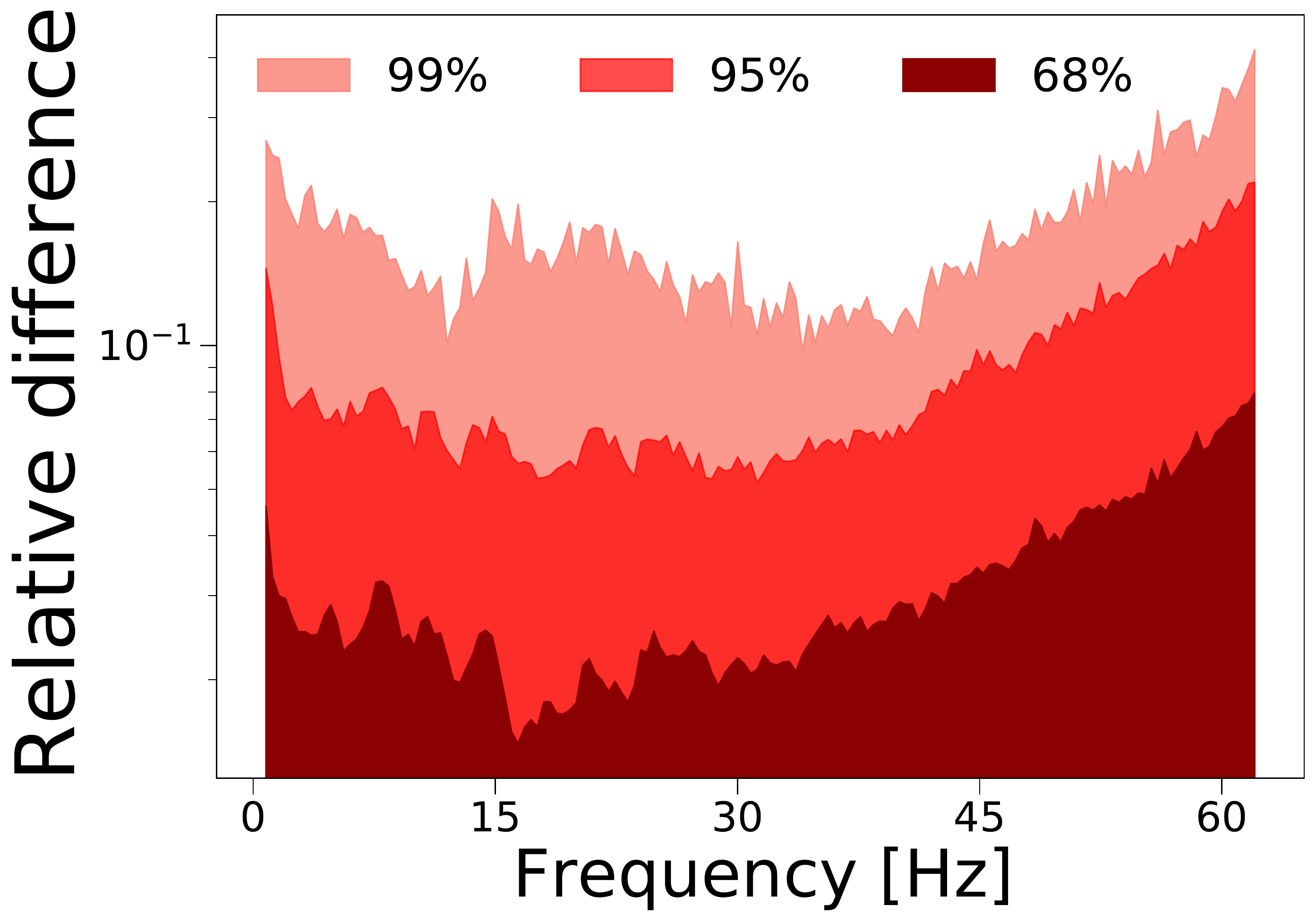}}
\subfloat[][DC]{\includegraphics[scale=0.2]{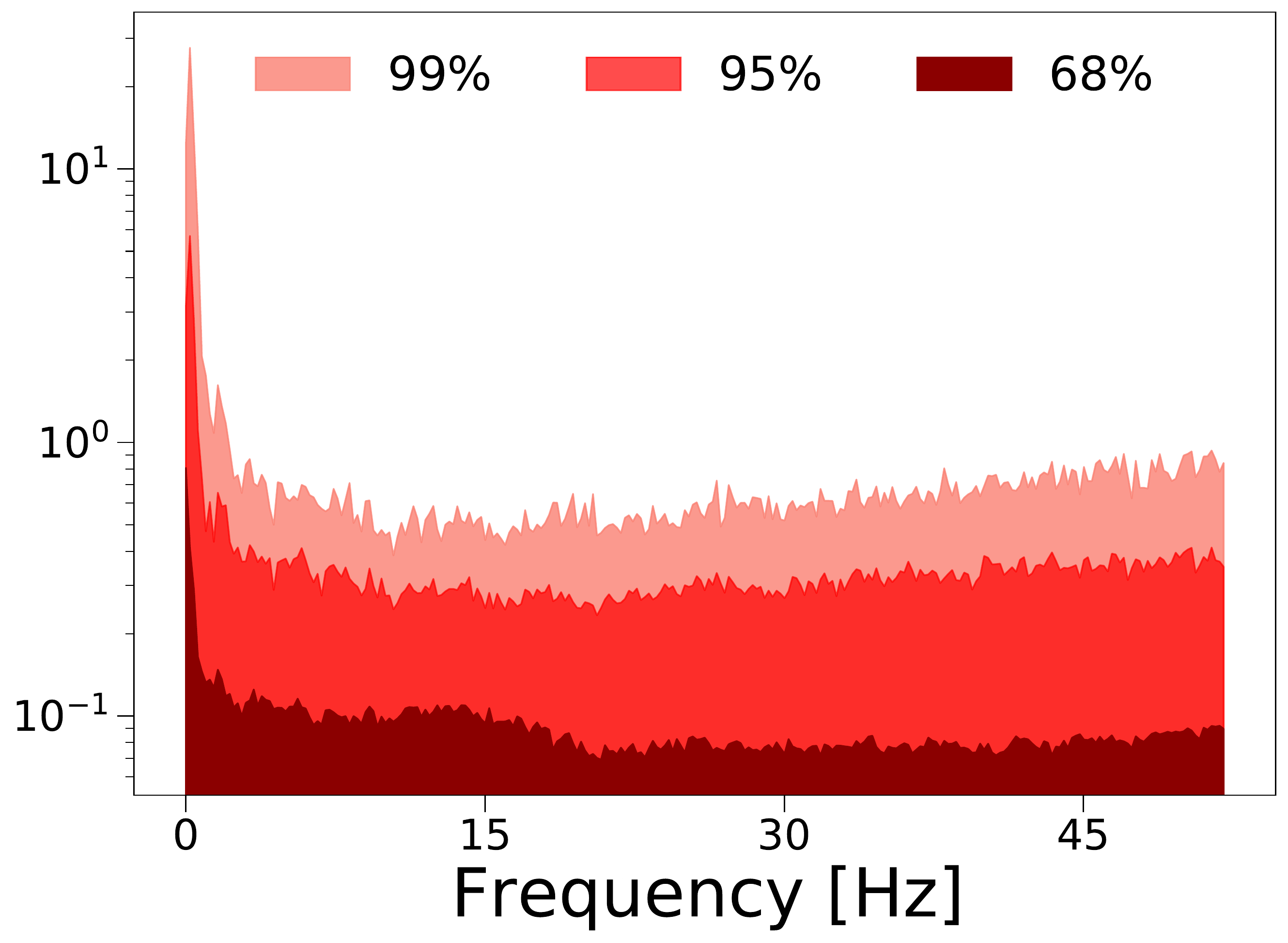}}
\subfloat[][CLVD]{\includegraphics[scale=0.2]{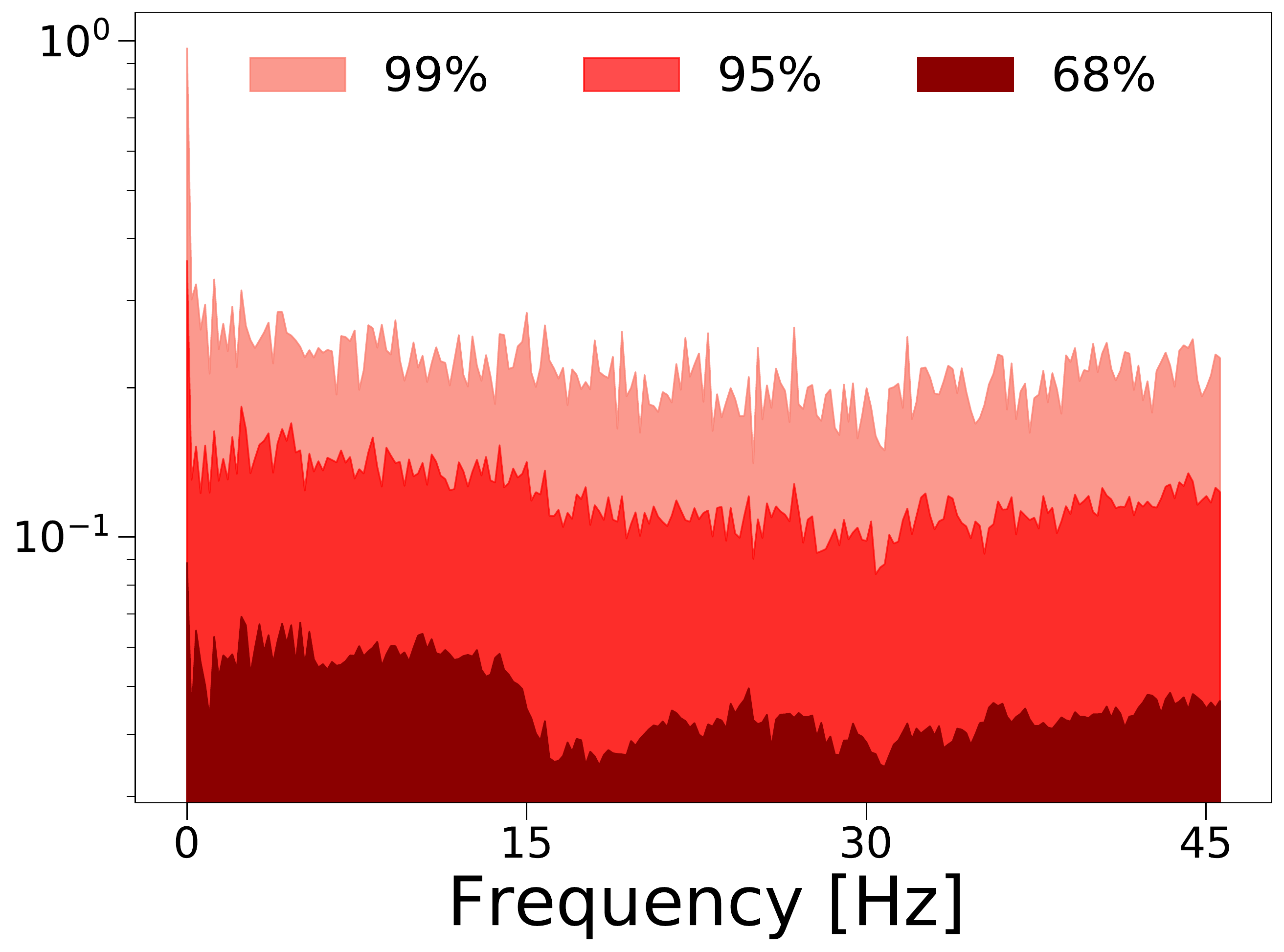}}
\caption{Accuracy of the generative model described in Sect.~\ref{sec:gen_model}, for isotropic (ISO), double couple (DC) and compensated linear vector dipole (CLVD). The dark red, red and salmon areas enclose the 68, 95 and 99 percentiles of the absolute value of the relative error between the predictions of the model and the target waveforms from the test set.}
\label{fig:gen_model_accuracy}
\end{figure*}

\subsection{Model selection}
\label{sec:evidence}
As detailed in Sect.~\ref{sec:inference}, we ignore the denominator in Eq.~\ref{eq:bayes} when inferring the coordinates of a microseismic event. However, we can use the evidence $\mathrm{Pr} \left( \boldsymbol{D} | \mathcal{H}\right)$ to perform model selection, thus showing another advantage of our proposed Bayesian approach \citep[see e.g.][]{Knuth15}. The quantity $\mathrm{Pr} \left( \boldsymbol{D} | \mathcal{H}\right)$ can be interpreted as the likelihood of a given signal under a certain hypothesis, with the constraint that the hypotheses form a set of $n_{\rm{hyp}}$ pairwise disjoint events whose union is the entire possibility space - i.e.\ $\mathrm{Pr} \left( \mathcal{H}_i \cap \mathcal{H}_j \right)$ = 0 for $i \neq j, \ \forall i, j = \{ 1, \dots, n_{\rm{hyp}} \}$, and $\sum_{i=1}^{n_{\rm{hyp}}} \mathrm{Pr} \left( \mathcal{H}_i \right) = 1$. 

To compare two hypotheses, and thus perform model selection, we define the Bayes factor BF as:

\begin{align}
\label{eq:bayes_f}
    \mathrm{BF} \equiv \frac{\mathrm{Pr} \left( \boldsymbol{D} | \mathcal{H}_i\right)}{\mathrm{Pr} \left( \boldsymbol{D} | \mathcal{H}_j\right)} = \frac{\mathrm{Pr} \left( \mathcal{H}_i |  \boldsymbol{D}\right) \mathrm{Pr} \left( \mathcal{H}_j \right) }{\mathrm{Pr} \left( \mathcal{H}_j |  \boldsymbol{D}\right) \mathrm{Pr} \left( \mathcal{H}_i \right)} \ ,
\end{align}
where the second equality has been obtained using Bayes' theorem, $\mathrm{Pr} \left( \mathcal{H}_i \right)$ is the prior distribution of the hypothesis $\mathcal{H}_i$, and $\mathrm{Pr} \left(  \mathcal{H}_i|\boldsymbol{D}\right)$ is the posterior distribution of hypothesis $\mathcal{H}_i$ given $\boldsymbol{D} $. If the hypotheses are equiprobable \textit{a priori}, we can write $\mathrm{Pr} \left( \mathcal{H}_i \right) = \mathrm{Pr} \left( \mathcal{H}_j \right)$, which allows us to express the Bayes factor as the ratio of the posterior distribution of one hypothesis over the other. Hence, if the Bayes factor as defined in Eq.~\ref{eq:bayes_f} is greater than 1, we can interpret it as hypothesis $\mathcal{H}_i$ being more favoured than hypothesis $\mathcal{H}_j$ under the observed data $\boldsymbol{D}$ \citep{Knuth15}.

Translating this into practice, after training the emulators, given an observed signal $\boldsymbol{D}$ as described in Sect.~\ref{sec:inference}, we can compare the three following equiprobable hypotheses: the source mechanism is isotropic ($\mathcal{H}_{\rm{ISO}}$), the source mechanism is double couple ($\mathcal{H}_{\rm{DC}}$), or the source mechanism is compensated linear vector dipole ($\mathcal{H}_{\rm{CLVD}}$). The advantage of using nested sampling is that the evidence is calculated while sampling the posterior distribution. Consequently, by feeding the observation $\boldsymbol{D}$ to each emulator it is straightforward to obtain the evidences $\mathrm{Pr} \left( \boldsymbol{D} | \mathcal{H}_{\rm{ISO}}\right)$, $\mathrm{Pr} \left( \boldsymbol{D} | \mathcal{H}_{\rm{DC}}\right)$ and $\mathrm{Pr} \left( \boldsymbol{D} | \mathcal{H}_{\rm{CLVD}}\right)$. 
By looking at the hypothesis that maximises the evidence, we can select the model that best describes the given observation, thus identifying the source type for a given observation. We show the results in Sect.~\ref{sec:model_sel_res} and Table~\ref{tab:evidence}.

\section{Results}
\label{sec:results}



\setlength{\tabcolsep}{6pt}

\subsection{Speed performance}
\label{sec:speed}

We first report on the speed performance of our method. We recall here that if we were to solve the elastic wave equation at each likelihood evaluation, inference would be severely compromised, as a single event's source inversion would take thousands of hours on a High Performance Computing (HPC) cluster, if at all possible. In contrast, our method requires only $\sim10^4$ simulations to be produced once, and the emulator to be trained once - an overhead of $\mathcal{O}(100 \ \rm{h})$ and $\mathcal{O}(1 \ \rm{h})$, respectively - and then it allows for the complete source inversion of any event in $\mathcal{O}(0.1 \ \rm{h})$ on a commercial laptop. In other words, most of the time taken by our approach is spent training the emulator, which needs to be done only once, provided the density and velocity models remain unchanged\footnote{In general, the stability of a given velocity model is not well known - see e.g.\ \citet{Thornton13, Usher13, Gesret13, Gesret14, Das18} and references therein for a discussion on the uncertainties of velocity models, and their consequences on location errors.}; after the training is complete, performing inference on a given recorded seismogram takes less than 10 minutes on a commercial laptop. As a reference, conventional full-waveform inversion techniques can take several hours on CPUs, and a comparable amount of time to our method when running on GPUs \citep[see e.g.][]{AbreoCarrillo15}. 



\subsection{Generative model accuracy}
\label{sec:gen_model_res}
In this section, we briefly look at the performance of the generative model that we described in Sect.~\ref{sec:gen_model}. We consider the 1000 events in the test set, and measure the relative difference between the prediction of our model and the target waveforms. We show the 68, 95 and 99 percentiles of the relative difference in Fig.~\ref{fig:gen_model_accuracy} for all source mechanisms. We observe that there are some significant discrepancies, especially at low frequency for the DC and CLVD, which however we attribute to low values in the logarithmic power spectra, and which do not seem to compromise inference, as we show in the next section.

\subsection{Inference results}
\label{sec:inference_results}

\begin{table*}
\centering
 \caption{Prior range and marginalised mean and 68 percent credibility intervals on the coordinates $(x,y,z)$ for each source mechanism - isotropic (ISO), double couple (DC) and compensated linear vector dipole (CLVD). The three events are randomly sampled from the test set. These results are obtained by considering all 23 receivers, and training on 8000 simulated events. The noise level is set to 10.0, 0.3 and 3.5 respectively, which corresponds to a signal-to-noise ratio of $33$ dB.}
 \label{tab:inference}
 \begin{tabular}{@{}ccccccc}
  \textbf{Event}    &\textbf{Coordinate}        & \textbf{Prior range [km]}    & \textbf{Ground truth [km]}               & \textbf{ISO [km]}                        & \textbf{DC [km]}      & \textbf{CLVD [km]     }     \\
    \hline \hline
  \multirow{3}{*}{1} & $x$ & $[0, 1]$ &  $0.71$                           & $0.63^{+0.13}_{-0.12}$              &  $0.72^{+0.13}_{-0.11}$  & $0.71^{+0.11}_{-0.10}$ \\&$y$& $[0, 1]$ &  $0.25$                           &  $0.28^{+0.13}_{-0.14}$            &  $0.25^{+0.10}_{-0.11}$  & $0.24^{+0.11}_{-0.11}$ \\&$z$  & $[0,2.43]$ &  $2.10$                           & $2.04^{+0.21}_{-0.47}$             &  $2.09^{+0.11}_{-0.13}$  & $2.10^{+0.22}_{-0.25}$                      \\
    \hline
  \multirow{3}{*}{2} & $x$ & $[0, 1]$ &  $0.46$                           &  $0.48^{+0.19}_{-0.19}$              &  $0.43^{+0.12}_{-0.13}$  &$0.46^{+0.15}_{-0.12}$ \\&$y$& $[0, 1]$ &  $0.34$                           &  $0.32^{+0.16}_{-0.19}$              &  $0.33^{+0.09}_{-0.08}$  &$0.33^{+0.15}_{-0.14}$ \\&$z$  & $[0, 2.43]$ &  $1.48$                           &  $1.60^{+0.23}_{-0.30}$             &  $1.54^{+0.18}_{-0.15}$  & $1.44^{+0.22}_{-0.24}$                    \\
    \hline
  \multirow{3}{*}{3} & $x$ & $[0, 1]$ &  $0.20$                           & $0.28^{+0.20}_{-0.15}$              &  $0.20^{+0.10}_{-0.10}$  & $0.22^{+0.25}_{-0.13}$ \\&$y$& $[0, 1]$ &  $0.43$                           &  $0.46^{+0.18}_{-0.16}$            &  $0.41^{+0.10}_{-0.13}$  & $0.54^{+0.17}_{-0.21}$ \\&$z$  & $[0, 2.43]$ &  $0.99$                           & $0.93^{+0.17}_{-0.19}$             &  $1.09^{+0.28}_{-0.17}$  & $0.98^{+0.21}_{-0.20}$                      \\
 \end{tabular}
\end{table*}

\begin{figure}%
    \begin{center}
    \includegraphics[width=0.41\textwidth]{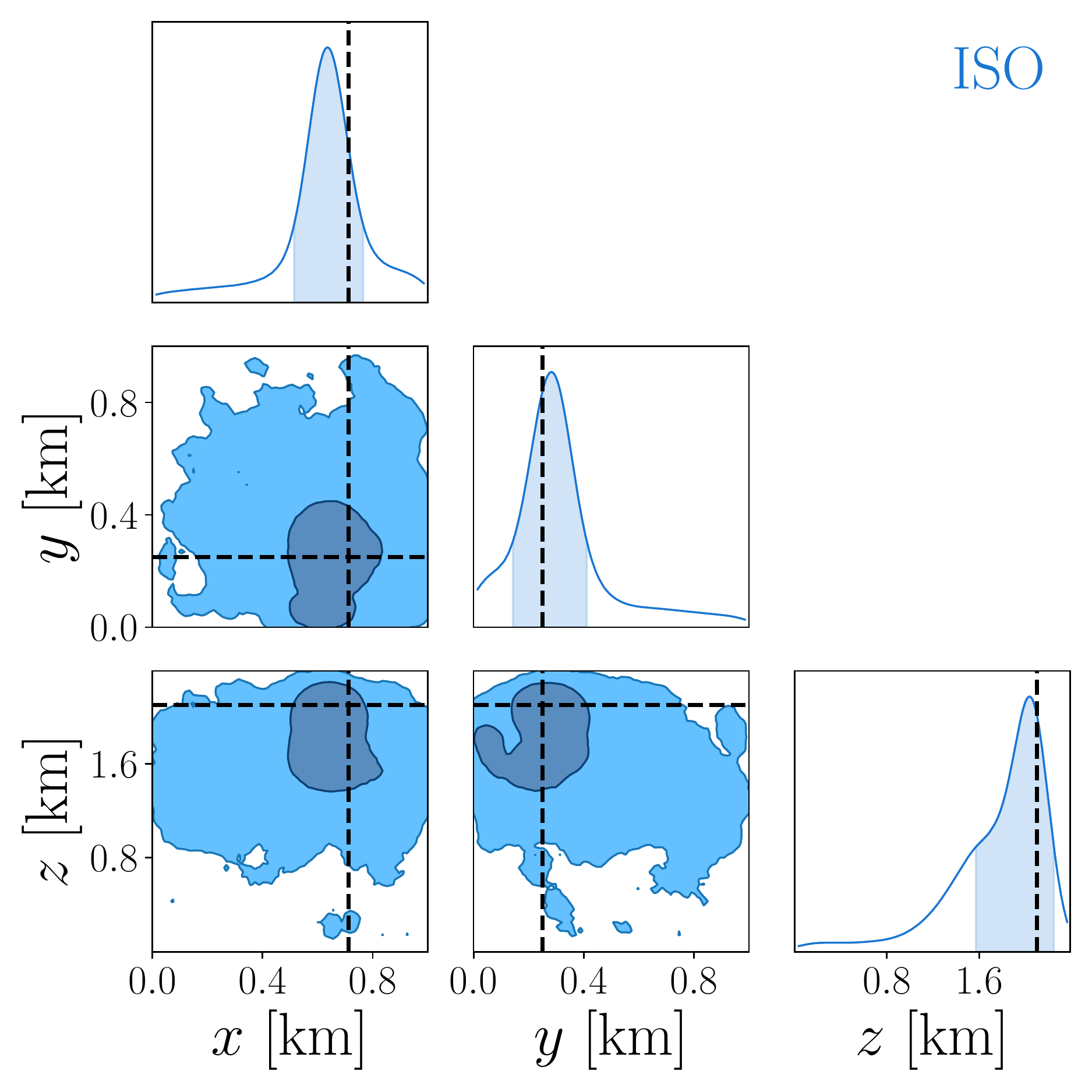}
    \includegraphics[width=0.41\textwidth]{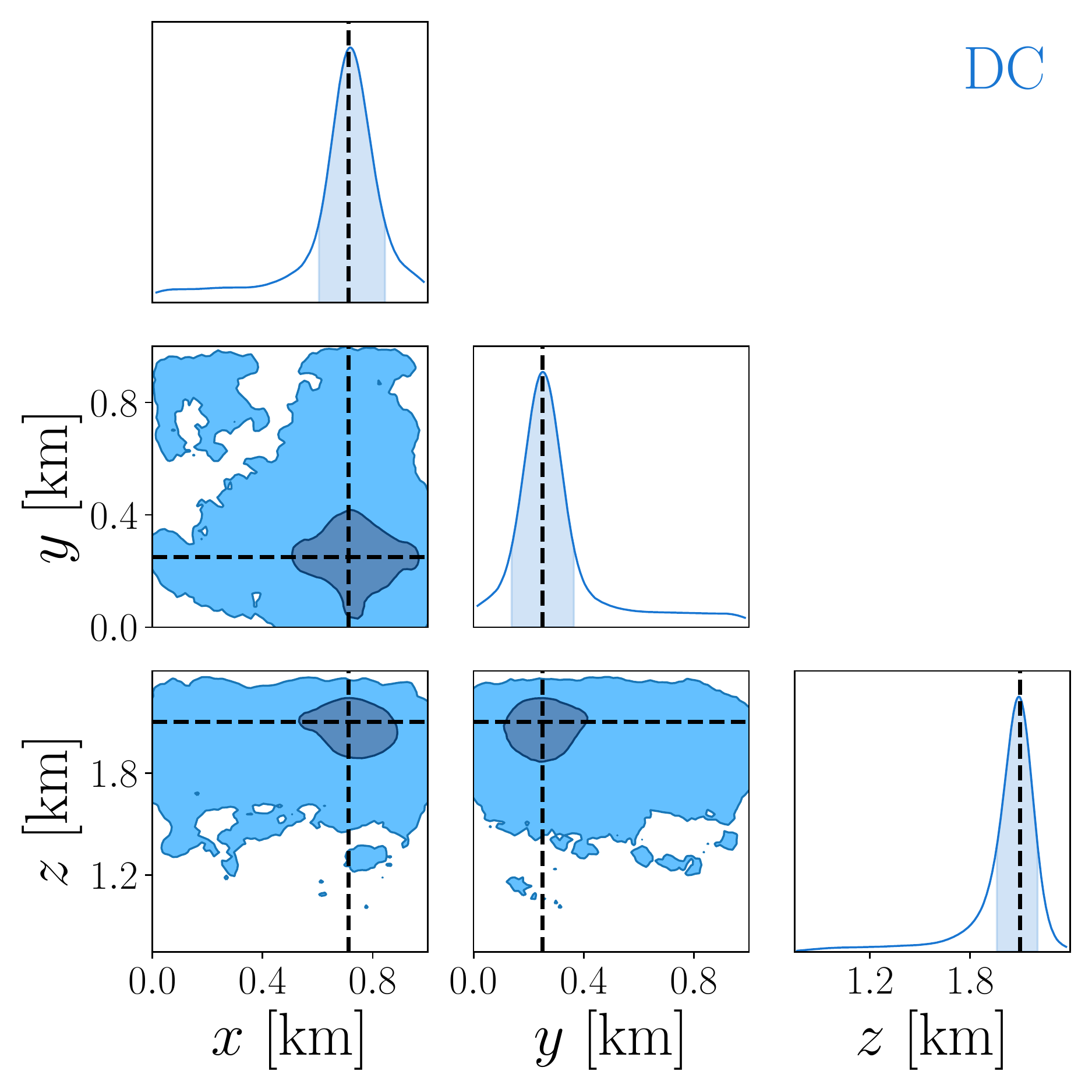}
    \end{center}
    \begin{center}
        \includegraphics[width=0.41\textwidth]{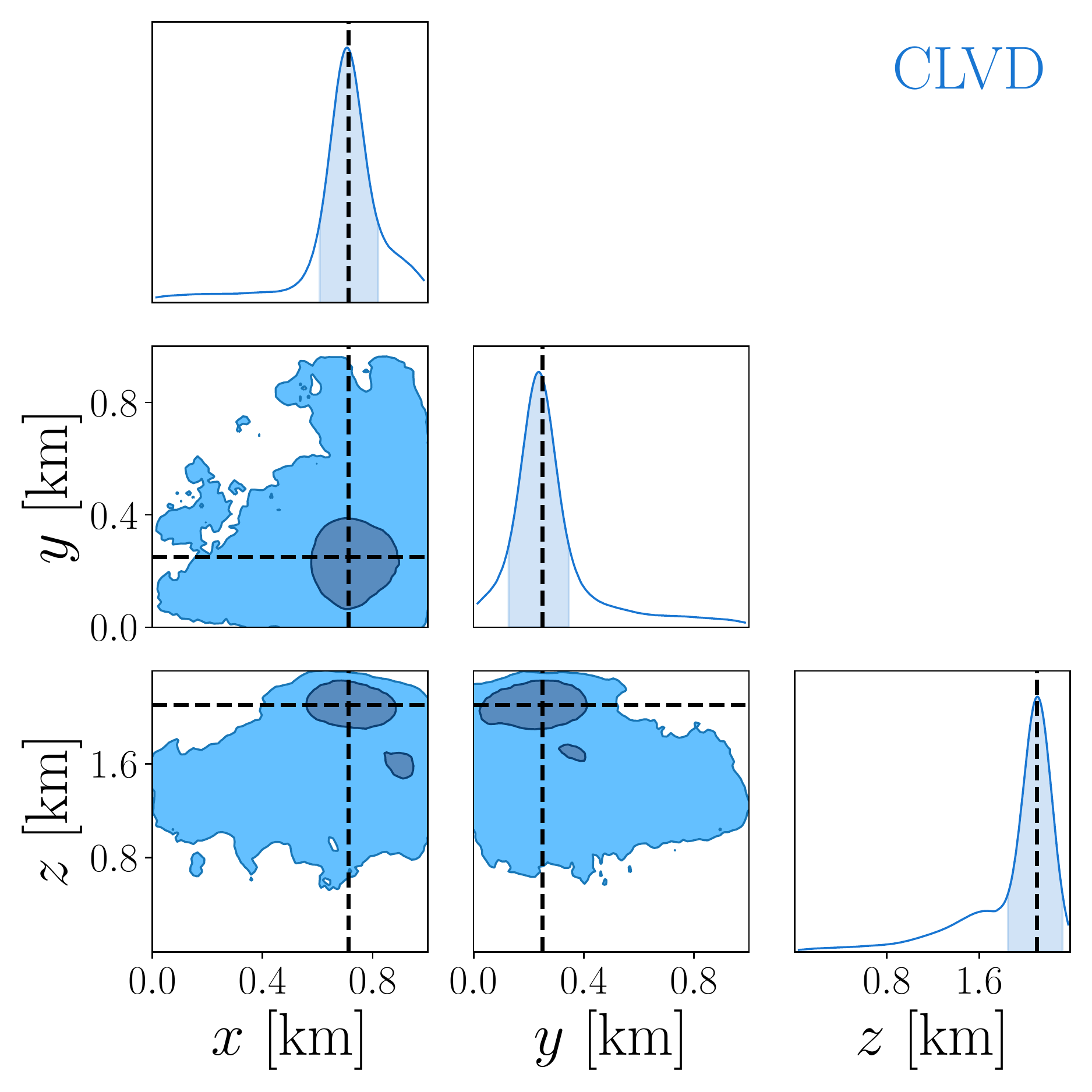}
    \end{center}    
\vspace{-2.4em}\caption{Marginalised 68 and 95 per cent credibility contours obtained with our method for a source located at $(x,y,z)=(0.71 \ \rm{km},0.25 \ \rm{km},2.10 \ \rm{km})$, indicated by the dashed black lines. We compare 3 source mechanisms: isotropic (ISO), double couple (DC), and compensated linear vector dipole (CLVD). The event corresponds to event 1 in Table~\ref{tab:inference}. Note that we are considering 23 receivers, the signal-to-noise ratio is 33 dB and the emulator was trained on 8000 training points.}
\label{fig:inference_1}
\end{figure}

\begin{figure}%
    \begin{center}
    \includegraphics[width=0.41\textwidth]{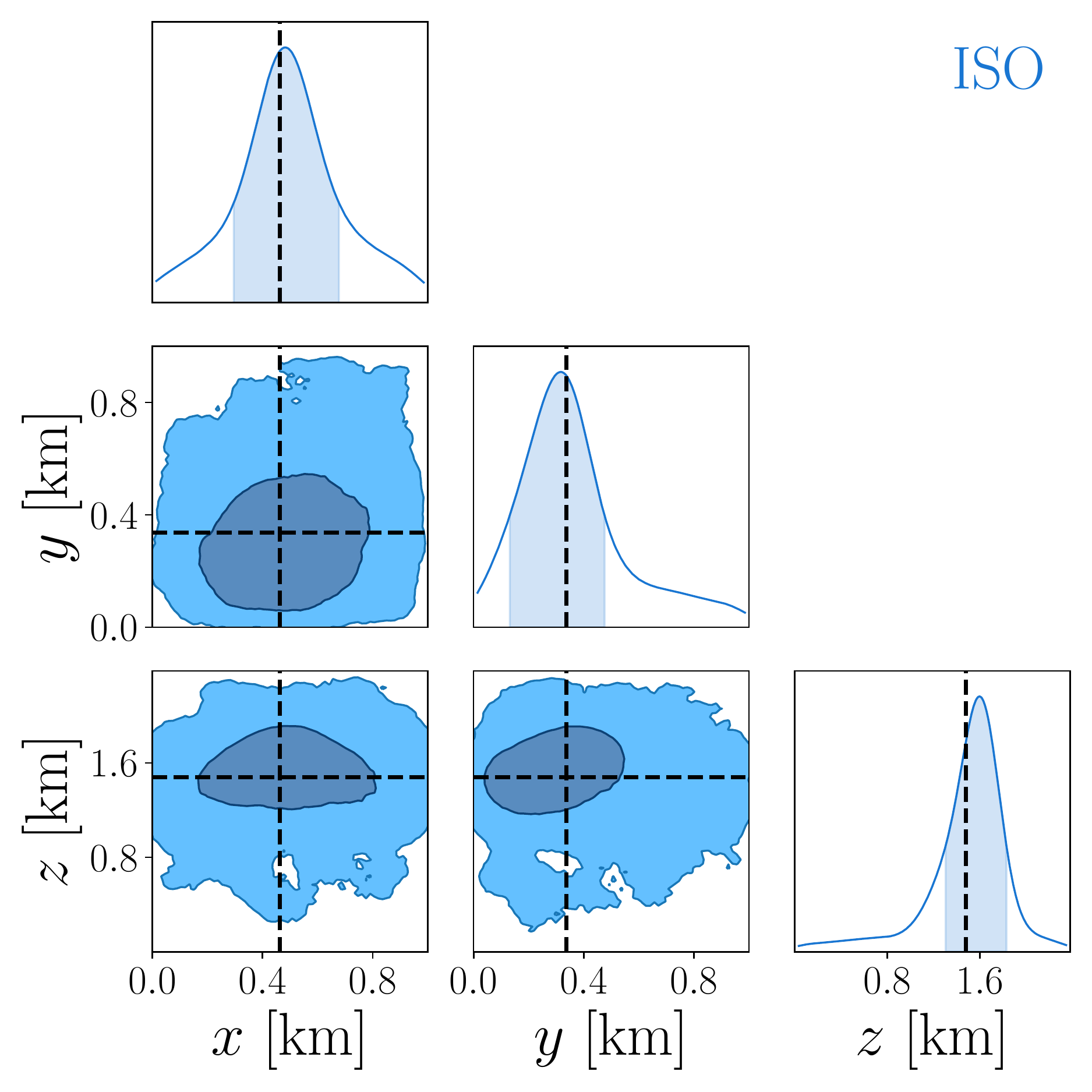}
    \includegraphics[width=0.41\textwidth]{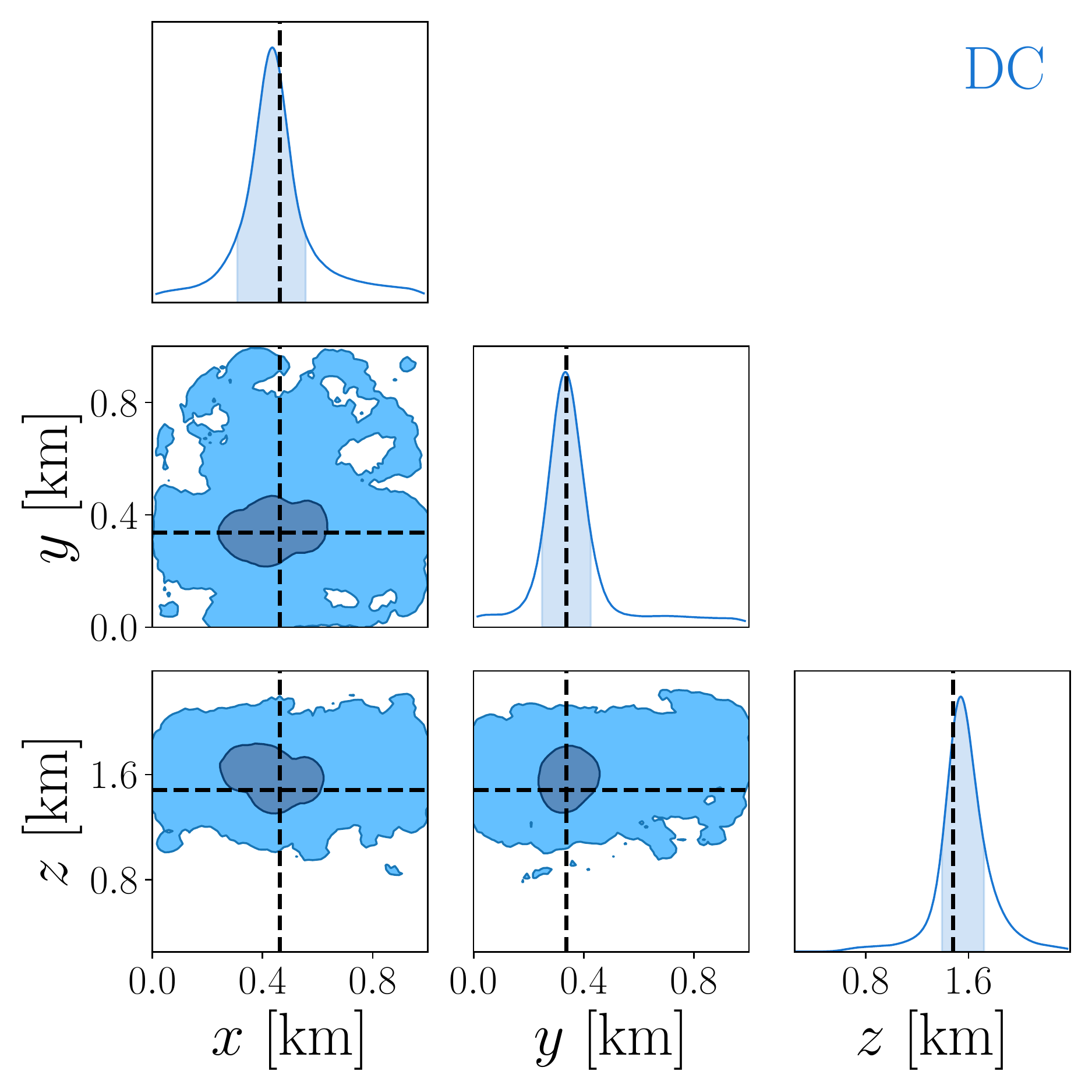}
    \end{center}
    \begin{center}
        \includegraphics[width=0.41\textwidth]{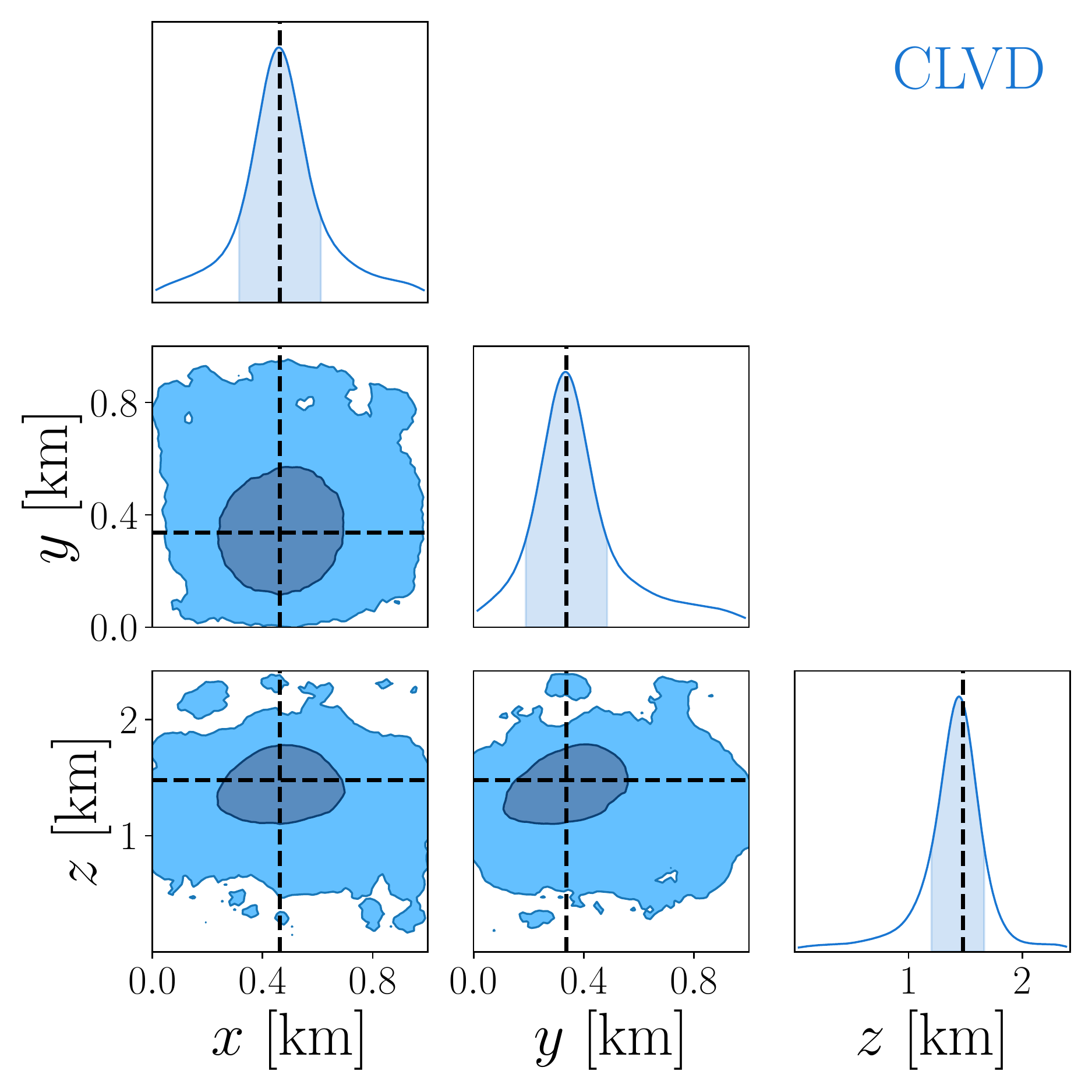}
    \end{center}
\vspace{-2.4em}\caption{Same as Fig.~\ref{fig:inference_1} for a source located at $(x,y,z)=(0.46 \ \rm{km},0.34 \ \rm{km},1.48 \ \rm{km})$. The event corresponds to event 2 in Table~\ref{tab:inference}.}
\label{fig:inference_2}
\end{figure}

\begin{figure}%
    \begin{center}

    \includegraphics[width=0.41\textwidth]{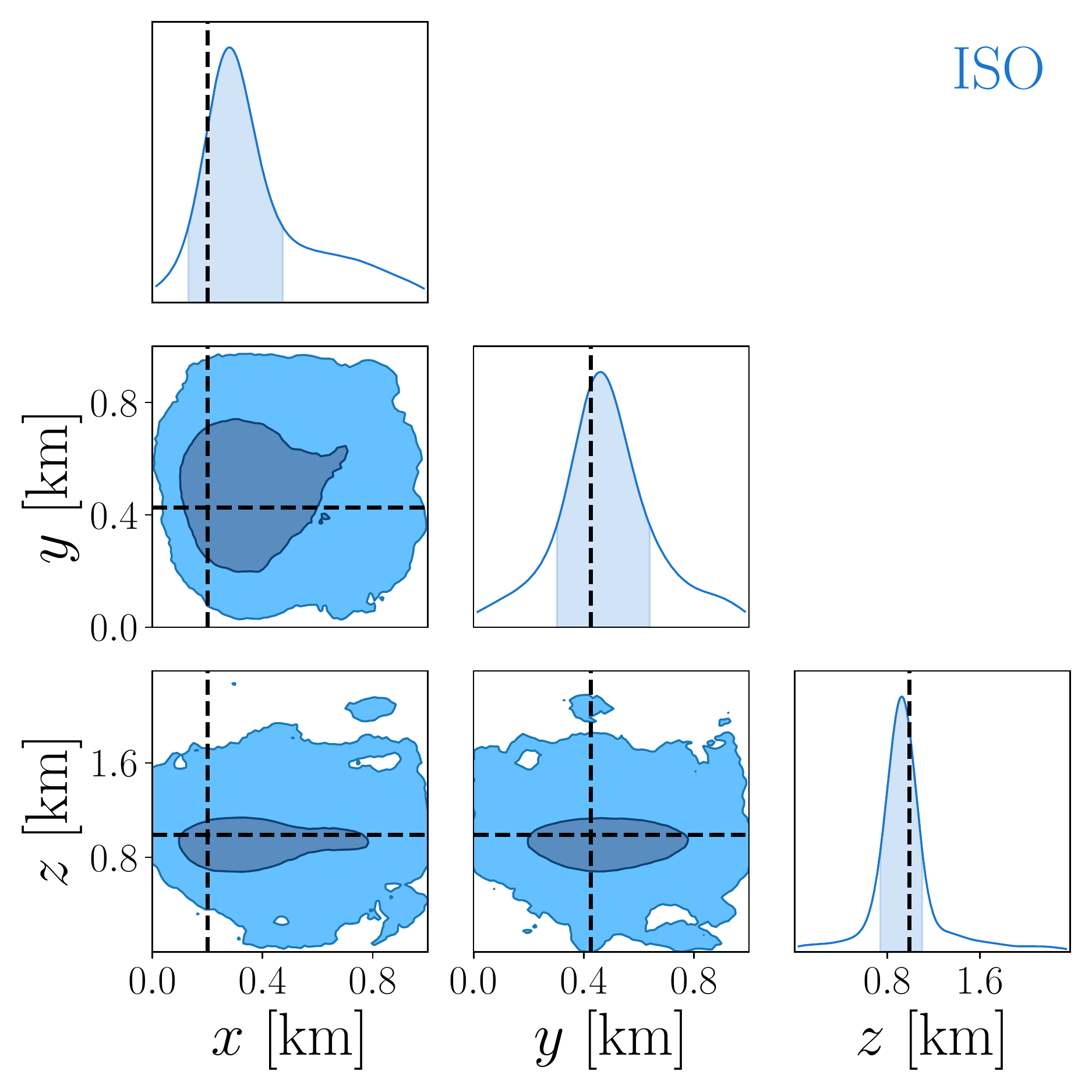}
    \includegraphics[width=0.41\textwidth]{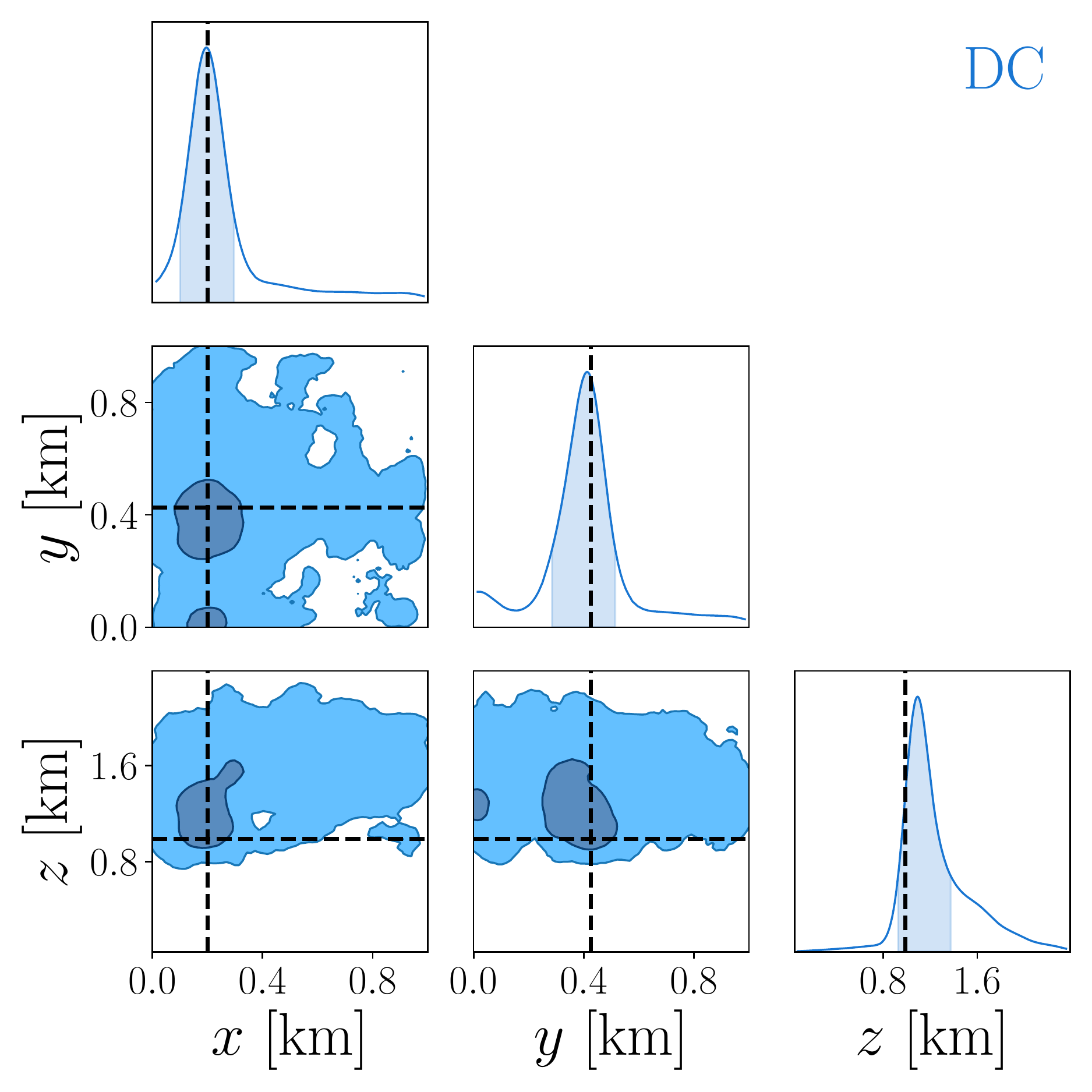}
    \end{center}

    \begin{center}
        \includegraphics[width=0.41\textwidth]{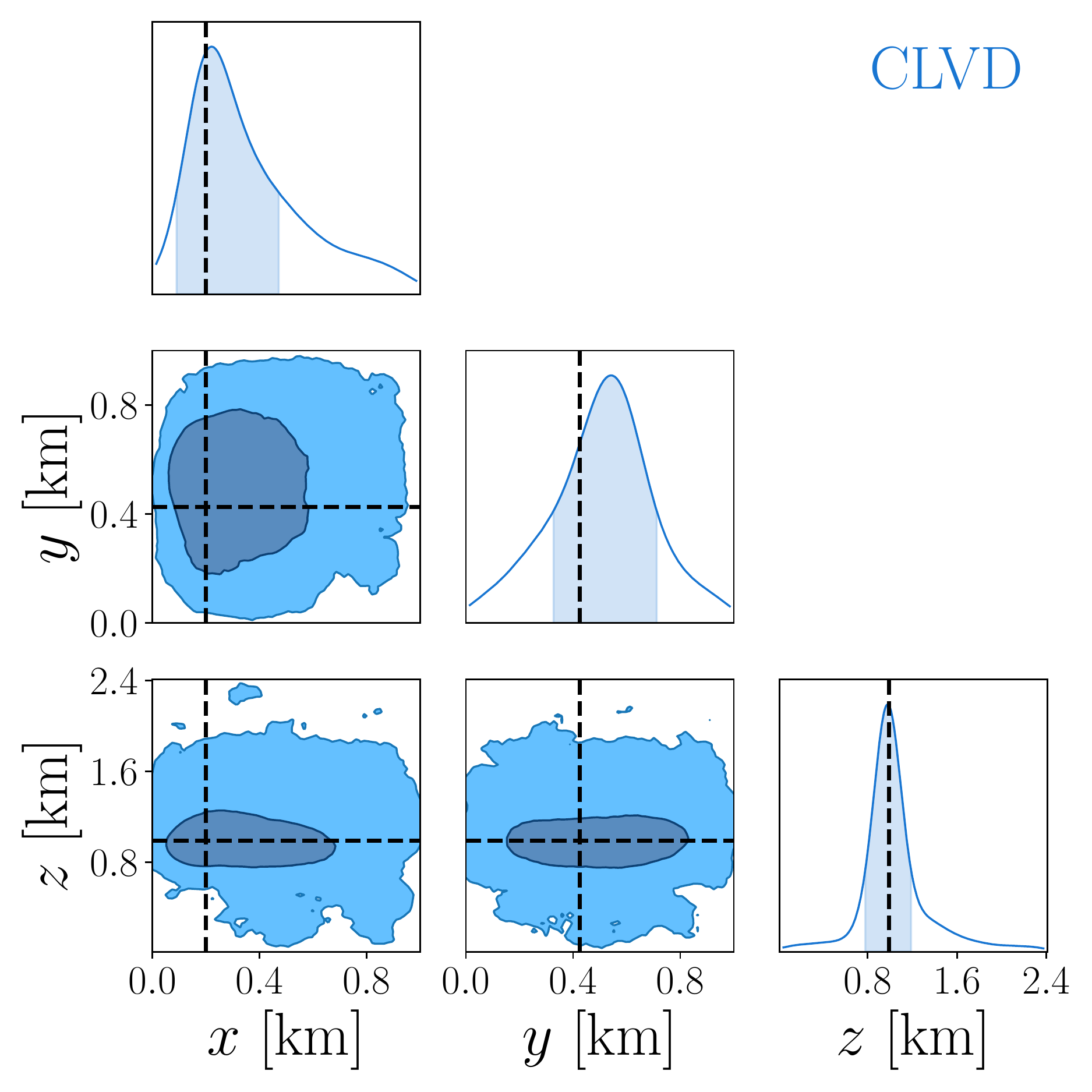}
    \end{center}
\caption{Same as Fig.~\ref{fig:inference_1} for a source located at $(x,y,z)=(0.20 \ \rm{km},0.43  \ \rm{km},0.99 \ \rm{km})$. The event corresponds to event 3 in Table~\ref{tab:inference}.}
\label{fig:inference_3}
\end{figure}

We then turn our attention to the accuracy of the inferred posterior distribution of the coordinates. For each source mechanism, we report the inference results for 3 different coordinates in Figs.~\ref{fig:inference_1}, \ref{fig:inference_2} and \ref{fig:inference_3}: each shows the posterior contour plots obtained with our methodology, considering all 23 receivers, SNR=33 dB and using 8000 seismograms as the training set for the emulator. The numerical results are summarised in Table~\ref{tab:inference}, reporting the prior ranges and marginalised mean and 68 percent credibility interval on the coordinates. We additionally report the mean and the standard deviation of the absolute value of the difference between the ground truth and the maximum of the retrieved posterior distribution for 100 randomly-picked test events: this is $(0.17 \pm 0.29)$ km. This difference indicates good agreement, despite being skewed by a small fraction (less than 5\%) of test cases for which the inference results were totally unconstrained, or for which the width of the posterior distribution was particularly broad: we remark that these occurrences generally corresponded to events for which the ground truth had at least one coordinate lying at the border of the prior space, thus complicating the sampling procedure. We argue that it is safe to ignore these limit cases, given their particular location; in a real scenario, one would train the neural network on events coming from a larger prior volume than the one that is being investigated here, thus avoiding this problem. We therefore conclude that with our method we can accurately retrieve the correct value of the coordinates across almost the entire prior parameter space.

Additionally, we observe that the $x$ and $y$ source coordinates are usually less constrained than the $z$ coordinate for ISO and CLVD, while this behaviour is less prominent in the DC case, for which all coordinates are always tightly constrained. We attribute this effect to two possible causes. On one hand, it can be related to the specific density model we are considering in this work, which has a layered structure. On the other hand, we note that when translating the seismic traces to the Fourier domain we ignored the phase signal (since we only considered the amplitude power spectra), thus possibly losing useful information for the source location purpose \citep[e.g.,][]{Ferreira07}; we argue that this also results in larger uncertainties in the retrieved posterior contours with respect to a full-waveform approach, which however would be too computationally expensive to run. We additionally note that retaining the phase information would yield a generative model, as by combining the power spectra with the phase one could in principle reconstruct a full seismogram from the source coordinates (after training the emulator). We defer the study of phase information to future work, as we anticipate that given the oscillatory behaviour of the phase signals it will be harder to train an emulator on them.

\begin{figure*}
$\begin{array}{cc}
    \includegraphics[width=0.41\textwidth]{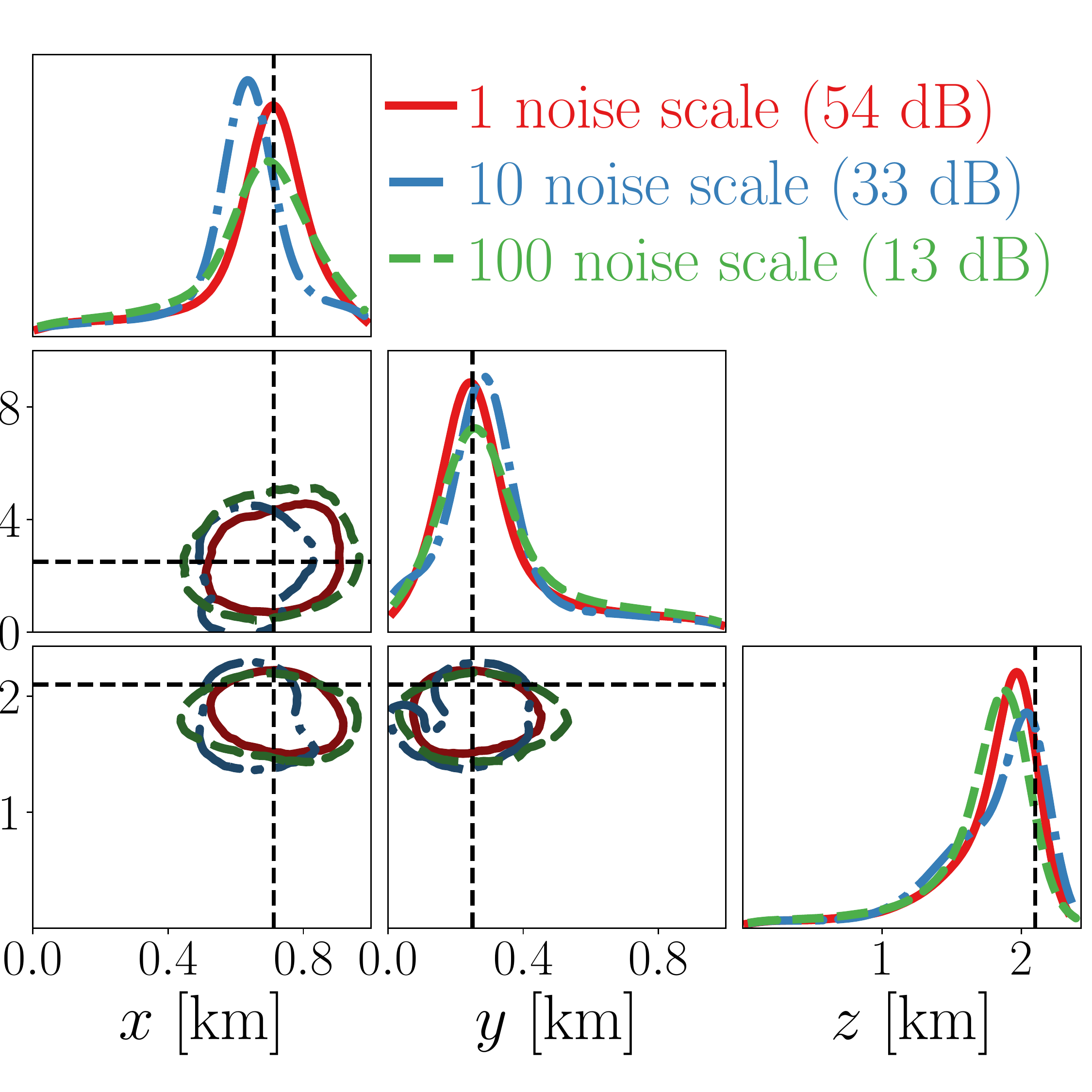} &
    \includegraphics[width=0.41\textwidth]{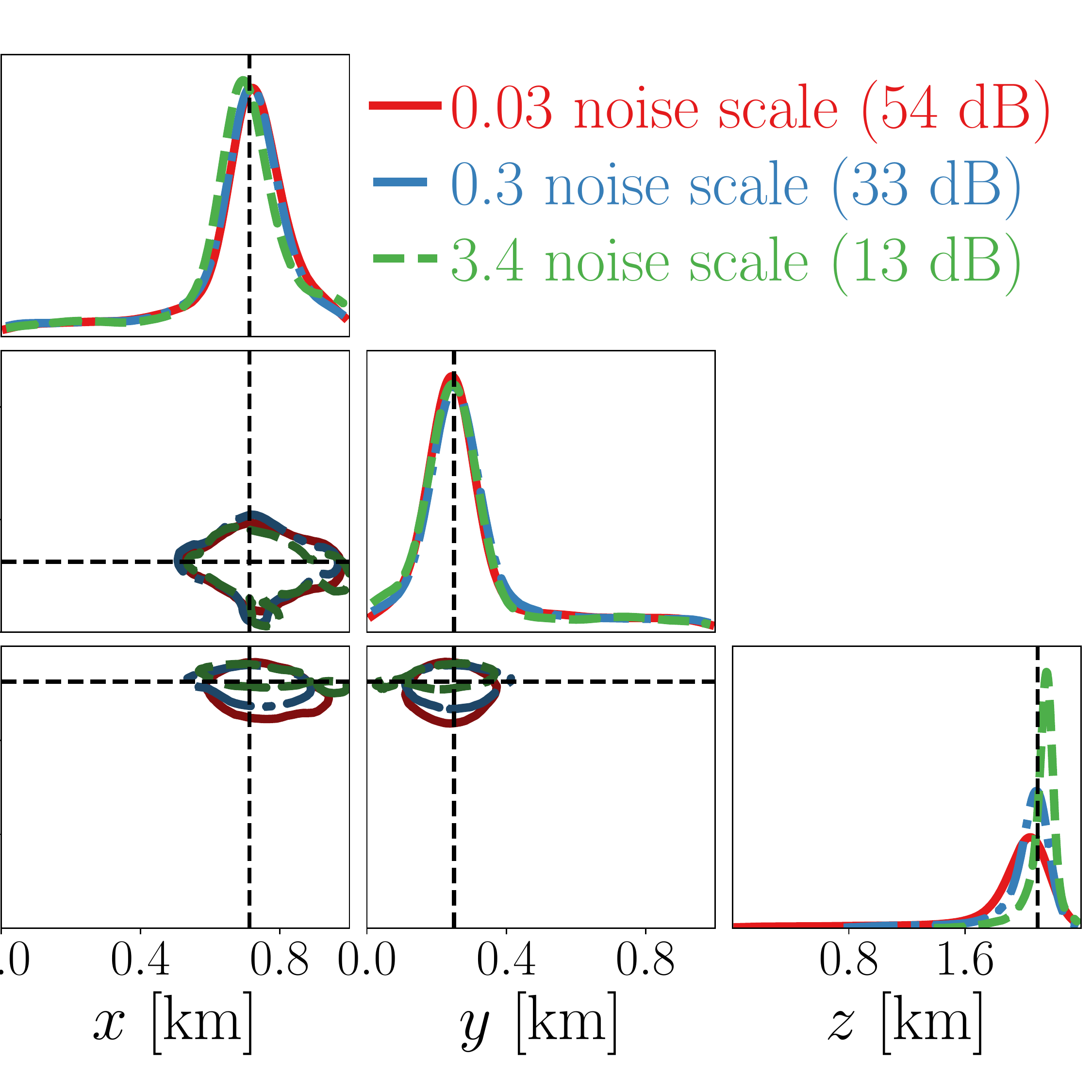}   \\
    \rm{ISO}& \rm{DC} 
\end{array}$
$\begin{array}{c}
    \includegraphics[width=0.41\textwidth]{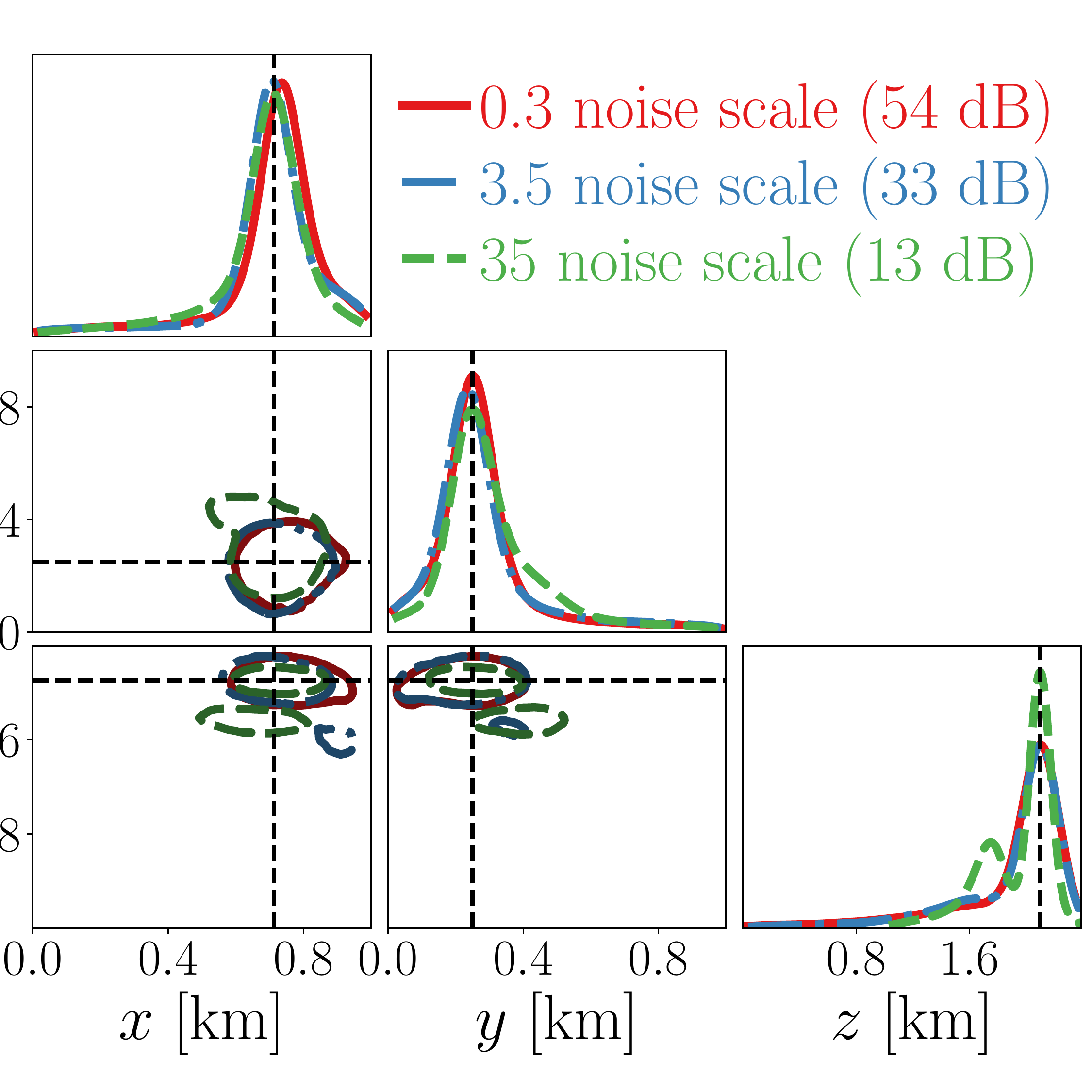}   \\
    \rm{CLVD}
\end{array}$
\caption[]{\label{fig:compare_noise}Marginalised 68 per cent credibility contours obtained with our method for a source located at $(x,y,z)=(0.71 \ \rm{km},0.25 \ \rm{km},2.10 \ \rm{km})$, indicated by the dashed black lines, comparing different levels of noise. In particular, in each panel we show signal-to-noise ratios of 13 dB, 33 dB and 54 dB for the 3 source mechanisms: isotropic (ISO), double couple (DC), and compensated linear vector dipole (CLVD). Note that we are considering 8000 training data for the emulator and 23 receivers. This figure is best viewed in colour.}
\end{figure*}

We then study the dependence of the posterior contours on the SNR, the number of training data and the number of receivers used. In Fig.~\ref{fig:compare_noise}, we first show the effect of different noise levels; in particular, for each source mechanism we vary the SNR from 13 dB to 54 dB. We note that, in order to increase the robustness to noise, we cut different frequency windows based on the noise levels: given the Gaussian noise model we assume in this work, a smaller SNR corresponds to a higher noise power, and hence to a smaller number of retained power spectrum principal components. We observe that while higher noise levels can lead to small biases in some of the 1-D marginalised coordinates' distributions, in general no significant variations in the shape of the 2-D posterior contours are present.

In Fig.~2 in the supplementary material we show the posterior contours when training the emulator with 2000, 5000 and 8000 data points. Again, we observe no significant differences for the ISO and CLVD sources, while very small deviations appear when using fewer training data in the DC case. In general, $\mathcal{O} (10^3)$ training data are enough to obtain accurate posterior contours for all source mechanisms. Finally, in Fig.~3 in the supplementary material we vary the geometry of the receivers used for recording the microseismic traces. While we are not interested in a full study of the optimal geometry of the receivers, we observe that using fewer receivers leads to broader posterior contours, while still allowing accurate event locations for all source mechanisms.

In summary, our results are very robust to the noise injected into the observed seismogram. Additionally, very few receivers - $\mathcal{O}(10)$ - are needed to obtain accurate results, and a number of training points of order $\mathcal{O} (10^3)$ is sufficient to locate any event.

\begin{figure}%
    \begin{center}
    \includegraphics[width=0.4\textwidth]{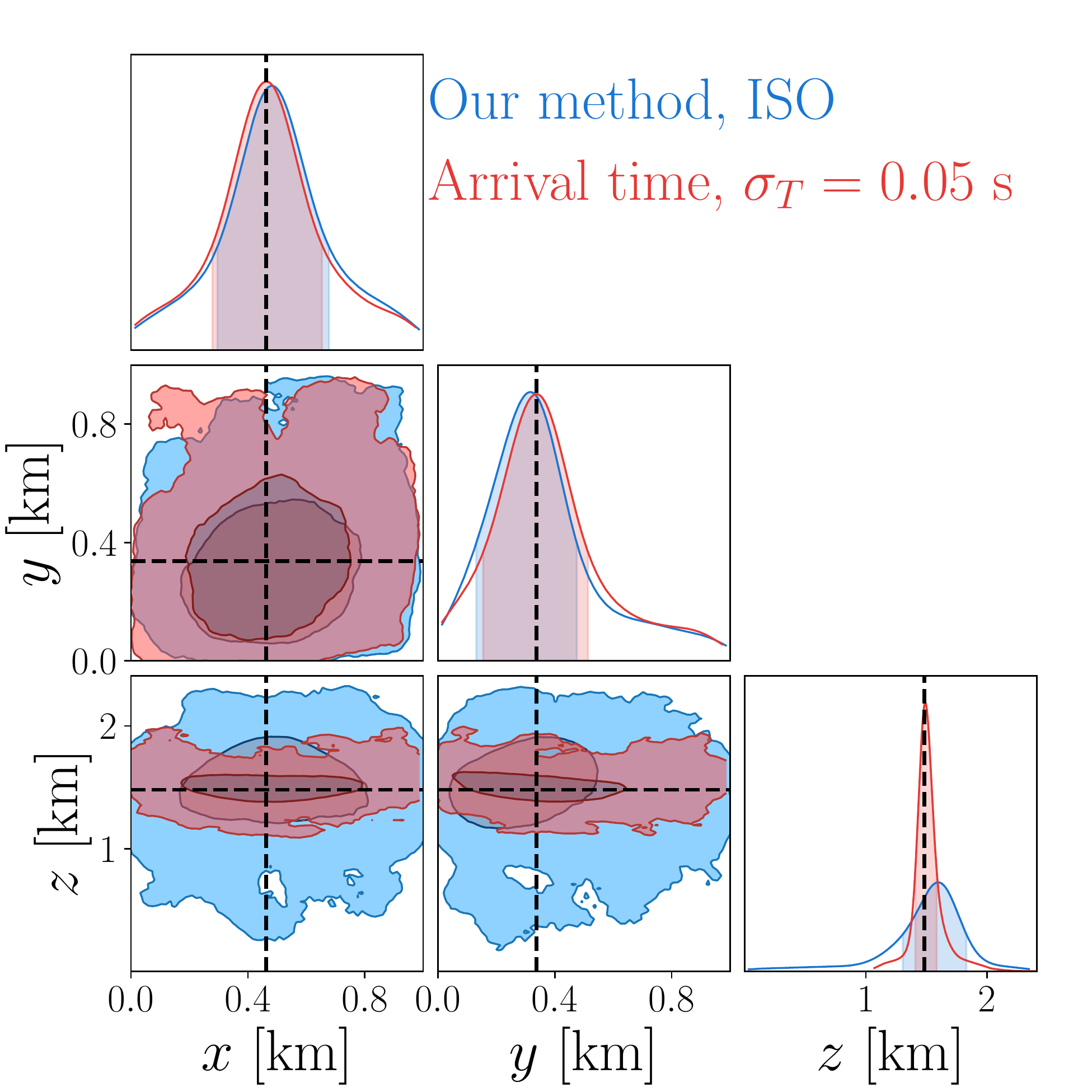}
    \includegraphics[width=0.4\textwidth]{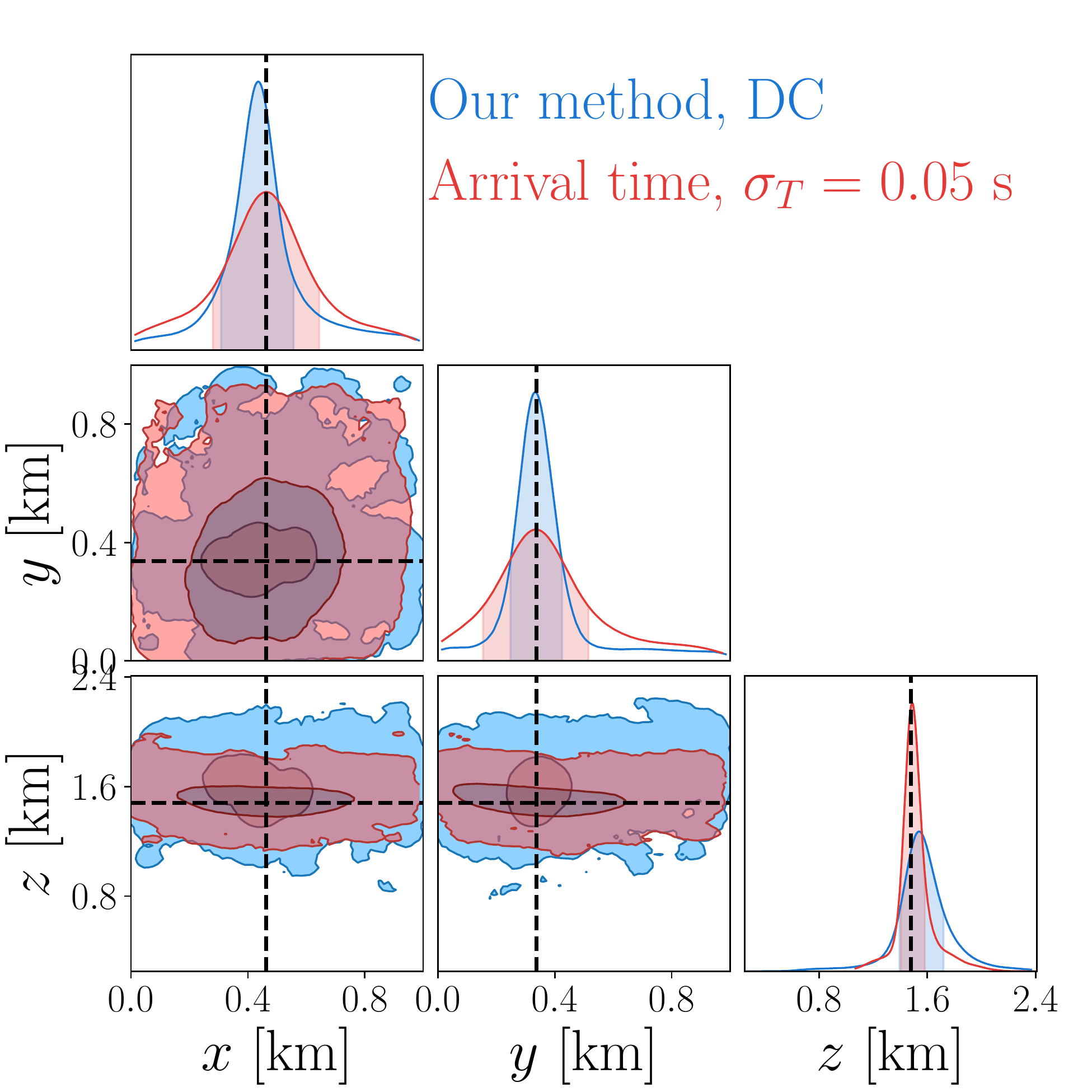}
    \end{center}
    \begin{center}
        \includegraphics[width=0.4\textwidth]{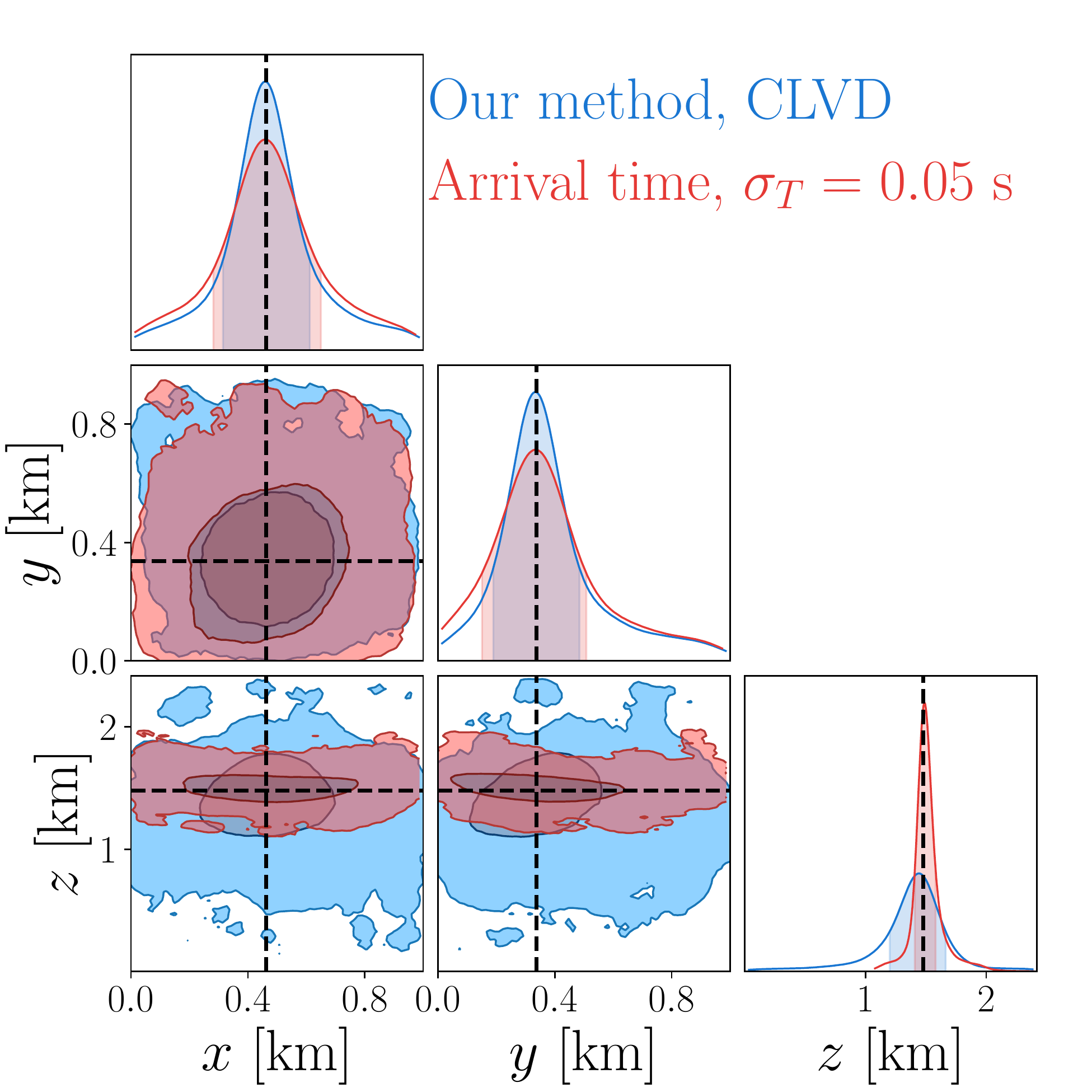}
    \end{center}    
\caption{Comparison of the marginalised 68 and 95 per cent credibility contours obtained with our method and with the arrival time approach described in Sect.~\ref{sec:arrival_time}. The event corresponds to event 2 in Table~\ref{tab:inference}. Note that we are considering 23 receivers in both cases, and for our approach the signal-to-noise ratio is 33 dB and the emulator was trained on 8000 training points.}
\label{fig:edt_compare}
\end{figure}

\subsection{Comparison with arrival time techniques}
\label{sec:arrival_time}
In order to establish a connection with existing standard earthquake location methods, we compare our results with a non-linear probabilistic location technique based on arrival times. We consider the \textsc{NonLinLoc} algorithm \citep{Lomax00}, which implements the LS-L2 approach \citep{Tarantola82, Moser92, Wittlinger93}. In this framework, the likelihood of the arrival time for a single observed event for a single receiver is:
\begin{equation}
     \mathcal{L}_{\rm{arrival \ time}} \propto \exp \left(-\frac{1}{2} \frac{\left(T_{\rm{obs}} - T_{\rm{calc}}\right)^2}{\sigma_T^2} \right) 
    \label{eq:edt}
\end{equation}
where $T_{\rm{obs}}$ is the observed arrival time, $T_{\rm{calc}}$ is the theoretical estimate for the travel time, and $\sigma_T$ is the error on the manual picking of arrival times. We use a fast marching method \citep{Sethian96} as implemented in \textsc{Pykonal} \citep{White20} to estimate theoretical arrival times given a set of coordinates for the geological model described in Sect.~\ref{sec:data}. We further replace the posterior sampler available in \textsc{NonLinLoc} with a \textsc{PyMultiNest} implementation for the likelihood in Eq.~\ref{eq:edt}.

We compare the location inference results for event 2 in Fig.~\ref{fig:edt_compare}. In this instance, we are interested in showing that there exists a reasonable value of the picking error $\sigma_T$ that yields comparable constraints to our method; this error is usually arbitrary and heavily depends on the SNR and prior information \citep{Smith19thesis, Abakumov20}. We remark that we are not interested in showing that the constraints obtained with our method are tighter than the ones obtained with the time-arrival information: while we start from the full waveform, we subsequently discard the phase information and part of the power spectrum, as we described in Sect.~\ref{sec:preprocessing}. We additionally note that the arrival time estimates are typically obtained an order of magnitude faster than our method, taking $\mathcal{O}(0.01 \ \rm{h})$ on the same commercial laptop used in Sect.~\ref{sec:speed}; this is not surprising, given the simple form of the likelihood in Eq.~\ref{eq:edt} and the efficiency of \textsc{Pykonal}.

Nevertheless, we show that with a picking error of $\sigma_T=0.05$ s, we obtain results which are generally in agreement with our approach. In particular, the constraints on the $z$ source coordinate are tighter in the arrival-time case, while the $x$ and $y$ coordinates are equally or better constrained by our approach; moreover, as in our method, the arrival-time analysis yields $z$ coordinates that are more constrained than the $x$ and $y$ ones. We speculate that this is due to the particular layered structure of the geological model described in Sect.~\ref{sec:data}, and, as anticipated in Sect.~\ref{sec:inference_results}, we expect the inference power of our approach to improve significantly by including the phase information into the analysis. We stress that our work is a step towards the fast joint inversion of coordinates and moment tensor components, for which arrival time techniques alone are not sufficient \citep{Dahm14, Pugh16, Alvizuri18}; therefore, we leave a full comparison to other techniques to future work. However, we further discuss the joint inversion approach and how to deal with more complicated noise types in Sect.~\ref{sec:conclusions}.

\subsection{Model selection results}
\label{sec:model_sel_res}
In general, a microseismic event can be described as a linear combination of the three source mechanisms described in Sect.~\ref{sec:data} \citep{Vavricuk15}. While we considered the three sources separately in this work, we show how the proposed Bayesian approach additionally allows for the identification of a source type given an observed seismogram. We consider event 1 as reported in Table~\ref{tab:inference}, and produce an observation for each source mechanism (ISO, DC and CLVD). As described in Sect.~\ref{sec:evidence}, we can run nested sampling for each event and for every trained emulator, and obtain 9 evidence values in total, whose logarithm we report in Table~\ref{tab:evidence}. As expected, the evidence is maximal in correspondence of the source type that generated the given event, which indicates that we are capable of correctly identifying the source type for a given observation. What is more, this selection is also very fast, as after training the emulators each evidence calculation takes less than 10 minutes on a commercial laptop.

\setlength{\tabcolsep}{3pt}

\begin{table}
\centering
\caption{Natural logarithm of the evidence for a source located at $(x,y,z)=(0.71 \ \rm{km},0.25 \ \rm{km},2.10 \ \rm{km})$ and source mechanism isotropic (ISO), double couple (DC) and compensated linear vector dipole (CLVD), as described in Sect.~\ref{sec:evidence} and Sect.~\ref{sec:model_sel_res}. The hypotheses correspond to an ISO ($\mathcal{H}_{\rm{ISO}}$), DC ($\mathcal{H}_{\rm{DC}}$) and CLVD ($\mathcal{H}_{\rm{CLVD}}$) source mechanism, respectively. We highlighted in bold the highest evidence in each line, which correctly corresponds to the known source mechanism. Note that the natural logarithm of the evidence is returned by \textsc{PyMultiNest}, and in this instance we ignored its associated error (which is very small).}
 \begin{tabular}{@{}cccc}
  \textbf{Event type}    &$\ln{ \mathrm{Pr} \left( \boldsymbol{D} | \mathcal{H}_{\rm{ISO}}\right)} $    & $\ln{ \mathrm{Pr} \left( \boldsymbol{D} | \mathcal{H}_{\rm{DC}}\right)} $     & $\ln{ \mathrm{Pr} \left( \boldsymbol{D} | \mathcal{H}_{\rm{CLVD}}\right)} $          \\
    \hline \hline 
  ISO & \textbf{2601} & $-$608 &  $-$79 \\
    \hline
  DC & $-$17688  & \textbf{$-$1165} & $-$2232                    \\
    \hline
  CLVD &  $-$6620 & $-$1052 & \textbf{$-$394}                    \\
 \end{tabular}
 \label{tab:evidence}
\end{table}

\subsection{Realistic noise and network configuration}
\label{sec:real}

\begin{figure}%
    \begin{center}
    \includegraphics[width=0.41\textwidth]{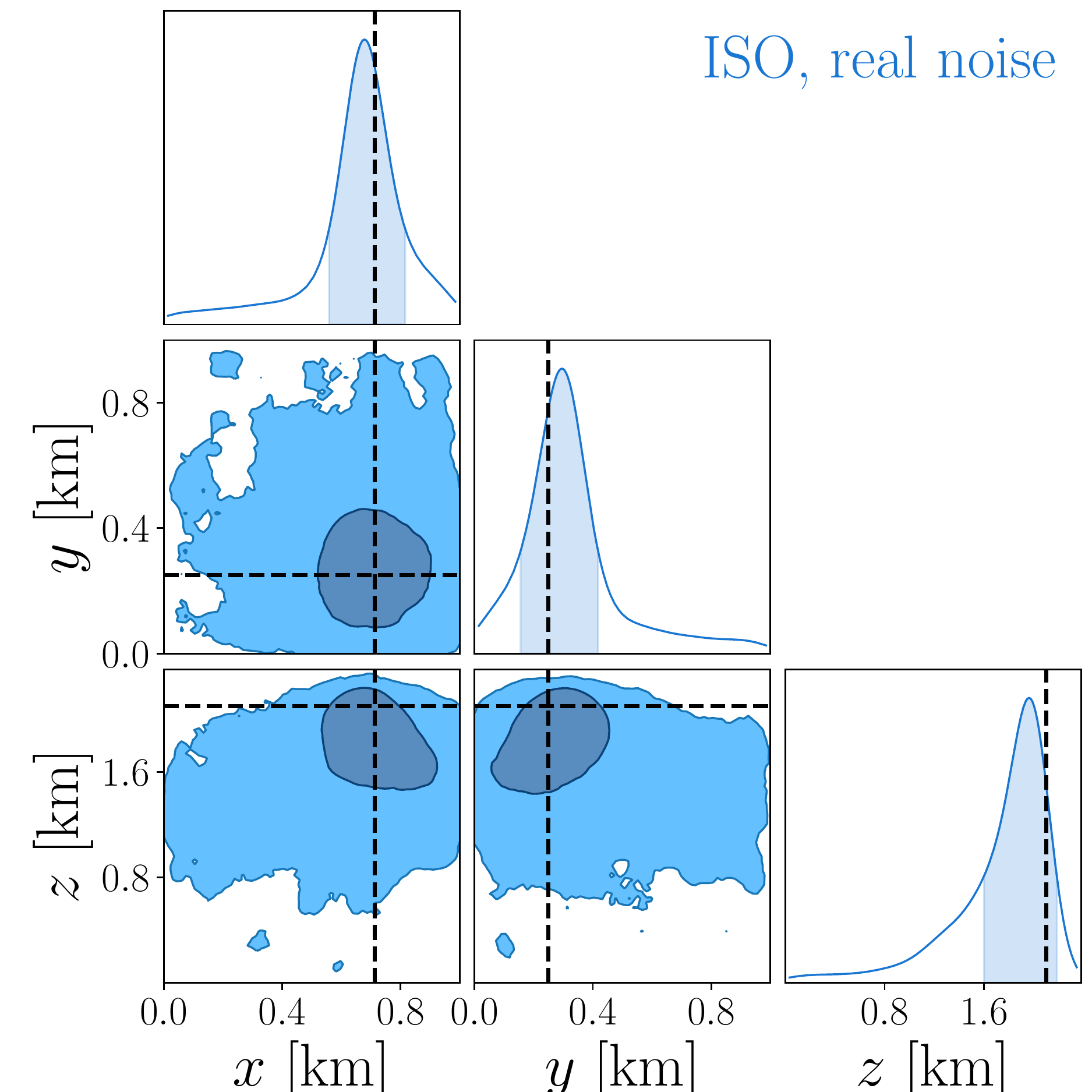}
    \includegraphics[width=0.41\textwidth]{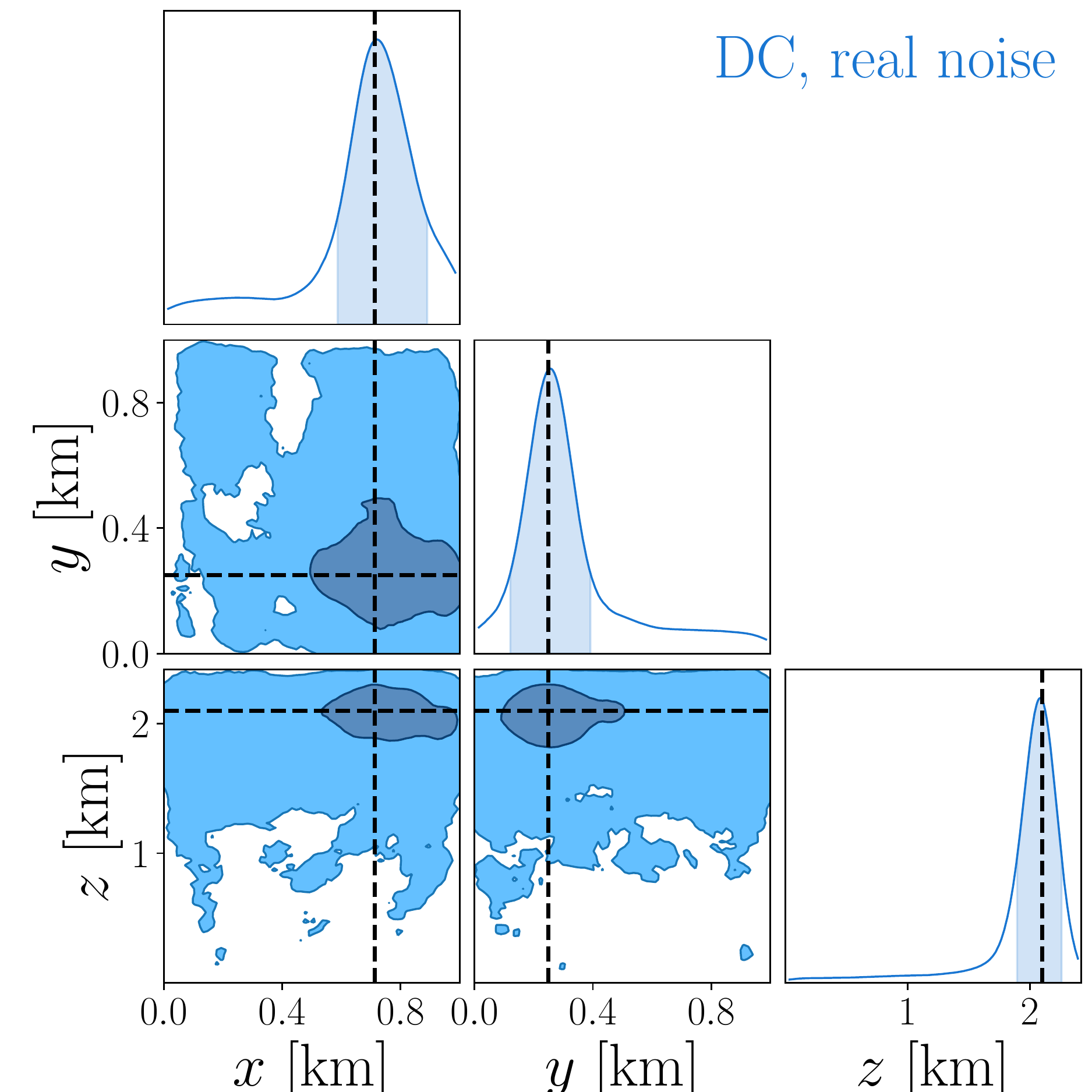}
    \end{center}
    \begin{center}
        \includegraphics[width=0.41\textwidth]{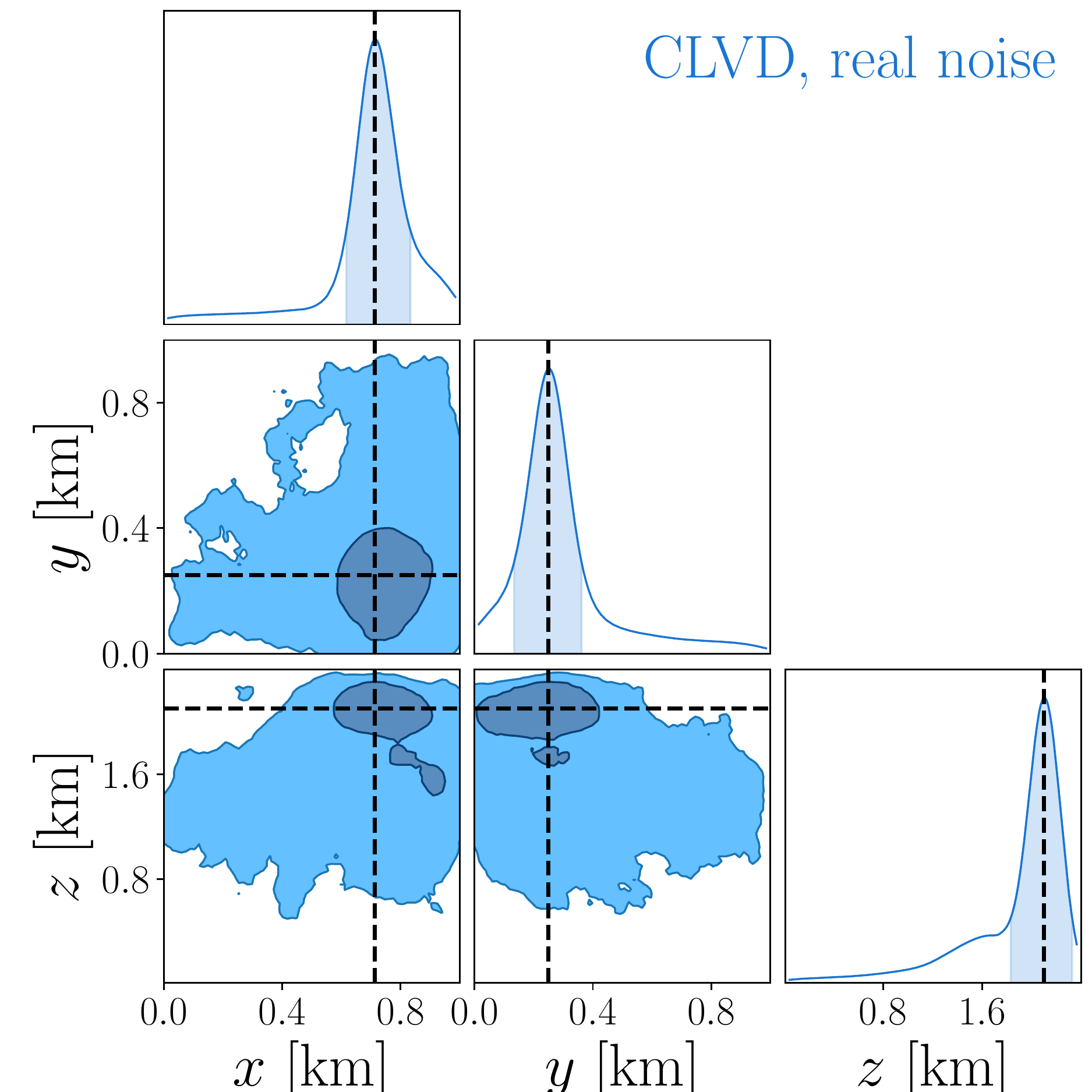}
    \end{center}    
\caption{Same as Fig.~\ref{fig:inference_1}, considering real noise from Groningen field data instead of Gaussian noise. As in our fiducial analysis, the signal-to-noise ratio is 33 dB, and we consider 23 receivers.}
\label{fig:real_noise}
\end{figure}

In this section, we explore the application of our approach to a more realistic scenario. We consider the same 23-receiver configuration, which is uniformly spread on the plane at $z=2.43$ km; however, instead of Gaussian noise, we add real noise traces extracted from the freely available Groningen field data set \citep{Smith19thesis}.

Our approach to obtain real noise traces is as follows. We cluster the receivers and synthetic traces in cubic volumes of side 0.5 km; this returns 100 clusters in total. We therefore take 100 different real waveforms among Groningen data, and cut the first 5 s window, well before the first wave arrival. We interpolate the signal in these windows in order to match the sampling rate of the synthetic signals; then, in order to reach the required SNR of 33 dB, we tune the amplitude of the noise signal we add to each synthetic waveform, and subtract the average signal so that the resulting noisy waveform has zero mean. We found that in some cases the SNR of the resulting noisy signals is lower than 33 dB: these correspond to waveforms originating at the edge of our simulated field, which, however, do not impact the inference analysis, as we show in this section.

Real noise is added to event 1 in Table~\ref{tab:inference}, as well as to all training waveforms to calculate the covariance matrix in Eq.~\ref{eq:glike}. We re-train the neural networks in the DC and CLVD cases removing the zero-frequency value of the power spectrum, in order to be insensitive to the scale of the noisy signals in the time domain. In Fig.~\ref{fig:real_noise} we show the posterior contours of our inference analysis using real noise: the results show little difference with respect to the case with Gaussian noise in Fig~\ref{fig:inference_1}, thus further demonstrating the robustness of our approach.

\section{Conclusions}
\label{sec:conclusions}
In this paper, we proposed a method that allows for the fast and accurate retrieval of the source coordinates for any microseismic source mechanism: isotropic (ISO), double couple (DC), and compensated linear vector dipole (CLVD). This offers an efficient technique to both locate an event and identify its source type, exploiting the power of machine learning and Bayesian tools to extract the information contained in seismic waveforms.

Our proposed method is based on a physically motivated preprocessing of the raw signals, using Fourier analysis and principal component compression, followed by the use of a neural network to learn the mapping between coordinates and principal components. Using the learnt forward model in combination with Bayesian techniques, we showed that we can retrieve an accurate estimate of any microseismic event coordinates, for any source mechanism, in less than $10$ minutes on a commercial laptop. Therefore, we demonstrated for the first time that machine learning techniques allow for a fast and accurate Bayesian analysis on microseismic traces, yielding competitive results on ISO sources and state-of-the-art results on DC and CLVD sources.

We showed that $\mathcal{O} (10^3)$ events for each source mechanism are enough to train a representative emulator, when using the data coming from $\mathcal{O}(10)$ receivers placed at the seabed as indicated in Fig.~\ref{fig:rec_pos}. We explored the effect of the noise level, and how the number of receivers and the number of training data for the emulator impact the accuracy of the coordinates' posterior distribution, demonstrating the robustness of our approach, which was also compared with a standard arrival-time inversion technique, showing good agreement for an appropriate choice of the picking error. We also showed that our method works in the presence of realistic noise statistics and network configuration. Finally, we demonstrated the utility of our Bayesian approach by calculating the Bayesian evidence for a given observation and three hypotheses, and showed that this correctly identifies the source type of any given event.

In conclusion, our work lays the foundations for the fast and reliable location of microseismic events with any source mechanism, given a minimal amount of computing resources. We also foresee that recent improvements in solving the forward model even outside the boundary of the training data, like physics-informed neural networks \citep[PINNs,][]{Raissi19, Xu19, Costa19, Moseley20}, could be combined with our proposed approach to make it even more robust, especially when different velocity models have to be employed.

Our work represents a step forward towards the fast Bayesian characterisation of microseismic events: after training our models on noiseless simulation data, it is possible to apply our method to each individual seismic trace as recorded by the receivers in order to obtain a fast and accurate source location estimate. We plan to integrate our approach into a joint inversion analysis, where both moment tensor components and coordinates are inferred: such analysis is usually slow and computationally expensive \citep[e.g.,][]{Pugh16}, and we therefore expect our approach to significantly accelerate it. Some straightforward extensions of our approach then include the following three points. First, a microseismic event is in general described by a linear mixture of components of the moment tensor, which we considered separately in this work. As a consequence, this implies that a larger set of parameters, including two more free parameters measuring the relative strength of the source types, has to be considered. We anticipate that this increase in the total number of parameters to infer 
will require a larger dataset to train the emulator; however, we note that our proposed method scales well with the number of training data, and therefore we anticipate that performing a Bayesian analysis with a larger parameter space is an attainable goal using our approach. Second we note that in order for this method to be deployed in a realistic scenario, the noise associated with the recorded seismograms should be modelled more carefully: a lower SNR may have to be considered, a more complicated likelihood distribution might have to be implemented, or a ``likelihood-free" approach should be investigated \citep{Sunnaker13}. Last, the errors in the 3-D density and velocity models should be incorporated into the analysis, in order to account for all sources of uncertainty \citep{Gesret13, Gesret14}. This will be addressed in future work.



\begin{acknowledgments}
We thank Saptarshi Das
for useful discussions, as well as Jonathan Smith and the anonymous reviewers for their helpful comments. DP was supported by the STFC UCL Centre for Doctoral Training in Data Intensive Science. Generation of the synthetic data used in this work has been performed in part on the Wilkes High Performance GPU computer cluster at the University of Cambridge, and in part on the Beaker cluster at UCL. 
This work was partially enabled by funding from Royal Dutch Shell plc, and used facilities provided by the UCL Cosmoparticle Initiative. We acknowledge the use of $\textsc{NumPy}$ \citep{Harris20}, $\textsc{Matplotlib}$ \citep{Hunter07}, $\textsc{TensorFlow}$ \citep{Abadi15}, $\textsc{SciPy}$ \citep{Virtanen20}, and $\textsc{ChainConsumer}$ \citep{Hinton16}.
\end{acknowledgments}

\begin{dataavailability}
The code to reproduce this work is publicly available in the GitHub repository at this link: \url{https://github.com/alessiospuriomancini/seismoML}. The data underlying this article will be shared on reasonable request to the corresponding author.
\end{dataavailability}

\newcommand{\newblock}{}
\bibliographystyle{apalike}
\bibliography{references}

\begin{thebibliography}{}

\bibitem[Abadi et~al., 2015]{Abadi15}
Abadi, M., Agarwal, A., Barham, P., Brevdo, E., Chen, Z., Citro, C., Corrado,
  G.~S., Davis, A., Dean, J., Devin, M., Ghemawat, S., Goodfellow, I., Harp,
  A., Irving, G., Isard, M., Jia, Y., Jozefowicz, R., Kaiser, L., Kudlur, M.,
  Levenberg, J., Man\'{e}, D., Monga, R., Moore, S., Murray, D., Olah, C.,
  Schuster, M., Shlens, J., Steiner, B., Sutskever, I., Talwar, K., Tucker, P.,
  Vanhoucke, V., Vasudevan, V., Vi\'{e}gas, F., Vinyals, O., Warden, P.,
  Wattenberg, M., Wicke, M., Yu, Y., and Zheng, X. (2015).
\newblock {TensorFlow}: Large-scale machine learning on heterogeneous systems.
\newblock Software available from tensorflow.org.

\bibitem[Abakumov et~al., 2020]{Abakumov20}
Abakumov, I., Roeser, A., and Shapiro, S.~A. (2020).
\newblock Arrival-time picking uncertainty: Theoretical estimations and their
  application to microseismic data.
\newblock {\em GEOPHYSICS}, 85(4):U65--U76.

\bibitem[Abreo-Carrillo et~al., 2015]{AbreoCarrillo15}
Abreo-Carrillo, S.~A., Ramirez, A.~B., Reyes, O., Abreo-Carrillo, D.~L., and
  Gonz\'{a}lez-Alvarez, H. (2015).
\newblock A practical implementation of acoustic full waveform inversion on
  graphical processing units.
\newblock {\em {CT\&F - Ciencia, Tecnolog\'{i}a y Futuro}}, 6(2):5 -- 16.

\bibitem[{Allison} and {Dunkley}, 2014]{Allison14}
{Allison}, R. and {Dunkley}, J. (2014).
\newblock {Comparison of sampling techniques for Bayesian parameter
  estimation}.
\newblock {\em Mon. Not. Roy. Astron. Soc.}, 437(4):3918--3928.

\bibitem[Alsing et~al., 2018]{Alsing18}
Alsing, J., Wandelt, B., and Feeney, S. (2018).
\newblock {Massive optimal data compression and density estimation for
  scalable, likelihood-free inference in cosmology}.
\newblock {\em Mon. Not. Roy. Astron. Soc.}, 477(3):2874--2885.

\bibitem[Alvizuri et~al., 2018]{Alvizuri18}
Alvizuri, C., Silwal, V., Krischer, L., and Tape, C. (2018).
\newblock Estimation of full moment tensors, including uncertainties, for
  nuclear explosions, volcanic events, and earthquakes.
\newblock {\em Journal of Geophysical Research: Solid Earth},
  123(6):5099--5119.

\bibitem[Angus et~al., 2014]{Angus14}
Angus, D., Aljaafari, A., Usher, P., and Verdon, J. (2014).
\newblock Seismic waveforms and velocity model heterogeneity: Towards a
  full-waveform microseismic location algorithm.
\newblock {\em Journal of Applied Geophysics}, 111:228 -- 233.

\bibitem[Auld et~al., 2007]{Auld07}
Auld, T., Bridges, M., Hobson, M., and Gull, S. (2007).
\newblock {Fast cosmological parameter estimation using neural networks}.
\newblock {\em Mon. Not. Roy. Astron. Soc.}, 376:L11--L15.

\bibitem[Auld et~al., 2008]{Auld08}
Auld, T., Bridges, M., and Hobson, M.~P. (2008).
\newblock cosmonet: fast cosmological parameter estimation in non-flat models
  using neural networks.
\newblock {\em Monthly Notices of the Royal Astronomical Society},
  387(4):1575–1582.

\bibitem[Bergen et~al., 2019]{Bergen19}
Bergen, K.~J., Johnson, P.~A., de~Hoop, M.~V., and Beroza, G.~C. (2019).
\newblock Machine learning for data-driven discovery in solid earth geoscience.
\newblock {\em Science}, 363(6433):eaau0323.

\bibitem[Bishop, 2006]{Bishop06}
Bishop, C.~M. (2006).
\newblock {\em Pattern Recognition and Machine Learning (Information Science
  and Statistics)}.
\newblock Springer-Verlag, Berlin, Heidelberg.

\bibitem[Brueckl et~al., 2008]{Brueckl08}
Brueckl, E., Binder, D., Hausmann, H., and Mertl, S. (2008).
\newblock Hazard {E}stimation of {D}eep {S}eated {M}ass {M}ovements by
  {M}icroseismic {M}onitoring.
\newblock In {\em International Strategy for Disaster Reduction}. Austrian
  Academy of Sciences.

\bibitem[{Buchner} et~al., 2014]{Buchner14}
{Buchner}, J., {Georgakakis}, A., {Nandra}, K., {Hsu}, L., {Rangel}, C.,
  {Brightman}, M., {Merloni}, A., {Salvato}, M., {Donley}, J., and {Kocevski},
  D. (2014).
\newblock {X-ray spectral modelling of the AGN obscuring region in the CDFS:
  Bayesian model selection and catalogue}.
\newblock {\em \rm{AAP}}, 564:A125.

\bibitem[{Chenthamarakshan} et~al., 2020]{Chenthamarakshan20}
{Chenthamarakshan}, V., {Das}, P., {Hoffman}, S.~C., {Strobelt}, H., {Padhi},
  I., {Lim}, K.~W., {Hoover}, B., {Manica}, M., {Born}, J., {Laino}, T., and
  {Mojsilovic}, A. (2020).
\newblock {CogMol: Target-Specific and Selective Drug Design for COVID-19 Using
  Deep Generative Models}.
\newblock {\em arXiv e-prints}, page arXiv:2004.01215.

\bibitem[{Clevert} et~al., 2015]{Clevert16}
{Clevert}, D.-A., {Unterthiner}, T., and {Hochreiter}, S. (2015).
\newblock {Fast and Accurate Deep Network Learning by Exponential Linear Units
  (ELUs)}.
\newblock {\em arXiv e-prints}, page arXiv:1511.07289.

\bibitem[Collettini and Barchi, 2002]{Collettini02}
Collettini, C. and Barchi, M.~R. (2002).
\newblock A low-angle normal fault in the {U}mbria region ({C}entral {I}taly):
  a mechanical model for the related microseismicity.
\newblock {\em Tectonophysics}, 359(1):97 -- 115.

\bibitem[Conrad et~al., 2016]{Conrad16}
Conrad, P.~R., Marzouk, Y.~M., Pillai, N.~S., and Smith, A. (2016).
\newblock Accelerating {A}symptotically {E}xact {MCMC} for {C}omputationally
  {I}ntensive {M}odels via {L}ocal {A}pproximations.
\newblock {\em Journal of the American Statistical Association},
  111(516):1591--1607.

\bibitem[Cooley and Tukey, 1965]{Cooley65}
Cooley, J.~W. and Tukey, J.~W. (1965).
\newblock {An Algorithm for the Machine Calculation of Complex Fourier Series}.
\newblock {\em Math. Comput.}, 19:297--301.

\bibitem[Costa Nogueira~Junior et~al., 2019]{Costa19}
Costa Nogueira~Junior, A., De~Sousa~Almeida, J., Paredes~Quiñones, M., and
  de~Albuquerque~Martins, L.~S. (2019).
\newblock Physics-based machine learning inversion of subsurface elastic
  properties.
\newblock {\em Conference Proceedings, 81st EAGE Conference and Exhibition},
  2019(1):1--5.

\bibitem[Craiu and Rosenthal, 2014]{Craiu14}
Craiu, R.~V. and Rosenthal, J.~S. (2014).
\newblock Bayesian {C}omputation via {M}arkov {C}hain {M}onte {C}arlo.
\newblock {\em Annual Review of Statistics and Its Application}, 1(1):179--201.

\bibitem[Dahm and Kr{\"u}ger, 2014]{Dahm14}
Dahm, T. and Kr{\"u}ger, F. (2014).
\newblock Moment tensor inversion and moment tensor interpretation.
\newblock In Bormann, P., editor, {\em New Manual of Seismological Observatory
  Practice 2 (NMSOP-2)}. Deutsches GeoForschungsZentrum GFZ, 1-37.

\bibitem[{Das} et~al., 2017]{Das17}
{Das}, S., {Chen}, X., and {Hobson}, M.~P. (2017).
\newblock Fast {GPU}-{B}ased {S}eismogram {S}imulation {F}rom {M}icroseismic
  {E}vents in {M}arine {E}nvironments {U}sing {H}eterogeneous {V}elocity
  {M}odels.
\newblock {\em IEEE Transactions on Computational Imaging}, 3(2):316--329.

\bibitem[Das et~al., 2018]{Das18}
Das, S., Chen, X., Hobson, M.~P., Phadke, S., van Beest, B., Goudswaard, J.,
  and Hohl, D. (2018).
\newblock {Surrogate regression modelling for fast seismogram generation and
  detection of microseismic events in heterogeneous velocity models}.
\newblock {\em Geophysical Journal International}, 215(2):1257--1290.

\bibitem[Ellsworth, 2013]{Ellsworth13}
Ellsworth, W.~L. (2013).
\newblock Injection-induced earthquakes.
\newblock {\em Science}, 341(6142).

\bibitem[Feroz et~al., 2009]{Feroz08}
Feroz, F., Hobson, M., and Bridges, M. (2009).
\newblock {MultiNest: an efficient and robust Bayesian inference tool for
  cosmology and particle physics}.
\newblock {\em Mon. Not. Roy. Astron. Soc.}, 398:1601--1614.

\bibitem[Ferreira and Woodhouse, 2007]{Ferreira07}
Ferreira, A. M.~G. and Woodhouse, J.~H. (2007).
\newblock {Source, path and receiver effects on seismic surface waves}.
\newblock {\em Geophysical Journal International}, 168(1):109--132.

\bibitem[Geiger, 1910]{Geiger10}
Geiger, L. (1910).
\newblock {H}erdbestimmung bei {E}rdbeben aus den {A}nkunftszeiten,
  {N}achrichten der {K}. {G}esellschaft der {W}issenschaften zu {G}ottingen.
\newblock {\em Math.-Phys. Klasse}, 1910:331--349.

\bibitem[Gesret et~al., 2014]{Gesret14}
Gesret, A., Desassis, N., Noble, M., Romary, T., and Maisons, C. (2014).
\newblock {Propagation of the velocity model uncertainties to the seismic event
  location}.
\newblock {\em Geophysical Journal International}, 200(1):52--66.

\bibitem[Gesret et~al., 2013]{Gesret13}
Gesret, A., Noble, M., Desassis, N., and Romary, T. (2013).
\newblock Microseismic {M}onitoring - {C}onsequences of {V}elocity {M}odel
  {U}ncertainties on {E}vent {L}ocation {U}ncertainties.
\newblock {\em Proceedings of the Third Passive Seismic Workshop, Eur. Ass. of
  Geoscientists and Engineers, Athens - Greece (2011)}.

\bibitem[Goodfellow et~al., 2014]{Goodfellow14}
Goodfellow, I., Pouget-Abadie, J., Mirza, M., Xu, B., Warde-Farley, D., Ozair,
  S., Courville, A., and Bengio, Y. (2014).
\newblock Generative {A}dversarial {N}ets.
\newblock In Ghahramani, Z., Welling, M., Cortes, C., Lawrence, N.~D., and
  Weinberger, K.~Q., editors, {\em Advances in Neural Information Processing
  Systems 27}, pages 2672--2680. Curran Associates, Inc.

\bibitem[{Gulrajani} et~al., 2016]{Gulrajani16}
{Gulrajani}, I., {Kumar}, K., {Ahmed}, F., {Taiga}, A.~A., {Visin}, F.,
  {Vazquez}, D., and {Courville}, A. (2016).
\newblock {PixelVAE: A Latent Variable Model for Natural Images}.
\newblock {\em arXiv e-prints}, page arXiv:1611.05013.

\bibitem[Harris et~al., 2020]{Harris20}
Harris, C.~R., Millman, K.~J., van~der Walt, S.~J., Gommers, R., Virtanen, P.,
  Cournapeau, D., Wieser, E., Taylor, J., Berg, S., Smith, N.~J., Kern, R.,
  Picus, M., Hoyer, S., van Kerkwijk, M.~H., Brett, M., Haldane, A., del
  R{'{\i}}o, J.~F., Wiebe, M., Peterson, P., G{'{e}}rard-Marchant, P.,
  Sheppard, K., Reddy, T., Weckesser, W., Abbasi, H., Gohlke, C., and Oliphant,
  T.~E. (2020).
\newblock Array programming with {NumPy}.
\newblock {\em Nature}, 585(7825):357--362.

\bibitem[Haykin, 2001]{Haykin01}
Haykin, S. (2001).
\newblock {\em Communication Systems}.
\newblock Wiley.

\bibitem[{Hinton}, 2016]{Hinton16}
{Hinton}, S.~R. (2016).
\newblock {ChainConsumer}.
\newblock {\em The Journal of Open Source Software}, 1:00045.

\bibitem[Hornik et~al., 1989]{Hornik89}
Hornik, K., Stinchcombe, M., and White, H. (1989).
\newblock Multilayer feedforward networks are universal approximators.
\newblock {\em Neural Networks}, 2(5):359 -- 366.

\bibitem[Hunter, 2007]{Hunter07}
Hunter, J.~D. (2007).
\newblock Matplotlib: A 2d graphics environment.
\newblock {\em Computing in Science \& Engineering}, 9(3):90--95.

\bibitem[Jakobsen and Ursin, 2015]{Jakobsen15}
Jakobsen, M. and Ursin, B. (2015).
\newblock Full waveform inversion in the frequency domain using direct
  iterative {T}-matrix methods.
\newblock {\em Journal of Geophysics and Engineering}, 12:400--418.

\bibitem[{Kasim} et~al., 2020]{Kasim20}
{Kasim}, M.~F., {Watson-Parris}, D., {Deaconu}, L., {Oliver}, S., {Hatfield},
  P., {Froula}, D.~H., {Gregori}, G., {Jarvis}, M., {Khatiwala}, S.,
  {Korenaga}, J., {Topp-Mugglestone}, J., {Viezzer}, E., and {Vinko}, S.~M.
  (2020).
\newblock {Building high accuracy emulators for scientific simulations with
  deep neural architecture search}.
\newblock {\em arXiv e-prints}, page arXiv:2001.08055.

\bibitem[{Kingma} and {Ba}, 2014]{Kingma14}
{Kingma}, D.~P. and {Ba}, J. (2014).
\newblock Adam: {A} {M}ethod for {S}tochastic {O}ptimization.
\newblock {\em arXiv e-prints}, page arXiv:1412.6980.

\bibitem[Knopoff and Randall, 1970]{Knopoff70}
Knopoff, L. and Randall, M.~J. (1970).
\newblock The compensated linear-vector dipole: A possible mechanism for deep
  earthquakes.
\newblock {\em Journal of Geophysical Research (1896-1977)}, 75(26):4957--4963.

\bibitem[Knuth et~al., 2015]{Knuth15}
Knuth, K.~H., Habeck, M., Malakar, N.~K., Mubeen, A.~M., and Placek, B. (2015).
\newblock Bayesian evidence and model selection.
\newblock {\em Digital Signal Processing}, 47:50 -- 67.
\newblock Special Issue in Honour of William J. (Bill) Fitzgerald.

\bibitem[{Kolen} and {Kremer}, 2001]{Kolen01}
{Kolen}, J.~F. and {Kremer}, S.~C. (2001).
\newblock {\em Gradient Flow in Recurrent Nets: The Difficulty of Learning
  LongTerm Dependencies}, pages 237--243.
\newblock Wiley-IEEE Press.

\bibitem[LeNail, 2019]{LeNail19}
LeNail, A. (2019).
\newblock {NN-SVG}: {P}ublication-{R}eady {N}eural {N}etwork {A}rchitecture
  {S}chematics.
\newblock {\em Journal of Open Source Software}, 4(33):747.

\bibitem[Li et~al., 2015]{Li15}
Li, H., Wang, R., and Cao, S. (2015).
\newblock {Microseismic forward modeling based on different focal mechanisms
  used by the seismic moment tensor and elastic wave equation}.
\newblock {\em Journal of Geophysics and Engineering}, 12(2):155--166.

\bibitem[Li, 2013]{Li13}
Li, J. (2013).
\newblock {\em Study of induced seismicity for reservoir characterization}.
\newblock PhD thesis, Massachusetts Institute of Technology. Department of
  Earth, Atmospheric, and Planetary Sciences.

\bibitem[Li et~al., 2018]{Li18}
Li, J., Ji, S., Li, Y., Qian, Z., and Lu, W. (2018).
\newblock {Downhole microseismic signal-to-noise ratio enhancement via strip
  matching shearlet transform}.
\newblock {\em Journal of Geophysics and Engineering}, 15(2):330--337.

\bibitem[Li et~al., 2016]{Li16}
Li, J., Kuehl, H., Droujinine, A., and Blokland, J.-W. (2016).
\newblock {\em Microseismic and induced seismicity simultaneous location and
  moment tensor inversion: Moving beyond picks with a robust full-waveform
  method}, pages 2535--2539.
\newblock Society of Exploration Geophysicists.

\bibitem[Li et~al., 2020]{Li20}
Li, L., Tan, J., Schwarz, B., Staněk, F., Poiata, N., Shi, P., Diekmann, L.,
  Eisner, L., and Gajewski, D. (2020).
\newblock {R}ecent {A}dvances and {C}hallenges of {W}aveform-{B}ased {S}eismic
  {L}ocation {M}ethods at {M}ultiple {S}cales.
\newblock {\em Reviews of Geophysics}, 58(1):e2019RG000667.

\bibitem[Li et~al., 2022]{Li22}
Li, Z., Zhu, L., Officer, T., Shi, F., Yu, T., and Wang, Y. (2022).
\newblock {A machine-learning-based method of detecting and picking the first
  P-wave arrivals of acoustic emission events in laboratory experiments}.
\newblock {\em Geophysical Journal International}, 230(3):1818--1823.

\bibitem[{Liu} et~al., 2017]{Liu17}
{Liu}, E., {Zhu}, L., {Govinda Raj}, A., {McClellan}, J.~H., {Al-Shuhail}, A.,
  {Kaka}, S.~I., and {Iqbal}, N. (2017).
\newblock {Microseismic events enhancement and detection in sensor arrays using
  autocorrelation-based filtering}.
\newblock {\em Geophysical Prospecting}, 65(6):1496--1509.

\bibitem[{Liu} et~al., 2018]{Liu18}
{Liu}, H., {Ong}, Y.-S., {Shen}, X., and {Cai}, J. (2018).
\newblock {When Gaussian Process Meets Big Data: A Review of Scalable GPs}.
\newblock {\em arXiv e-prints}, page arXiv:1807.01065.

\bibitem[Lomax et~al., 2009]{Lomax09}
Lomax, A., Michelini, A., and Curtis, A. (2009).
\newblock {\em Earthquake Location, Direct, Global-Search Methods}, pages
  1--33.
\newblock Springer New York, New York, NY.

\bibitem[Lomax et~al., 2000]{Lomax00}
Lomax, A., Virieux, J., Volant, P., and Berge-Thierry, C. (2000).
\newblock {\em Probabilistic Earthquake Location in 3D and Layered Models},
  pages 101--134.
\newblock Springer Netherlands, Dordrecht.

\bibitem[Maas et~al., 2013]{Maas13}
Maas, A.~L., Hannun, A.~Y., Ng, A.~Y., et~al. (2013).
\newblock Rectifier nonlinearities improve neural network acoustic models.
\newblock In {\em Proc. ICML}, volume~30, page~3.

\bibitem[Majer et~al., 2007]{Majer07}
Majer, E.~L., Baria, R., Stark, M., Oates, S., Bommer, J., Smith, B., and
  Asanuma, H. (2007).
\newblock Induced seismicity associated with {E}nhanced {G}eothermal {S}ystems.
\newblock {\em Geothermics}, 36(3):185 -- 222.

\bibitem[{Moseley} et~al., 2020]{Moseley20}
{Moseley}, B., {Markham}, A., and {Nissen-Meyer}, T. (2020).
\newblock {Solving the wave equation with physics-informed deep learning}.
\newblock {\em arXiv e-prints}, page arXiv:2006.11894.

\bibitem[Moser et~al., 1992]{Moser92}
Moser, T.~J., van Eck, T., and Nolet, G. (1992).
\newblock Hypocenter determination in strongly heterogeneous earth models using
  the shortest path method.
\newblock {\em Journal of Geophysical Research: Solid Earth},
  97(B5):6563--6572.

\bibitem[Mousavi et~al., 2020]{Mousavi20}
Mousavi, S.~M., Ellsworth, W.~L., Zhu, W., Chuang, L.~Y., and Beroza, G.~C.
  (2020).
\newblock Earthquake transformer—an attentive deep-learning model for
  simultaneous earthquake detection and phase picking.
\newblock {\em Nature Communications}, 11(1):1--12.

\bibitem[Mukuhira et~al., 2016]{Mukuhira16}
Mukuhira, Y., Asanuma, H., Ito, T., and H\"{a}ring, M.~O. (2016).
\newblock Physics-based seismic evaluation method: Evaluating possible seismic
  moment based on microseismic information due to fluid stimulation.
\newblock {\em \textsc{Geophysics}}, 81(6):KS195--KS205.

\bibitem[Noack and Clark, 2017]{Noack17}
Noack, M.~M. and Clark, S. (2017).
\newblock Acoustic wave and eikonal equations in a transformed metric space for
  various types of anisotropy.
\newblock {\em Heliyon}, 3(3):e00260.

\bibitem[Papoulis et~al., 2002]{Papoulis02}
Papoulis, A., Pillai, S., and Pillai, S. (2002).
\newblock {\em Probability, Random Variables, and Stochastic Processes}.
\newblock McGraw-Hill electrical and electronic engineering series.
  McGraw-Hill.

\bibitem[Pratt, 1999]{Pratt99a}
Pratt, R. (1999).
\newblock Seismic waveform inversion in the frequency domain; {P}art 1,
  {T}heory and verification in a physical scale model.
\newblock {\em Geophysics}, 64:888--901.

\bibitem[Pratt and Shipp, 1999]{Pratt99b}
Pratt, R. and Shipp, R. (1999).
\newblock Seismic waveform inversion in the frequency domain, {P}art 2: {F}ault
  delineation in sediments using crosshole data.
\newblock {\em Geophysics}, 64:902--914.

\bibitem[Pugh et~al., 2016]{Pugh16}
Pugh, D.~J., White, R.~S., and Christie, P. A.~F. (2016).
\newblock {A Bayesian method for microseismic source inversion}.
\newblock {\em Geophysical Journal International}, 206(2):1009--1038.

\bibitem[{Raissi} et~al., 2019]{Raissi19}
{Raissi}, M., {Perdikaris}, P., and {Karniadakis}, G.~E. (2019).
\newblock {Physics-informed neural networks: A deep learning framework for
  solving forward and inverse problems involving nonlinear partial differential
  equations}.
\newblock {\em Journal of Computational Physics}, 378:686--707.

\bibitem[{Rajaratnam} and {Sparks}, 2015]{Rajaratnam15}
{Rajaratnam}, B. and {Sparks}, D. (2015).
\newblock {MCMC-Based Inference in the Era of Big Data: A Fundamental Analysis
  of the Convergence Complexity of High-Dimensional Chains}.
\newblock {\em arXiv e-prints}, page arXiv:1508.00947.

\bibitem[Rasmussen and Williams, 2005]{Rasmussen05}
Rasmussen, C.~E. and Williams, C. K.~I. (2005).
\newblock {\em Gaussian Processes for Machine Learning (Adaptive Computation
  and Machine Learning)}.
\newblock The MIT Press.

\bibitem[Ross et~al., 2018]{Ross18}
Ross, Z.~E., Meier, M.-A., and Hauksson, E. (2018).
\newblock P wave arrival picking and first-motion polarity determination with
  deep learning.
\newblock {\em Journal of Geophysical Research: Solid Earth},
  123(6):5120--5129.

\bibitem[Sethian, 1996]{Sethian96}
Sethian, J.~A. (1996).
\newblock A fast marching level set method for monotonically advancing fronts.
\newblock {\em Proceedings of the National Academy of Sciences},
  93(4):1591--1595.

\bibitem[Shapiro et~al., 2010]{Shapiro10}
Shapiro, S.~A., Dinske, C., Langenbruch, C., and Wenzel, F. (2010).
\newblock Seismogenic index and magnitude probability of earthquakes induced
  during reservoir fluid stimulations.
\newblock {\em The Leading Edge}, 29(3):S. 304--309.

\bibitem[Skilling, 2006]{Skilling06}
Skilling, J. (2006).
\newblock {Nested sampling for general Bayesian computation}.
\newblock {\em Bayesian Analysis}, 1(4):833--859.

\bibitem[Smith, 2019]{Smith19thesis}
Smith, J. (2019).
\newblock {\em Geomechanical properties of the Groningen reservoir}.
\newblock PhD thesis, University of Cambridge, Department of Earth Sciences -
  Bullard Laboratories.

\bibitem[{Smith} et~al., 2020]{Smith20}
{Smith}, J.~D., {Azizzadenesheli}, K., and {Ross}, Z.~E. (2020).
\newblock {EikoNet: Solving the Eikonal equation with Deep Neural Networks}.
\newblock {\em arXiv e-prints}, page arXiv:2004.00361.

\bibitem[Song and Toksöz, 2011]{Song11}
Song, F. and Toksöz, M.~N. (2011).
\newblock Full-waveform based complete moment tensor inversion and source
  parameter estimation from downhole microseismic data for hydrofracture
  monitoring.
\newblock {\em \textsc{Geophysics}}, 76(6):WC103--WC116.

\bibitem[Spurio~Mancini et~al., 2022]{SpurioMancini21}
Spurio~Mancini, A., Piras, D., Alsing, J., Joachimi, B., and Hobson, M.~P.
  (2022).
\newblock {CosmoPower: emulating cosmological power spectra for accelerated
  Bayesian inference from next-generation surveys}.
\newblock {\em Monthly Notices of the Royal Astronomical Society},
  511(2):1771--1788.

\bibitem[Spurio~Mancini et~al., 2021]{ASM20}
Spurio~Mancini, A., Piras, D., Ferreira, A. M.~G., Hobson, M.~P., and Joachimi,
  B. (2021).
\newblock Accelerating {B}ayesian microseismic event location with deep
  learning.
\newblock {\em Solid Earth Discussions}, 2021:1--36.

\bibitem[St\"ahler and Sigloch, 2014]{Staehler14}
St\"ahler, S.~C. and Sigloch, K. (2014).
\newblock Fully probabilistic seismic source inversion -- part 1: Efficient
  parameterisation.
\newblock {\em Solid Earth}, 5(2):1055--1069.

\bibitem[St\"ahler and Sigloch, 2016]{Staehler16}
St\"ahler, S.~C. and Sigloch, K. (2016).
\newblock Fully probabilistic seismic source inversion -- part 2: Modelling
  errors and station covariances.
\newblock {\em Solid Earth}, 7(6):1521--1536.

\bibitem[Sunnåker et~al., 2013]{Sunnaker13}
Sunnåker, M., Busetto, A.~G., Numminen, E., Corander, J., Foll, M., and
  Dessimoz, C. (2013).
\newblock {A}pproximate {B}ayesian {C}omputation.
\newblock {\em PLOS Computational Biology}, 9(1):1--10.

\bibitem[Tao and Sen, 2013]{Tao13}
Tao, Y. and Sen, M.~K. (2013).
\newblock Frequency-domain full waveform inversion with a scattering-integral
  approach and its sensitivity analysis.
\newblock {\em Journal of Geophysics and Engineering}, 10:065008.

\bibitem[Tarantola, 2005]{Tarantola05}
Tarantola, A. (2005).
\newblock {\em Inverse Problem Theory and Methods for Model Parameter
  Estimation}.
\newblock Society for Industrial and Applied Mathematics.

\bibitem[Tarantola and Valette, 1982]{Tarantola82}
Tarantola, A. and Valette, B. (1982).
\newblock Generalized nonlinear inverse problems solved using the least squares
  criterion.
\newblock {\em Reviews of Geophysics}, 20(2):219--232.

\bibitem[Thornton, 2013]{Thornton13}
Thornton, M. (2013).
\newblock Velocity uncertainties in surface and downhole monitoring.
\newblock {\em Conference Proceedings, 4th EAGE Passive Seismic Workshop}.

\bibitem[{Treeby} et~al., 2014]{Treeby14}
{Treeby}, B.~E., {Jaros}, J., {Rohrbach}, D., and {Cox}, B.~T. (2014).
\newblock Modelling elastic wave propagation using the k-{W}ave {MATLAB}
  {T}oolbox.
\newblock In {\em 2014 IEEE International Ultrasonics Symposium}, pages
  146--149.

\bibitem[Usher et~al., 2013]{Usher13}
Usher, P., Angus, D., and Verdon, J. (2013).
\newblock Influence of a velocity model and source frequency on microseismic
  waveforms: some implications for microseismic locations.
\newblock {\em Geophysical Prospecting}, 61(s1):334--345.

\bibitem[Vasco et~al., 2019]{Vasco19}
Vasco, D.~W., Nakagawa, S., Petrov, P., and Newman, G. (2019).
\newblock {Rapid estimation of earthquake locations using waveform
  traveltimes}.
\newblock {\em Geophysical Journal International}, 217(3):1727--1741.

\bibitem[{Vavry{\v{c}}uk}, 2015]{Vavricuk15}
{Vavry{\v{c}}uk}, V. (2015).
\newblock {Moment tensor decompositions revisited}.
\newblock {\em Journal of Seismology}, 19(1):231--252.

\bibitem[Vavryčuk, 2001]{Vavrycuk01}
Vavryčuk, V. (2001).
\newblock Inversion for parameters of tensile earthquakes.
\newblock {\em Journal of Geophysical Research: Solid Earth},
  106(B8):16339--16355.

\bibitem[Vavryčuk, 2005]{Vavrycuk05}
Vavryčuk, V. (2005).
\newblock {Focal mechanisms in anisotropic media}.
\newblock {\em Geophysical Journal International}, 161(2):334--346.

\bibitem[Virtanen et~al., 2020]{Virtanen20}
Virtanen, P., Gommers, R., Oliphant, T.~E., Haberland, M., Reddy, T.,
  Cournapeau, D., Burovski, E., Peterson, P., Weckesser, W., Bright, J., {van
  der Walt}, S.~J., Brett, M., Wilson, J., Millman, K.~J., Mayorov, N., Nelson,
  A. R.~J., Jones, E., Kern, R., Larson, E., Carey, C.~J., Polat, {\.I}., Feng,
  Y., Moore, E.~W., {VanderPlas}, J., Laxalde, D., Perktold, J., Cimrman, R.,
  Henriksen, I., Quintero, E.~A., Harris, C.~R., Archibald, A.~M., Ribeiro,
  A.~H., Pedregosa, F., {van Mulbregt}, P., and {SciPy 1.0 Contributors}
  (2020).
\newblock {{SciPy} 1.0: Fundamental Algorithms for Scientific Computing in
  Python}.
\newblock {\em Nature Methods}, 17:261--272.

\bibitem[White et~al., 2020]{White20}
White, M. C.~A., Fang, H., Nakata, N., and Ben‐Zion, Y. (2020).
\newblock {PyKonal: A Python Package for Solving the Eikonal Equation in
  Spherical and Cartesian Coordinates Using the Fast Marching Method}.
\newblock {\em Seismological Research Letters}, 91(4):2378--2389.

\bibitem[Willacy et~al., 2019]{Willacy19}
Willacy, C., van Dedem, E., Minisini, S., Li, J., Blokland, J.-W., Das, I., and
  Droujinine, A. (2019).
\newblock Full-waveform event location and moment tensor inversion for induced
  seismicity.
\newblock {\em \textsc{Geophysics}}, 84(2):KS39--KS57.

\bibitem[Wittlinger et~al., 1993]{Wittlinger93}
Wittlinger, G., Herquel, G., and Nakache, T. (1993).
\newblock {Earthquake location in strongly heterogeneous media}.
\newblock {\em Geophysical Journal International}, 115(3):759--777.

\bibitem[{Wuestefeld} et~al., 2018]{Wuestfeld18}
{Wuestefeld}, A., {Greve}, S.~M., {N{\"a}sholm}, S.~P., and {Oye}, V. (2018).
\newblock {Benchmarking earthquake location algorithms: A synthetic
  comparison}.
\newblock {\em \textsc{Geophysics}}, 83(4):KS35--KS47.

\bibitem[Xu et~al., 2019]{Xu19}
Xu, Y., Li, J., and Chen, X. (2019).
\newblock {\em Physics informed neural networks for velocity inversion}, pages
  2584--2588.
\newblock Society of Exploration Geophysicists.

\bibitem[Yao et~al., 2007]{Yao07}
Yao, Y., Rosasco, L., and Caponnetto, A. (2007).
\newblock On {E}arly {S}topping in {G}radient {D}escent {L}earning.
\newblock {\em Constructive Approximation}, 26:289--315.

\bibitem[Zhang et~al., 2020]{Zhang20}
Zhang, J., Dong, L., and Xu, N. (2020).
\newblock {N}oise {S}uppression of {M}icroseismic {S}ignals via {A}daptive
  {V}ariational {M}ode {D}ecomposition and {A}kaike {I}nformation {C}riterion.
\newblock {\em Applied Sciences}, 10(11):3790.

\bibitem[{Zheng} et~al., 2018]{Zheng18}
{Zheng}, J., {Lu}, J., {Peng}, S., and {Jiang}, T. (2018).
\newblock {An automatic microseismic or acoustic emission arrival
  identification scheme with deep recurrent neural networks}.
\newblock {\em Geophysical Journal International}, 212(2):1389--1397.

\bibitem[{Zhu} et~al., 2019]{Zhu19}
{Zhu}, L., {Peng}, Z., {McClellan}, J., {Li}, C., {Yao}, D., {Li}, Z., and
  {Fang}, L. (2019).
\newblock {Deep learning for seismic phase detection and picking in the
  aftershock zone of 2008 M$_{w}$7.9 Wenchuan Earthquake}.
\newblock {\em Physics of the Earth and Planetary Interiors}, 293:106261.

\bibitem[Zhu and Beroza, 2018]{Zhu18}
Zhu, W. and Beroza, G.~C. (2018).
\newblock {PhaseNet: a deep-neural-network-based seismic arrival-time picking
  method}.
\newblock {\em Geophysical Journal International}, 216(1):261--273.

\bibitem[{Zhu} et~al., 2015]{Zhu15}
{Zhu}, X., {Vondrick}, C., {Fowlkes}, C., and {Ramanan}, D. (2015).
\newblock {Do We Need More Training Data?}
\newblock {\em arXiv e-prints}, page arXiv:1503.01508.

\end{thebibliography}

\label{lastpage}

\end{document}


\label{firstpage}

\maketitle

\section{Loss curves}
\label{app:loss}
\begin{figure*}
    \centering
    \includegraphics[width=\textwidth, keepaspectratio]{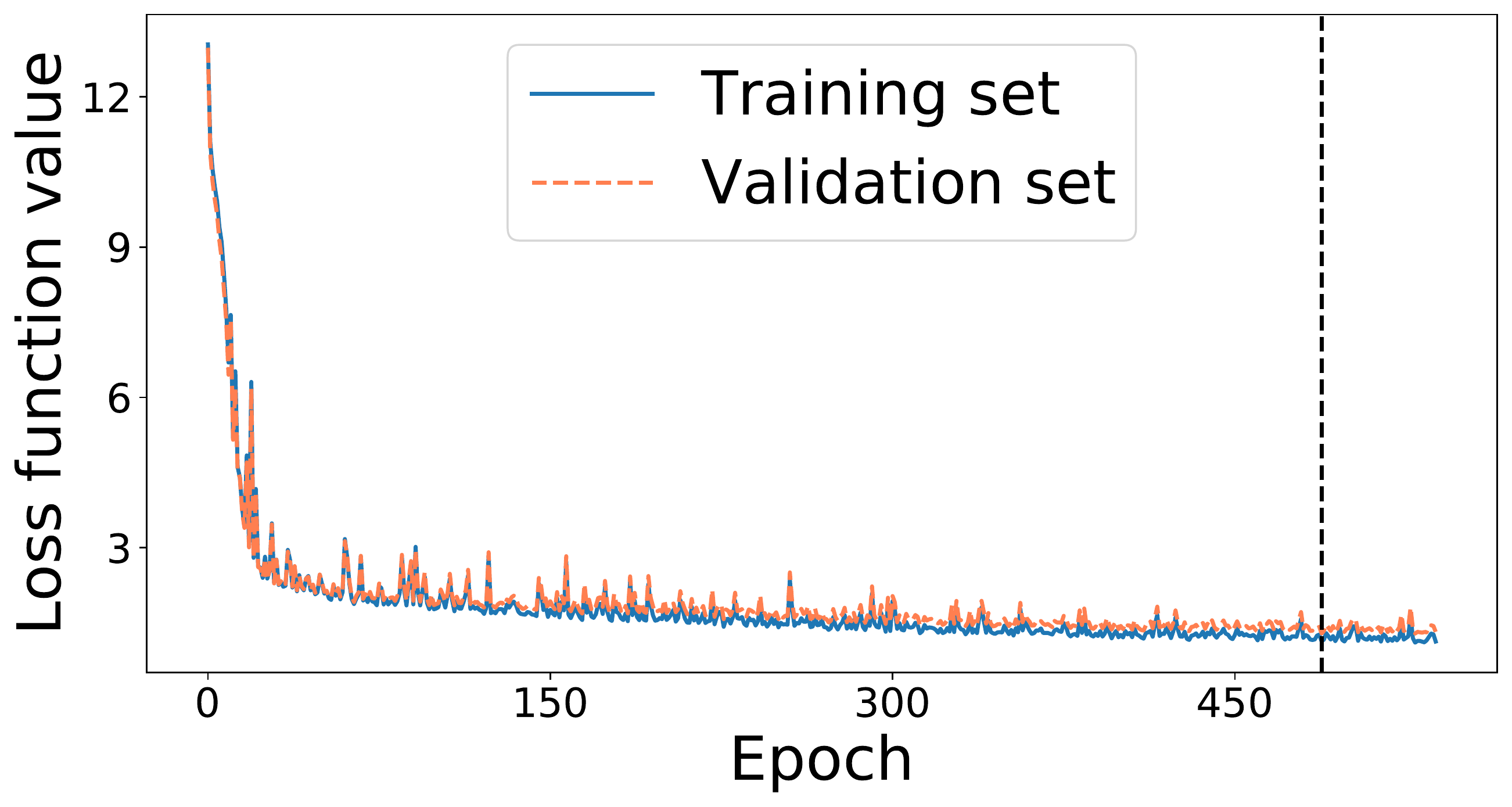}
    \caption{Typical example of training (solid blue) and validation (dashed coral) loss curves for one receiver in the CLVD case. The vertical dashed black line indicates the epoch with minimum validation loss, which we interpret as corresponding to the best model.}
    \label{fig:loss}
\end{figure*}

In Fig.~\ref{fig:loss} we report the training and validation loss curves for one receiver in the CLVD case. Note this is a typical example, as the trend in the loss curves is similar for all receivers and all source types.

\section{Additional studies on the posterior distribution}
\label{app:more_studies}
\begin{figure*}
$\begin{array}{cc}
    \includegraphics[width=0.41\textwidth]{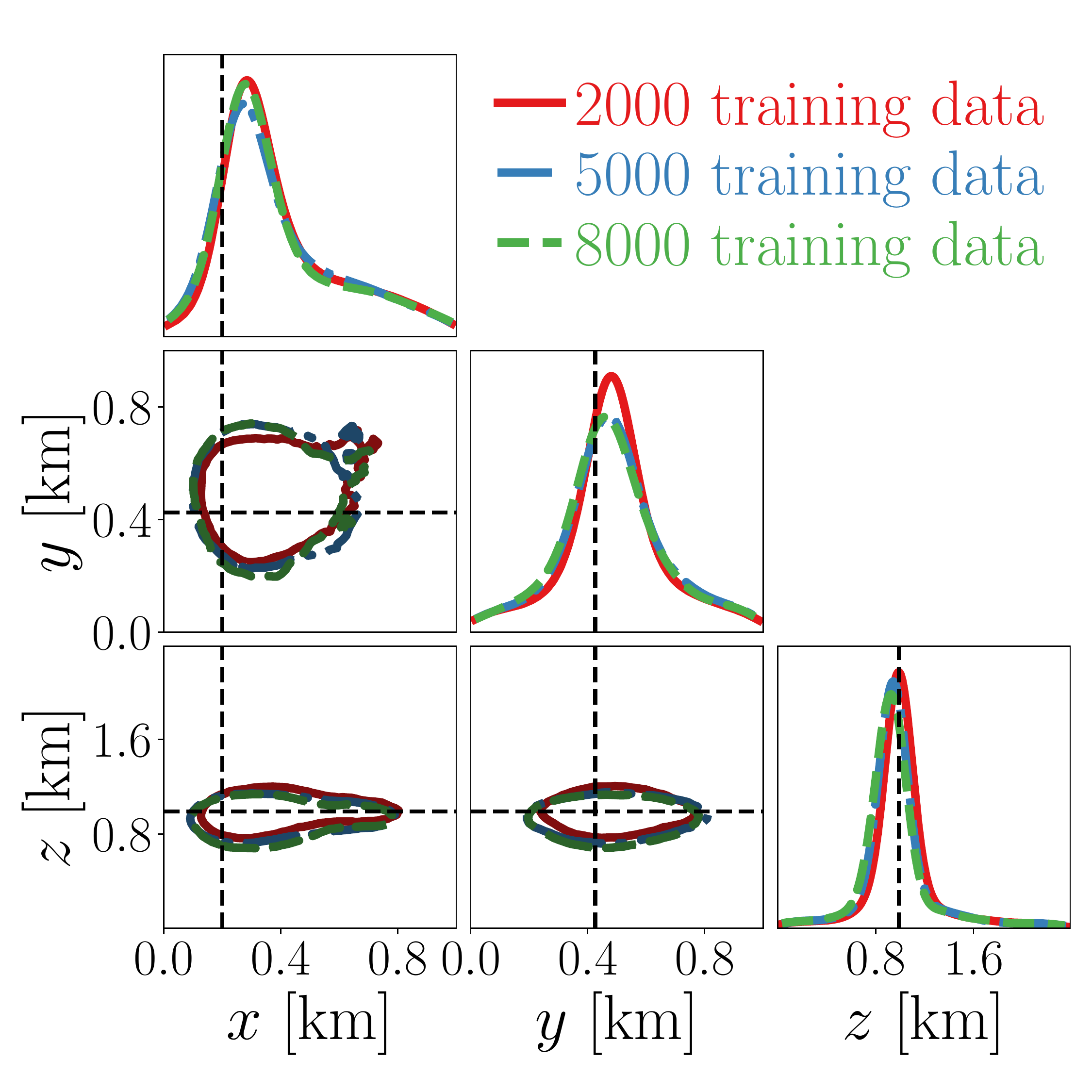} &
    \includegraphics[width=0.41\textwidth]{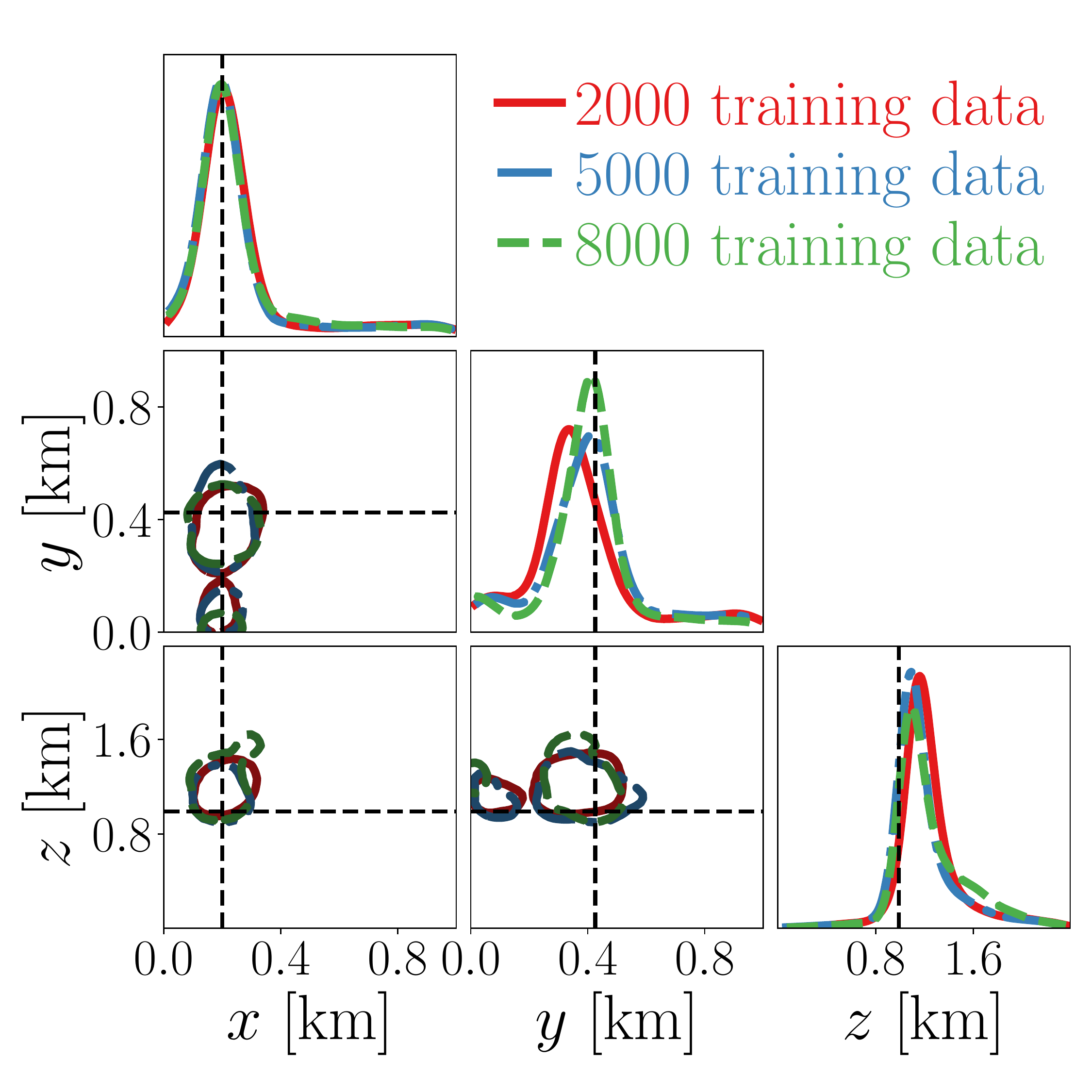}   \\
    \rm{ISO}& \rm{DC} 
\end{array}$
$\begin{array}{c}
    \includegraphics[width=0.41\textwidth]{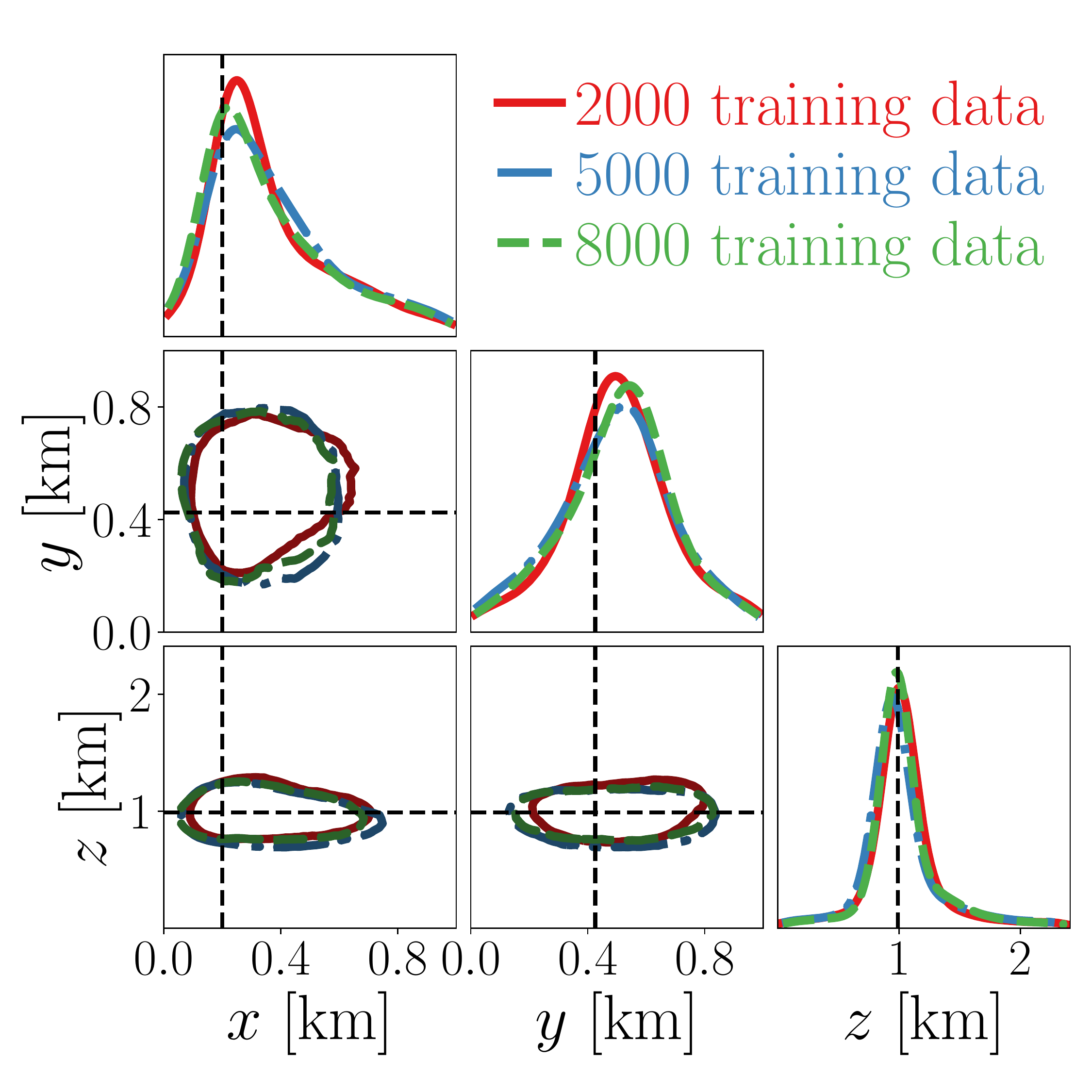}   \\
    \rm{CLVD}
\end{array}$
\caption[]
{\label{fig:compare_training}Marginalised 68 per cent credibility contours obtained with our method for a source located at $(x,y,z)=(0.20 \ \rm{km},0.43  \ \rm{km},0.99 \ \rm{km})$, indicated by the dashed black lines, comparing different numbers of training data for the emulator (2000, 5000 and 8000). We show these results for 3 source mechanisms: isotropic (ISO), double couple (DC), and compensated linear vector dipole (CLVD). Note that we are considering 23 receivers and a signal-to-noise ratio of 33 dB. This figure is best viewed in colour.}
\end{figure*}

\begin{figure*}
$\begin{array}{cc}
    \includegraphics[width=0.41\textwidth]{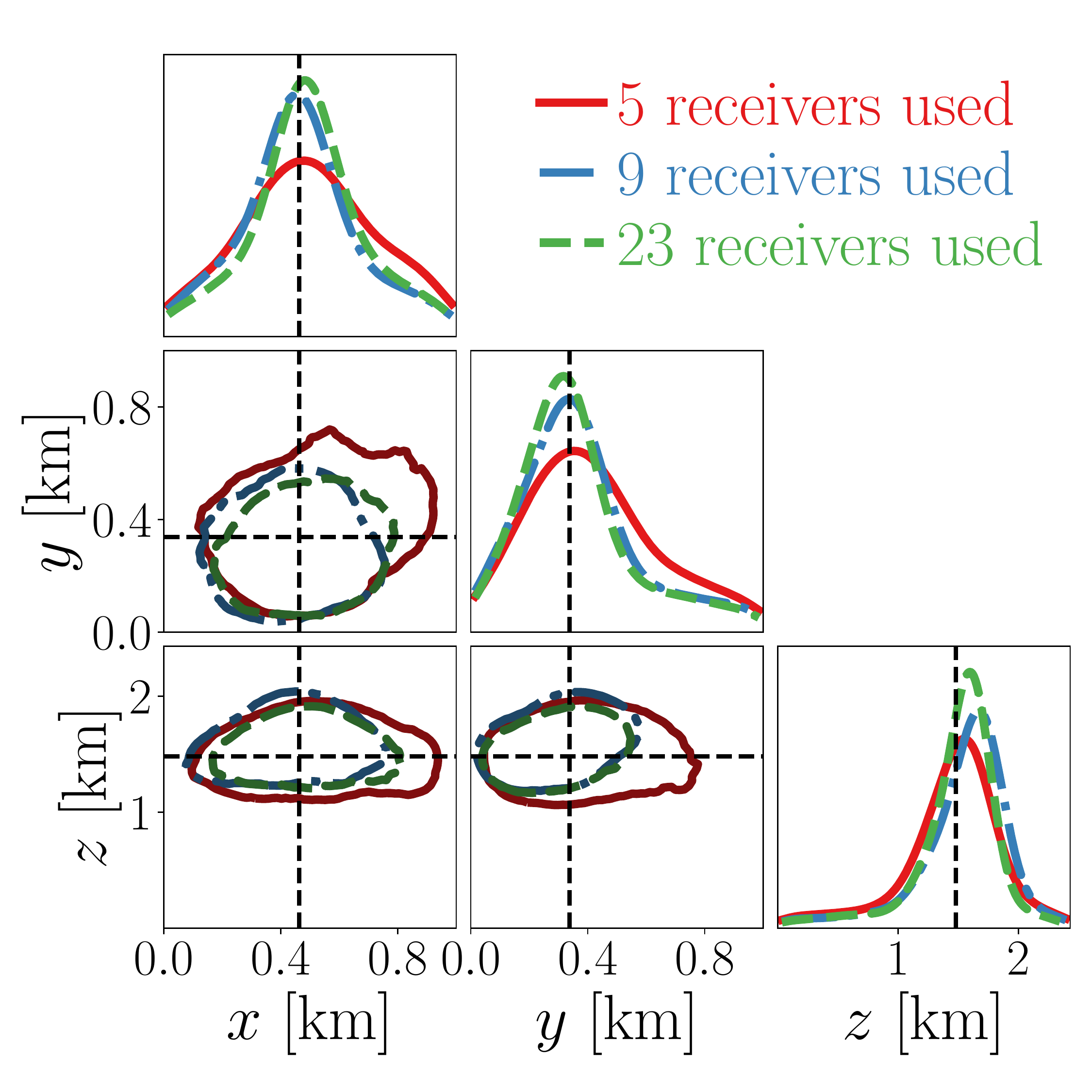} &
    \includegraphics[width=0.41\textwidth]{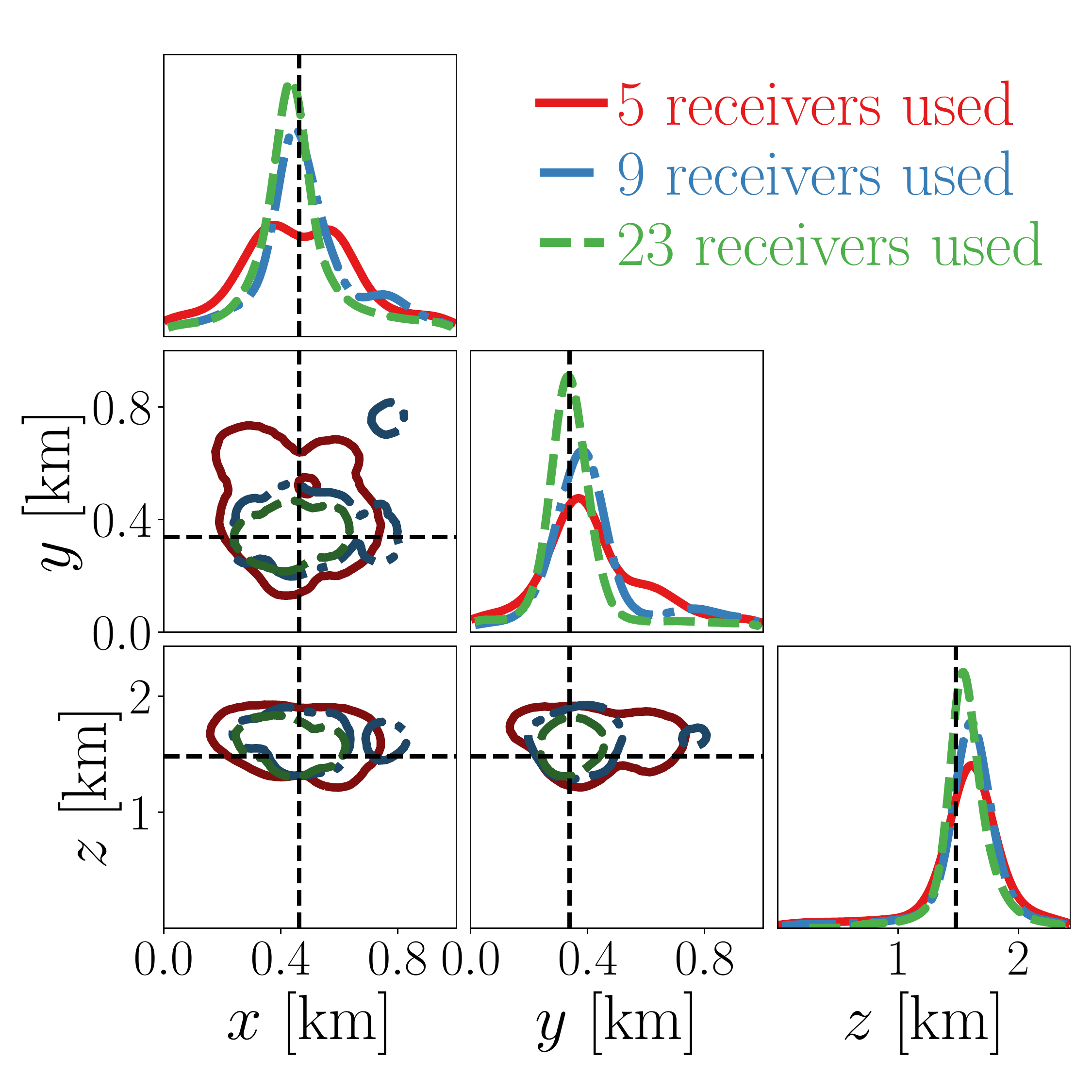}   \\
    \rm{ISO}& \rm{DC} 
\end{array}$
$\begin{array}{c}
    \includegraphics[width=0.41\textwidth]{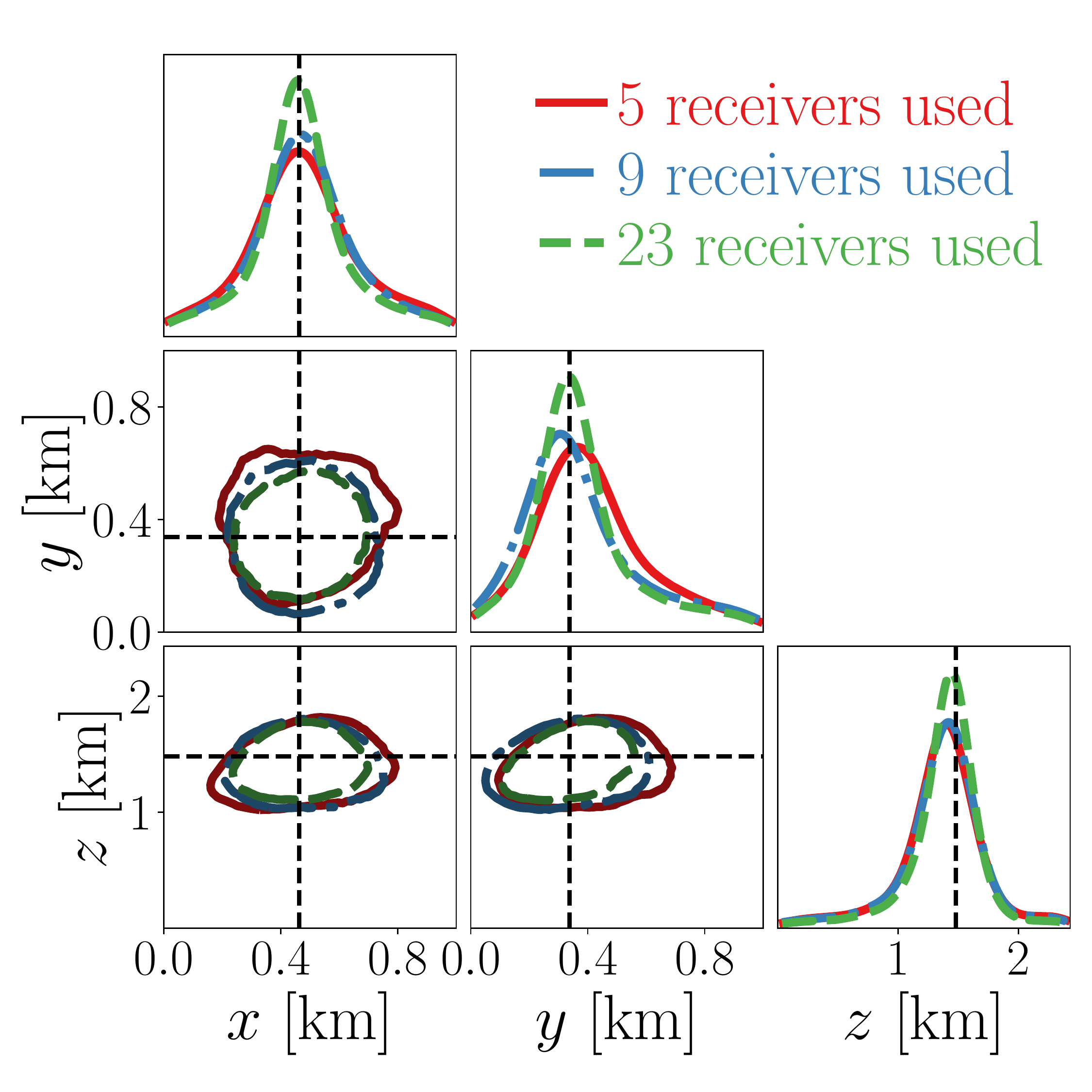}   \\
    \rm{CLVD}
\end{array}$
\caption[]
{\label{fig:compare_recs}Marginalised 68 per cent credibility contours obtained with our method for a source located at $(x,y,z)=(0.46 \ \rm{km},0.34 \ \rm{km},1.48 \ \rm{km})$, indicated by the dashed black lines, comparing different dispositions of the receivers. In particular, 5 receivers refer to the blue squares in Fig.~2 in the main text, and 9 receivers refer to the green circles in Fig.~2 in the main text. We show these results for 3 source mechanisms: isotropic (ISO), double couple (DC), and compensated linear vector dipole (CLVD). Note that we are considering 8000 training data for the emulator and a signal-to-noise ratio of 33 dB. This figure is best viewed in colour.}
\end{figure*}

In this section, we show complementary experiments to those presented at the end of Sect.~4.3 in the main text. In order to demonstrate the robustness of our inference pipeline, we show the dependence of the posterior distribution retrieved with our method as a function of the number of training points (Fig.~\ref{fig:compare_training}) and of the number of receivers (Fig.~\ref{fig:compare_recs}). We further comment on these results at the end of Sect.~4.3 in the main text.

\label{lastpage}